\documentclass[sigconf,nonacm]{acmart}

\usepackage[utf8]{inputenc}
\usepackage{subcaption}
\usepackage{mathtools} 
\usepackage{fancyvrb}
\usepackage{listings}
\usepackage{enumitem}
\usepackage{textcomp}
\usepackage[toc,page]{appendix}

\usepackage{amssymb}

\AtBeginDocument{%
  }

\copyrightyear{2025} 
\acmYear{2025} 
\setcopyright{acmlicensed}\acmConference[CHI '25]{Proceedings of the CHI Conference on Human Factors in Computing Systems}{April 26--May 1, 2025}{Yokohama, Japan}
\acmBooktitle{Proceedings of the CHI Conference on Human Factors in Computing Systems (CHI '25), April 26--May 1, 2025, Yokohama, Japan}
\acmDOI{}
\acmISBN{}

\begin{document}

\title{AIdeation: Designing a Human-AI Collaborative Ideation System for Concept Designers}

\thanks{This is a preprint of the paper accepted at CHI 2025. The final version will be available in the ACM Digital Library.}

% A Human-AI Collaborative Ideation System for Concept Designers

% Augmenting Pathologists with NaviPath:Design and Evaluation
% of a Human-AI Collaborative Navigation System

\author{Wen-Fan Wang}
\email{vann@cmlab.csie.ntu.edu.tw}
\orcid{0009-0001-1050-1170}
\affiliation{%
  \institution{National Taiwan University}
  \city{Taipei}
  \country{Taiwan}
}

\author{Chien-Ting Lu}
\email{B09902109@csie.ntu.edu.tw}
\orcid{0009-0000-8863-1277}
\affiliation{%
  \institution{National Taiwan University}
  \city{Taipei}
  \country{Taiwan}
}

\author{Nil Ponsa Campanyà}
\email{R12944063@ntu.edu.tw}
\orcid{0009-0009-7947-4834}
\affiliation{%
  \institution{National Taiwan University}
  \city{Taipei}
  \country{Taiwan}
}

\author{Bing-Yu Chen}
\email{robin@ntu.edu.tw}
\orcid{0000-0003-0169-7682}
\affiliation{%
  \institution{National Taiwan University}
  \city{Taipei}
  \country{Taiwan}
  }

\author{Mike Y. Chen}
\orcid{0000-0001-5410-652X}
\email{mikechen@csie.ntu.edu.tw}
\affiliation{
  \institution{National Taiwan University}
  \streetaddress{No. 1, Sec. 4, Roosevelt Rd.}
  \city{Taipei}
  \country{Taiwan}
  }

\renewcommand{\shortauthors}{Wang Lu Ponsa Chen Chen}

%% Generate the CCS concept using the tool at http://dl.acm.org/ccs.cfm
%===copy and replace the code generated from the website directly!===%


\begin{abstract}
  Concept designers in the entertainment industry create highly detailed, often imaginary environments for movies, games, and TV shows. Their early ideation phase requires intensive research, brainstorming, visual exploration, and combination of various design elements to form cohesive designs. However, existing AI tools focus on image generation from user specifications, lacking support for the unique needs and complexity of concept designers' workflows. Through a formative study with 12 professional designers, we captured their workflows and identified key requirements for AI-assisted ideation tools. Leveraging these insights, we developed AIdeation to support early ideation by brainstorming design concepts with flexible searching and recombination of reference images. A user study with 16 professional designers showed that AIdeation significantly enhanced creativity, ideation efficiency, and satisfaction (all \textit{p}<.01) compared to current tools and workflows. A field study with 4 studios for 1 week provided insights into AIdeation's benefits and limitations in real-world projects. After the completion of the field study, two studios, covering films, television, and games, have continued to use AIdeation in their commercial projects to date, further validating AIdeation's improvement in ideation quality and efficiency.
\end{abstract}


\begin{teaserfigure}
  \includegraphics[width=\textwidth]{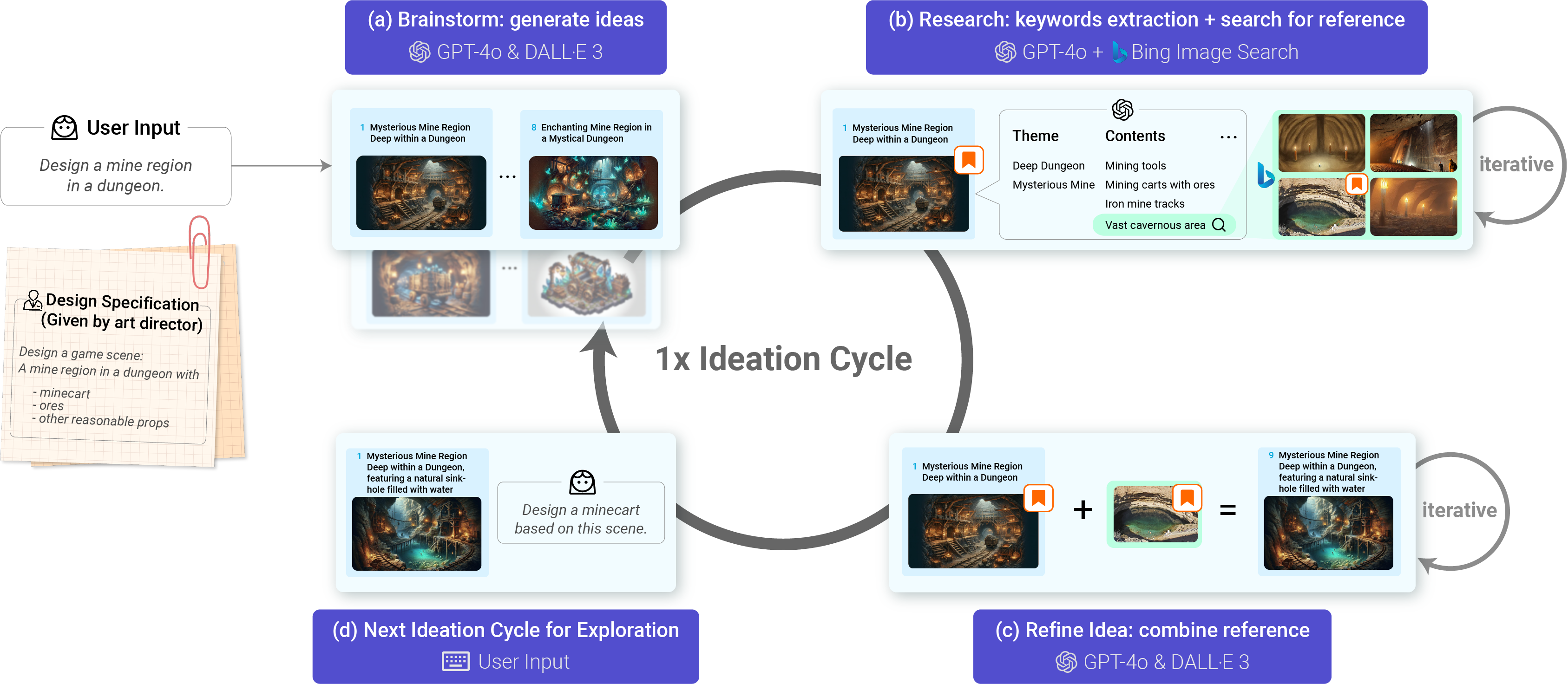}
  \caption{AIdeation, an ideation tool designed to support concept designers in exploring ideas in both breadth and depth, with flexible iterative refinement. The figure illustrates the ideation cycle using a real-world project example from our field study, showcasing AIdeation's key design components. The process starts with the designer’s input and moves through the ideation cycle: (a) Breadth by Brainstorming: AIdeation generates a variety of ideas based on the input; (b) Depth by Research: AIdeation provides keywords extracted from the design ideas, helping the designer understand the key elements of the generated concepts. The Designer can click on keywords to search for relevant references, enhancing their understanding of the elements; (c) Idea Refinement: The Designer iteratively refine the idea by combining searched references or giving instructions. (d) Once the desired result is achieved, the designer can begin a new ideation cycle using new input building on the current idea.}
  \Description{The figure illustrates AIdeation’s ideation cycle through a real-world project example, showing how designers explore and refine ideas. The process starts with a user input, "Design a mine region in a dungeon," followed by a design specification listing elements like a minecart and ores. The cycle consists of four stages: (a) Brainstorming, where AIdeation (GPT-4o and DALL·E 3) generates multiple design ideas displayed in image cards with titles such as "Mysterious Mine Region Deep within a Dungeon." (b) Research and Information Support, where AIdeation extracts key themes and contents like "Deep Dungeon" and "Mining Tools," and enables reference searches via Bing Image Search. (c) Idea Refinement, where designers refine concepts by combining references or giving instructions, represented by an equation-like structure merging a generated image with a reference. (d) Next Ideation Cycle, where the refined concept leads to new input, such as "Design a minecart based on this scene." The cycle is labeled "1x Ideation Cycle" in the center, with arrows indicating iterative refinements between research and refinement. Different background colors and icons visually distinguish each stage.}
  \label{fig:hero image}
\end{teaserfigure}


\maketitle

\section{INTRODUCTION}
\label{sec:introduction}

Concept design is the initial step in visual development within the entertainment industry, including films, TV shows, and video games~\cite{adobe2020, artofgame2008, bigbadWorld2015, randomguidebook2023, gameworkshop2018, levelup2014}.
Concept designers collaborate with art directors to visualize core ideas through initial sketches to
define the aesthetic and visual tone of the projects~\cite{artofgame2008}, along with detailed character, environment, and prop design to provide blueprints for computer graphics (CG) and set construction teams~\cite{gameworkshop2018, levelup2014}.
For fantasy and sci-fi settings, everything needs to be designed and created by concept designers~\cite{adobe2020}. 

The concept design workflow consists of two main phases~\cite{bigbadWorld2015, conceptart2018, randomguidebook2023}: 1) \textit{early ideation (or blue sky) phase}, where raw ideas are researched, brainstormed, and explored, followed by the 2) \textit{final concept phase}, where approved concepts are refined with detailed specifications for use by the production teams.
Literature has described several challenges during the early ideation phase~\cite{bigbadWorld2015, 80lv2020, artstation2021, conceptart2018}, including difficulties in finding references that align with their creative vision~\cite{son2024genquery, li2022analyzing, bigbadWorld2015} combined with extreme time constraints with designers being expected to deliver multiple designs daily~\cite{bigbadWorld2015, 80lv2020, artstation2021}, resulting in limited depth of research~\cite{bigbadWorld2015, conceptart2018} and limited unique designs incorporating diverse elements explored (i.e. limited breadth)~\cite{bigbadWorld2015, 80lv2020, interview2020}. 
%Futhermore, Inefficient
%communication between directors and designers, whether due to vague instructions or the challenge of translating
%complex ideas into visual formats, often results in multiple revisions, further complicating the process
%translating complex ideas into visual formats that meet the , often require in multiple revisions, and ~\cite{80lv2020, bigbadWorld2015, interview2020}.

With the rapid advancement of generative AI (GenAI), concept designers and studios have adopted them into their workflows~\cite{boucher2024resistance, ko2023large, filmhandbook2024, vimpari2023adapt, qin2023does}. Recent studies highlight challenges faced by artists and designers in integrating GenAI into their workflows. Crafting effective prompts often requires significant trial and error, as users struggle with optimal structures~\cite{mahdavi2024ai}. Additionally, many tools lack intuitive design, posing barriers for non-technical users~\cite{shi2023understanding}. GenAI's single-step generation contrasts with the iterative, reflective practices of human creativity~\cite{zhang2024confrontation}, and outputs often misalign with users’ visions, necessitating extensive fine-tuning and post-editing~\cite{mahdavi2024ai, vimpari2023adapt}. However, recent research in HCI shows promise in better integrating GenAI into traditional creative workflows, such as graphic design~\cite{choi2024creativeconnect, son2024genquery}, animation ~\cite{Tseng2024KeyframerEA}, fashion design~\cite{davis2024fashioning, rw29}, and interior design~\cite{wang2024roomdreaming, hou2024c2ideas}. Motivated by the success of these advancements, this work aims to develop a flexible system that leverages concept designers' domain knowledge, focusing on facilitating iterative ideation and enhancing their workflows.

We structure this work into four parts: 1) a formative study to understand the workflow, ideation processes, and needs of concept designers using both traditional methods and AI tools; 2) the design and implementation of a human-AI collaborative ideation system; 3) a summative study focusing on the ideation process; and 4) a field study in real-world commercial projects to assess its quality and efficiency of the final design outcomes.

We conducted a formative study with 12 professional environment concept designers. Among character, prop, and environment concept designs, we focused on environment concept design for the scope of this paper because it typically requires the most concept design resources and involves designing complex spatial and visual elements across both large (macro) and detailed (micro) scales. Through in-depth interviews and workflow analyses, we examined participants' design processes, reference-gathering strategies, and their use of GenAI tools.
Designers often struggle to gather diverse, relevant references, especially for unique or poorly defined topics. Traditional search tools often do not align with the creative intentions of designers or provide sufficient material to blend different styles and themes.
Additionally, generating multiple unique design variations under tight deadlines is a major obstacle, with participants citing limited time, insufficient reference diversity, and the extensive effort required to create complex designs. While GenAI tools offer potential, participants noted critical limitations, including difficulties in formulating prompts, low diversity in outputs, AI hallucinations, lack of detailed information, and limited flexibility for refining results. These findings highlight the need for tools that better support concept design workflows' iterative and exploratory nature.

Based on our observations, we designed AIdeation to bridge the gap between GenAI and concept design, enhancing the early ideation phase of concept designers. The key components of AIdeation include: 
\begin{enumerate}
    \item \textbf{Brainstorming: Supporting Breadth Exploration:} AIdeation generates a wide variety of diverse design ideas based on user input, which can be in the form of natural language, images, or both (Figure \ref{fig:hero image}-a). These design ideas are presented visually, providing an overview that helps designers quickly grasp different directions while offering high-diversity variations for further exploration. 
    \item \textbf{Research: Supporting Depth Exploration:} AIdeation extracts key design elements from the generated ideas and presents them as keywords to help designers explore further into the visual elements. These keywords also link to corresponding search results, supporting the reference-gathering process (Figure \ref{fig:hero image}-b).
    \item \textbf{Refining Idea: Supporting Flexible Iterative Exploration:} AIdeation allows designers to refine their ideas through an iterative process (Figure \ref{fig:hero image}-c). Users can refine their designs by combining them with additional references or issuing natural language instructions to adjust specific elements. This flexibility helps designers experiment with both broad and focused refinements, aligning design ideas with their creative intent while maintaining design diversity.
\end{enumerate}

% Evaluation and Result
To evaluate AIdeation, we conducted a summative study focusing on the ideation process and a field study to examine its impact in real-world settings and the final design outcomes. The summative study employed a within-subjects design with 16 professional environment concept designers, using their original workflow as the baseline. The study simulated real-world tasks where designers were assigned topics involving both exterior and interior scenes. Findings showed that participants significantly preferred AIdeation for enhancing creativity (\textit{p} = 0.001), found it more efficient for generating diverse ideas (\textit{p} = 0.003) while maintaining comparable quality, and reported higher satisfaction (\textit{p} = 0.005) and enjoyment (\textit{p} = 0.005) with AIdeation.

%Designers were asked to explore visuals and gather references based on the topic, using either their original workflow or AIdeation. There were no restrictions for the baseline; designers could use any preferred tools, including traditional or AI-based tools. They were required to collect reference sets for both interior and exterior scenes to present to the client and support future design work. Findings from our study showed that compared to their original workflows, participants significantly preferred AIdeation for enhancing breadth (\textit{p} = 0.014) and depth (\textit{p} = 0.033) of idea exploration, flexibility (\textit{p} = 0.046), and creativity (\textit{p} = 0.001). They also found AIdeation more efficient for generating diverse ideas (\textit{p} = 0.003) while maintaining comparable quality. Furthermore, participants reported higher satisfaction (\textit{p} = 0.005), greater enjoyment (\textit{p} = 0.005), and reduced task difficulty (\textit{p} = 0.005) with AIdeation, as well as improved support for gathering design information (\textit{p} = 0.009) and visually presenting ideas (\textit{p} = 0.004). While some participants highlighted limitations, such as restricted control over specific design modifications or slower image generation, AIdeation demonstrated the potential to improve ideation workflows and support creative processes effectively.

% field Study
For the field study, we collaborated with 4 design studios and 8 professional environment concept designers, who used AIdeation as part of their ongoing commercial projects for one week. 
All studios reported improved creativity, with 3 reporting improved efficiency and quality.
%Two studios reported significant efficiency gains: a Metroidvania game designer reduced their workload from 14 to 6 days, while a AAA studio designer completed a task in 2 days instead of 5. 
After the completion of the field study, 2 studios have continued using AIdeation for commercial projects to date.
%, with one studio reporting a significant quality boost based on feedback from their director.

%Participants from two studios reported a significant boost in efficiency; for instance, a designer working on a Metroidvania game reduced their workload from 14 days to 6 days, while another designer at a AAA game studio completed a task in 2 days instead of 5. In other cases, such as a matte painting project, AIdeation showed limitations, slightly increasing the time required due to challenges with aesthetics and spatial representation. Despite this, the tool demonstrated potential for early ideation in real-world scenarios, with designers expressing satisfaction and an interest in incorporating it into future projects.

% Contribution
In summary, our key contributions are as follows:
\begin{itemize}
    \item An in-depth understanding of concept designers' workflows in the early ideation stage and how AI design tools are currently used in practice. 
    \item The design and implementation of a system, AIdeation, that allows concept artists to rapidly explore creative ideas through a flexible, iterative approach. Designed with a human-centered AI process, it addresses key barriers to GenAI adoption, including AI factual accuracy and lack of transparency/creative control.
    \item Empirical evidence that AIdeation improves creativity, satisfaction, and efficiency of concept artists' workflow through: 1) a comparative study with 16 professional concept designers; 2) a field deployment to production use in 4 studios; and 3) continued usage to date by two studios after the completion of the field study.
\end{itemize}
%We d
%AIdeation bridges the gap between concept designers and AI design tools by incorporating domain knowledge and offering an intuitive design interface, making it easy to use and enhancing the designer's workflow in real-world projects.

% > 1AC:
% > Insufficient context of related work to differentiate contribution from previous work. See individual reviews for specific areas of weakness.
% > 2AC:    
% > The related work section covers the main themes of the paper, but it feels a little bit surface level, a list of related references, with no particular discussion, with the exception of section 2.3 where they do explore a few specific examples. It might be useful to explore some other examples of AI that are tailored to specific human workflows, even if they are outside the specifics of this setting, it would help the authors identify and set-out key concerns.

% In our work, we aim to support concept designers' ideation workflows, explore design ideas, supporting their design workflow through AI-enhanced tools. To this end, we review related work in (1) the existing concept of the design exploration process, (2) AI approaches for design idea exploration, and (3) AI tools that support and integrate into the design workflow.%  

%Insufficient context of related work to differentiate contribution from previous work.
%Emphasized More on Human-AI Collaboration, how AIdeation better support workflow compare to other AI tools
%Find more related works, add more discussion in related works.

% -----------------NEW PART1-----------
\section{RELATED WORK}
We aim to integrate GenAI into the ideation process of concept designers and enhance their workflows. To achieve this, we reviewed related work in three key areas: (1) ideation within the design process, (2) GenAI tools that support visual exploration and ideation for designers, and (3) human-centered approaches for integrating AI into workflows.

\subsection{Idea Exploration Process of Designers} %fundamental or traditional
% Concept designers as well as any other creative professionals utilize a variety of methods to explore and refine ideas during the idea exploration process.

Like many other creative professionals, concept designers engage in an iterative process throughout their ideation workflows ~\cite{adams1999cognitive}. The process starts with divergent thinking, where the designer explores various possibilities and generates diverse ideas without the burden of constraints~\cite{rw5,rw8,rw9,rw10,Imagination}. During this stage, designers conduct intensive visual exploration ~\cite{rw11,rw12}, accumulate a collection of references ~\cite{rw17}, and organize in reference boards ~\cite{rw18}. This visual process encourages designers to absorb visual elements, inspiring their future designs ~\cite{linsey2011experimental}. Similarly to concept design, in some other design fields, such as architecture ~\cite{newland1987understanding}, product design ~\cite{boston1998design}, and interactive design ~\cite{park1993empirically}, not only do these fields rely on visual references, but they also require extensive research to gather factual knowledge and data. A previous study highlights research methodologies tailored for designers, emphasizing the potential of integrating research into the iterative creative process ~\cite{Navarro2022Research}. 
Both visual exploration and research serve as core sources of inspiration ~\cite{eckert2000sources}, fostering innovation and preventing design fixation ~\cite{rw4, rw21}.

Once a variety of ideas are generated, convergent thinking helps designers identify the most effective solution ~\cite{rw6}. During this phase, designers utilize the resources collected earlier to sketch the evolving idea on paper ~\cite{rw14, rw15, rw16}. They continuously evaluate and iteratively refine their ideas, explore different aesthetics, and ensure clear communication with stakeholders until a satisfactory result is achieved ~\cite{johnson1997analysis, Stamps1999Demographic, Stigliani2018The}. 

Numerous studies have proposed frameworks based on similar concepts to support the iterative process, such as the Wizard of Oz approach ~\cite{dow2005wizard} and Muse ~\cite{muller2013muse}. AIdeation integrates these insights to enhance concept designers' design process, supporting flexible divergent and convergent thinking while bridging designers with the latest GenAI tools that preserve the core elements of creativity and exploration.
% ------------------------------------

% -----------------NEW PART2-----------
\subsection{GenAI as a Catalyst for Visual Exploration and Ideation}

% comments
% I think we should put general GenAI tool likes Image generation model(Like Dall-E) first. Talk about how those basic model effect ideation process. And then discuss why those basic model cannot integrate into designer's ideation workflow. Then we talk about some related works that try to integrate GenAI into workflow or boosting ideation process.
%let's see if this is well written.

With the advancement of GenAI tools, many design domains have already integrated them into creative processes ~\cite{ko2023large, qin2023does}. Designers and artists extensively use general image generation tools to transform text prompts into visuals ~\cite{Epstein2023Art, Rick2023Supermind}. However, these tools are not specifically adapted to designers' creative process ~\cite{boucher2024resistance, vimpari2023adapt}.
Recent research has increasingly focused on enhancing user experience with image-generation tools. Reprompt ~\cite{wang2023reprompt} automatically refines the text prompts for the generated images.
Promptify ~\cite{rw23}, PromptCharm~\cite{PromptCharm2024} and DesignPrompt~\cite{DesignPrompt2024} introduce interactive prompt refinement to improve text-to-image generation workflows. IntentTuner ~\cite{zeng2024intenttuner} combines fine-tuning and generation functionalities to support a flexible workflow for text-to-image generation. StyleFactory~\cite{zhou2024stylefactory}
facilitates style alignment in image creation. DreamSheets~\cite{almeda2024prompting} enables users to explore the relationship between input prompts and image outputs through a spreadsheet interface. Collectively, these tools reduce the burden on designers to craft intricate prompts and help generate visuals that better align with their intentions.

Additionally, recent research explores the potential of GenAI by closely examining designers' needs during the ideation process. Researchers designed systems and user interfaces specifically to address the challenges they face.
For visual exploration, GenQuery ~\cite{son2024genquery} addresses the challenge of reference search by supporting expressive visual searches and enabling iterative refinement of image-based queries. 
CreativeConnect ~\cite{choi2024creativeconnect} streamlines the traditionally time-consuming process of recombining references by providing automated suggestion options.  C2Ideas ~\cite{hou2024c2ideas} assisted interior designers in generating color schemes aligned with user intentions. 
For ideation, DesignAID ~\cite{rw22} and MuseTree\footnote{MuseTree, https://www.asus.com/proart/software-solutions/musetree/} combat creative blocks by using large language models (LLM) to deliver diverse prompts and generate visuals. These systems effectively integrate GenAI to address specific challenges in traditional ideation processes across various domains. 

Recent works have explored new possibilities for human-AI collaboration in creativity. A study found AI can foster novel communication, with designers curating and refining generated images ~\cite{DesigningChiou2023}. Optimuse ~\cite{OptiMuse} aligns with designers' nonlinear creative processes and proposes a human-AI co-design framework that supports iterative idea exploration and flexible communication. COFI ~\cite{rezwana2023designing} advocates for AI systems that balance divergent and convergent process, and calls for expanding AI's creative roles beyond generation and evaluation to include conceptual exploration. These works mentioned above provide valuable insights to integrate GenAI into the creative process, such as optimizing user experience, designing tools to address specific challenges, and exploring models of human-AI collaboration.

\subsection{Human-Centered AI for Workflow Support}%our range is outside the designers, find more papaers about AI.h Human centered AI focused collaboration workflow

With advancements in AI, human-centered AI (HCAI) has emerged as a crucial approach to enhance human abilities by fostering collaboration between humans and AI systems. It emphasizes a symbiotic relationship where AI tools enhance human capabilities and streamline workflows in various domains ~\cite{Venigandla2024Hybrid, shneiderman2022human, xu2023transitioning}. 
In alignment with user needs, these systems amplify human expertise while ensuring transparency and explainability, helping users understand the decisions and limitations of AI ~\cite{Ehsan2021Expanding, Kim2024Establishing}. Through effective communication, iterative feedback, and user control, these systems create dynamic collaborations to enhance workflows ~\cite{Hois2019How, Scharowski2023Exploring, Usmani2023Human-Centered}.

% Human-centered AI has emerged as a pivotal approach to fostering collaboration between humans and machines. It focus on fostering a symbiotic relationship between humans and AI, where AI tools amplify human capabilities and enhance experts' workflows across a wide range of domains~\cite{Venigandla2024Hybrid, shneiderman2022human, xu2023transitioning}.
% Human-centered AI enhances workflows by amplifying human expertise through systems designed to align with user needs. Transparency and explainability are crucial~\cite{Ehsan2021Expanding}~\cite{Kim2024Establishing}, ensuring users understand AI decisions and its limitations. Effective communication and iterative feedback create a dynamic collaboration, while user control and customization allow systems to adapt to specific workflows~\cite{Hois2019How}~\cite{Scharowski2023Exploring}. These principles ensure seamless and symbiotic integration, fostering trust and usability across diverse domains~\cite{Usmani2023Human-Centered}.
%
%usability, why sucess? becuse human ai guidelines 
% Overview: User Centered Workflow: Integration AI into workflow "consideration", following the workflow
% Typical Cases
% https://scholar.google.com/scholar?hl=en&as_sdt=0%2C5&q=human+centered+AI+system&btnG=

Recent research has applied these principles across various fields. In the creative industry, researchers have delved deeply into domain knowledge and workflows of different design disciplines, crafting systems thoughtfully tailored to align with user workflows ~\cite{Anantrasirichai2020Artificial, Knearem2023Exploring, Mccormack2020Design}. For example, 
RoomDreaming ~\cite{wang2024roomdreaming} generates photorealistic interior design alternatives and enables the user to clearly understand and iteratively refine their options, allowing designers to work collaboratively with their clients. 
MemoVis ~\cite{rw20} enables feedback providers to create companion reference images for 3D designs with real-time viewpoints, democratizing actionable feedback regardless of 3D expertise. Both works reduce the communication time between clients and designers.
PlantoGraphy ~\cite{PlantoGraphy2024} integrates iterative design processes into landscape rendering, offering users control and flexibility to better align with their unique workflows. 
Keyframer ~\cite{Tseng2024KeyframerEA} uses a natural language interface to make motion design intuitive and accessible, fostering a feedback loop that allows animators to explore and refine ideas with creative autonomy. 
In addition to these works, researchers have developed GenAI systems for fashion ~\cite{rw29}, UX and industrial design ~\cite{claytoplay2024}, and 3D scene design ~\cite{oh2024lumimood}. 
These works enhance design workflows by leveraging GenAI to reduce repetitive tasks, providing intuitive user interfaces that foster system understanding and enabling precise control to refine design outputs.
% Reframer ~\cite{lawton2023drawing} facilitates real-time user-AI collaboration in drawing, highlighting the importance of balancing fine-grained control with emergent creativity.

Research beyond the design field also offers valuable insights for developing human-centered AI systems. In medicine, NaviPath ~\cite{navipath2023} uses AI models to simplify the navigation of high-resolution tumor images, aligning with pathologists' workflows by enabling smooth transitions between low to high magnifications. It allows customization of AI recommendations and provides clear explanations, enhancing user engagement and improving overall accuracy. 
In aviation, the AI Support System for Pilots’ Decision-Making Process ~\cite{pilot} highlights the importance of transparent feedback loops, enabling pilots to understand AI recommendations. Its real-time guidance and customization enhance decision-making, safety, and efficiency, especially under information overload. Both approaches emphasize user control and refining AI contributions to effectively augment human expertise.

% With all the contributions of these works in mind, it was clear that we needed to create a system that allowed AI to be integrated to the concept designers' workflow. Thus we developed Aideation, a tool specifically designed for concept designers. Aideation enhances exploration, broadens and enhances the first stages of ideation, and provides the flexibility required to meet the complex and dynamic demands of concept designers' creative workflows. A system that allows a high degree of controllability, transparency and dynamic interaction between the Human and the AI. 

While many studies demonstrate how GenAI can enhance creative processes and design workflows, no tool fully addresses the complexities of concept designers' workflows. Concept design requires specialized support throughout the iterative process, from research and brainstorming to refining ideas. These threads of work offer valuable inspiration for developing AIdeation, paving the way for a solution tailored to the unique needs of concept designers.

\section{BACKGROUND: WORKFLOW OF THE ENTERTAINMENT INDUSTRY AND CONCEPT DESIGNERS}
The entertainment industry's production process, whether for films, TV shows, or video games, transforms creative ideas into final products through a series of four stages: 1) \textit{development}, where the initial concept and creative direction are set; 2) \textit{pre-production}, involving detailed planning and preparation; 3) \textit{production}, where the main content is created; and 4) \textit{post-production}, which includes editing, enhancing, and polishing of the final product~\cite{gameworkshop2018, bigbadWorld2015, directing2020, Singh2023Artificial, filmmaker2019}.

\begin{figure*}
    \centering
    \includegraphics[width=1\linewidth]{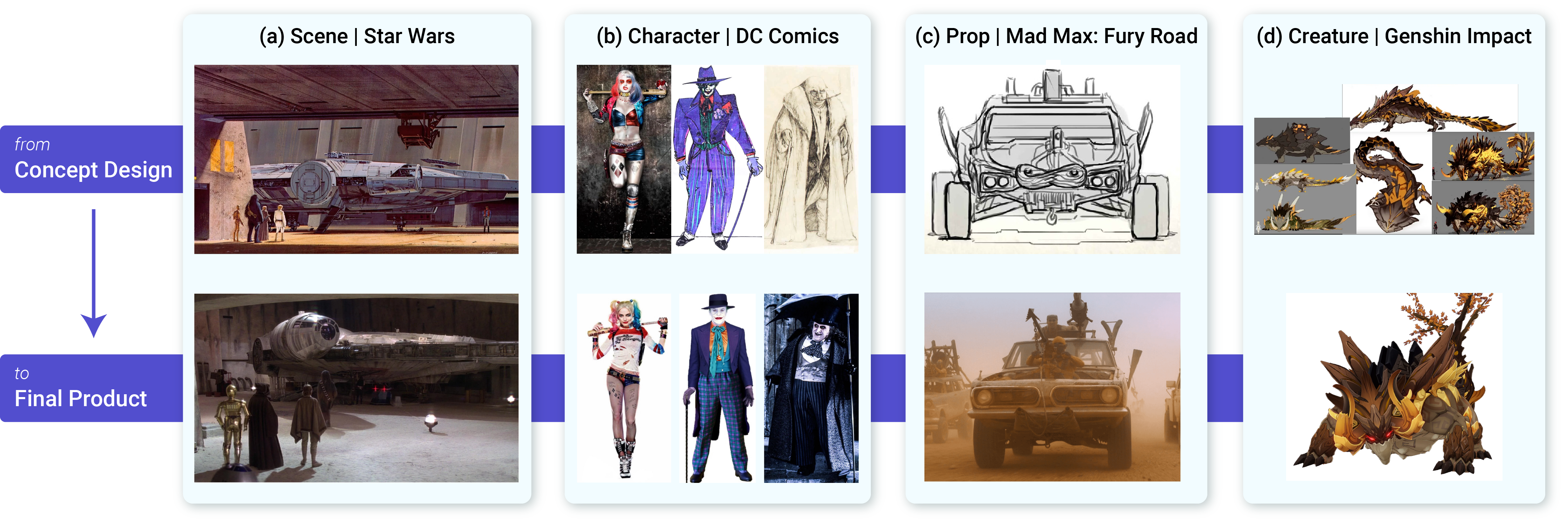}
    \caption{The figure showcases designs from concept to final product, including four well-known projects: (a) a scene from Star Wars, (b) characters from DC Comics (Harley Quinn, the Joker, and the Penguin), (c) a prop from Mad Max: Fury Road, and (d) a creature from Genshin Impact. This demonstrates the critical role of concept designers in shaping the creative vision from the earliest production stages to the final product}
    \Description{The figure illustrates the transformation from concept design to final production through four examples. A vertical arrow labeled "from Concept Design to Final Product" connects the two stages. (a) A scene from Star Wars shows an early illustration of a hangar with a spaceship, followed by a final film still of the same scene. (b) DC Comics characters, including Harley Quinn and the Joker, are depicted in initial sketches and concept art, transitioning to their final movie appearances. (c) A prop from Mad Max: Fury Road is shown as a vehicle design sketch, with the final version appearing in a film still. (d) A creature from Genshin Impact progresses from multiple concept sketches to a fully rendered in-game model. The figure highlights how concept designers shape creative vision from early ideas to production.}
    \label{fig:importance}
\end{figure*}

Concept designers are pivotal across the first three stages, particularly in the \textit{pre-production} stage. During \textit{development}, concept designers collaborate with art directors/clients to visualize core ideas through initial sketches and designs to define the project's aesthetic and visual tone~\cite{artofgame2008}. During \textit{pre-production}, they design scenes, characters, environments, and props to provide blueprints for computer graphics (CG) and set construction teams~\cite{gameworkshop2018, levelup2014}. During \textit{production}, their work ensures consistency as concepts are translated into tangible assets~\cite{bigbadWorld2015, randomguidebook2023}. Figure \ref{fig:importance} shows actual examples of concept designs that led to their final products in several well-known movies and games.

Concept designers undertake the majority of their work in \textit{pre-production} stage, with the workflow consisting of the following two phases~\cite{conceptart2018, bigbadWorld2015, randomguidebook2023, iterationandreference2023}:
\begin{enumerate}
    \item \textbf{Early ideation (or blue sky) phase:} 
    % After receiving a design specification from the art director/client (Figure~\ref{fig:workflow}-a), the designer research and gather information, brainstorm design ideas (Figure 3-b), explores visuals, gathering reference images from sources such as Pinterest\footnote{Pinterest, https://www.pinterest.com/}, search engines, or portfolio websites like Artstation\footnote{Artstation, https://www.artstation.com/} (Figure 3-c) that match their creative vision and generating preliminary sketches that explore various options, maintaining consistency across different settings while presenting ample variations~\cite{skillful2005}. These early concepts and references are shown to the director/client for feedback (Figure 3-d).

    This phase focuses on brainstorming and exploring initial ideas. Designers research the topic, perform visual searches, brainstorm ideas, and create preliminary sketches to propose creative options for feedback from art directors or clients. If they are not satisfied with the results, designers iterate the process until the direction of the concept is approved.
    %designer researches and explores visuals (Figure 3-b), gathering reference images (Figure 3-c) aligned with their creative vision and creating rough sketches exploring different possibilities, ensuring consistency across various environments while offering sufficient variation. These initial concepts and references are presented to the director or client for feedback (Figure 3-d).

    \item \textbf{Final concept phase:} Once initial concepts are approved, designers refine the sketches into detailed and polished designs. They enhance chosen concepts with depth, texture, and fine details to align with the project's vision. Approved final designs serve as comprehensive guides for the \textit{production} teams, which are realized through 3D modeling or set construction. The designer may provide ongoing support to ensure consistency throughout production~\cite{levelup2014}.
\end{enumerate}

This work focuses on the early ideation phase, establishing the project's creative vision and shaping its direction, style, and coherence~\cite{80lv2020,adobe2020,randomguidebook2023}. This stage demands intensive creativity and is often seen as the most exciting part of the workflow~\cite{iterationandreference2023,rassa2018concept}. 

% In early ideation phase, concept designers receive a briefing of design specification from art director/client, such as a description of what the project, the scene of the keyframe is about, and a set of references that fit their intention ~\cite{80lv2020}. Designers then analyse and disassemble the brief, into the core of the assignment, and what directions they should think of taking it in. Based on this, designers get into the \textit{ideation cycle}. They start to \textit{research} the topics extracted from the brief. During this step, they read about the subject, gather information, visual searching for a lot of image references to be knowledgable about the topic~\cite{iterationandreference2023}. When the designers consider having enough information and references gathered, they start to \textit{brainstorm} the idea. They write down the design elements they want to includes into their design, and then rough sketch multiple variation of the design ~\cite{skillful2005}. The research and brainstorming steps are often 交錯的 and iterative, the designer may think of the design idea when gathering reference, and they may want to gathering more information when they brainstorm some new idea. After multiple ideation cycle, the designer may be satisfied with few ideas (normally one to five) they come up with. They then 稍微 refine those idea對應到的sketches, and arrange the reference the sketch 參考的, as a set of design idea, to make those design idea 可以被清楚理解. and present to the director/client~\cite{80lv2020}. 

\begin{figure*}
    \centering
    \includegraphics[width=1\linewidth]{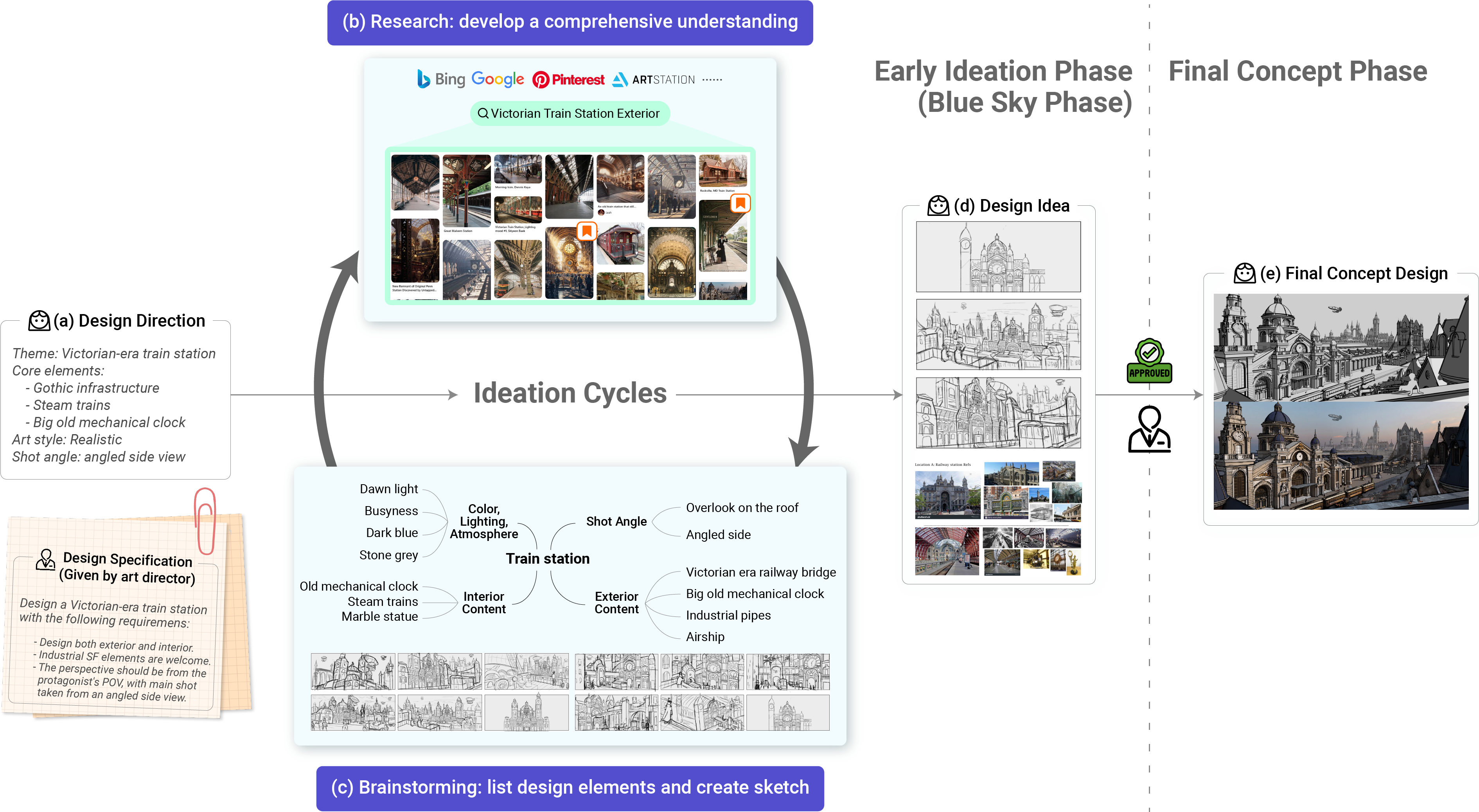}
    \caption{A typical workflow for an environment concept designer begins with receiving the design specification from the art director or client. The designer then (a) determines a potential design direction and enters the iterative ideation cycle, which includes (b) researching based on the task, and (c) brainstorming innovative ideas. Once some suitable design ideas are formed, (d) both sketches and references are presented to art directors or clients for feedback. Upon approval, (e) they refine the sketch into a polished, detailed design, which is then shared with other teams, such as the CG team.}
    \Description{The figure illustrates the iterative workflow of an environment concept designer, progressing from initial specifications to a final concept design. (a) The process begins with a design direction, defining a Victorian-era train station with key elements like Gothic infrastructure, steam trains, and a large mechanical clock. A design specification lists detailed requirements, including interior and exterior views and an angled perspective. (b) Research follows, with online searches for reference images to develop a comprehensive understanding. (c) Brainstorming organizes design elements, considering aspects like lighting, materials, and architectural details, with initial sketches. (d) Design ideas, including sketches and references, are reviewed for approval. (e) Once approved, the final concept design is refined with detailed rendering, transitioning from an outlined sketch to a fully developed environment ready for collaboration with other teams. The workflow is structured as iterative ideation cycles, emphasizing research, feedback, and refinement.}
    \label{fig:workflow}
\end{figure*}

In the early ideation phase, concept designers receive a design specification briefing from the art director or client, which includes a project description, keyframe scene details, and a set of relevant references~\cite{80lv2020}. Designers analyze the brief to identify its core elements and potential design directions (Figure \ref{fig:workflow}-a), then begin the \textit{ideation cycle}. The cycle starts with \textit{research}, where designers study the subject, gather information, explore visuals, and collect image references to develop a comprehensive understanding (Figure \ref{fig:workflow}-b)~\cite{iterationandreference2023}. This step ensures that future designs are coherent, such as maintaining historical accuracy, aligning with the period's style, or achieving mechanical and structural feasibility. With sufficient references and information, designers move to \textit{brainstorming}, where they list design elements and create rough sketches with multiple variations (Figure \ref{fig:workflow}-c)~\cite{skillful2005}. Research and brainstorming often intertwine as designers refine ideas while gathering references or seeking new material when generating fresh concepts. This iterative process helps designers gradually develop and refine their designs. After several ideation cycles, designers complete a small set of ideas they find most suitable (typically one to five), polishing the sketches and organizing the corresponding references into cohesive design ideas (Figure \ref{fig:workflow}-d). These finalized ideas are then presented to the art directors or clients for feedback~\cite{80lv2020}. Designers may repeat multiple ideation cycles until the art directors or clients are satisfied with the direction of the concept. Once approved, the process transitions to the final concept phase (Figure \ref{fig:workflow}-e).

% 細節的Workflow Stage在這邊講清楚
% Design spec generate很多idea, user可以做出更多detail research
% 以前是要分開去research再去combine，現在一次做完。
% 符合Sepc 做完research，然後去Design
% 現在的workflow是這樣: ---

\section{FORMATIVE STUDY}
We conducted a formative study to gain deeper insights into current concept designers' workflows and the challenges they face using traditional and AI-based ideation tools.
% In this work, we decided to focus on environment concept design as it involves complex spatial and visual elements that demand high levels of creativity, consistency, and attention to detail. 
\subsection{Participants}
We recruited 22 professional environment concept designers (15 males, 7 females; ages 23 to 45) across three studies. Each participant was assigned a unique ID. Participants were recruited through personal referrals and directly contacting studios by email to request collaboration. Detailed participant information, including their participation in each study, is provided in Table ~\ref{tab:demographics}. We will highlight the relevant details of the participants in each study.

In the formative study, we worked with 12 environment concept designers (P1-P12) from various industries, including Animation (P1, P4, P6, P8-P9), Game (P5, P10-P11), Art Outsourcing (P3, P12), and Freelancing (P2), with 3 to 15 years of experience (mean = 7.7, SD = 4.6). Participation in the study was voluntary, and uncompensated.

\begin{table*}[h!]
\centering
\small
\begin{tabular}{|c|c|c|c|c|c|}
\hline
\textbf{ID} & \textbf{Years of Experience} & \textbf{Industry} & \textbf{Formative} & \textbf{Summative} & \textbf{Field Study} \\ \hline
1 & 3 & Animation, Films, TV shows & \checkmark & \checkmark & \checkmark \\ \hline
2 & 4 & Freelancing & \checkmark & \checkmark & \\ \hline
3 & 7 & Art Outsourcing & \checkmark & \checkmark & \checkmark \\ \hline
4 & 3 & Animation, Films, TV shows & \checkmark & \checkmark & \\ \hline
5 & 3 & Game & \checkmark & \checkmark & \checkmark \\ \hline
6 & 5 & Animation, Films, TV shows & \checkmark & \checkmark & \\ \hline
7 & 3 & Game & \checkmark & & \\ \hline
8 & 8 & Animation, Films, TV shows & \checkmark & & \\ \hline
9 & 12 & Animation, Films, TV shows & \checkmark & & \\ \hline
10 & 15 & Game & \checkmark & & \\ \hline
11 & 13 & Game & \checkmark & & \\ \hline
12 & 14 & Art Outsourcing & \checkmark & & \\ \hline
13 & 5 & Art Outsourcing & & \checkmark & \checkmark \\ \hline
14 & 8 & Game & & \checkmark & \\ \hline
15 & 12 & Game & & \checkmark & \\ \hline
16 & 3 & Animation, Films, TV shows & & \checkmark & \checkmark \\ \hline
17 & 2 & Animation, Films, TV shows & & \checkmark & \checkmark \\ \hline
18 & 2 & Animation, Films, TV shows & & \checkmark & \checkmark \\ \hline
19 & 1 & Animation, Films, TV shows & & \checkmark & \\ \hline
20 & 1 & Animation, Films, TV shows & & \checkmark & \\ \hline
21 & 5 & Freelancing & & \checkmark & \\ \hline
22 & 11 & Game & & \checkmark & \checkmark \\ \hline
\end{tabular}
\caption{Demographic Details of Participants}
\Description{This table provides detailed demographic information about the 22 professional environment concept designers who participated in the study. Participants were identified by a unique ID and categorized based on their years of experience, industry affiliation, and participation in the formative study, summative study, and field study phases. The participants had a wide range of professional experience, from early-career designers with 1 year of experience to seasoned professionals with up to 15 years in the field. Detailed participant breakdown: ID 1: 3 years of experience, works in Animation, Films, and TV Shows, participated in the formative study. ID 2: 4 years of experience, works as a Freelancer, participated in both formative and summative studies. ID 3: 7 years of experience, works in Art Outsourcing, participated in formative and summative studies. ID 4: 3 years of experience, works in Animation, Films, and TV Shows, participated in the formative study. ID 5: 3 years of experience, works in Game Design, participated in the formative study. ID 6: 5 years of experience, works in Animation, Films, and TV Shows, participated in the summative study. ID 7: 3 years of experience, works in Game Design, participated in the summative study. ID 8: 8 years of experience, works in Animation, Films, and TV Shows, participated in formative and summative studies. ID 9: 12 years of experience, works in Animation, Films, and TV Shows, participated in formative and summative studies. ID 10: 15 years of experience, works in Game Design, participated in the summative study. ID 11: 13 years of experience, works in Game Design, participated in formative and summative studies. ID 12: 14 years of experience, works in Art Outsourcing, participated in the summative and field studies. ID 13: 5 years of experience, works in Art Outsourcing, participated in formative, summative, and field studies. ID 14: 8 years of experience, works in Game Design, participated in formative and summative studies. ID 15: 12 years of experience, works in Game Design, participated in formative and summative studies. ID 16: 3 years of experience, works in Animation, Films, and TV Shows, participated in the summative and field studies. ID 17: 2 years of experience, works in Animation, Films, and TV Shows, participated in the formative study. ID 18: 2 years of experience, works in Animation, Films, and TV Shows, participated in the formative study. ID 19: 1 year of experience, works in Animation, Films, and TV Shows, participated in the formative study. ID 20: 1 year of experience, works in Animation, Films, and TV Shows, participated in the formative study. ID 21: 5 years of experience, works as a Freelancer, participated in formative and field studies. ID 22: 11 years of experience, works in Game Design, participated in formative and field studies.}
\label{tab:demographics}
\end{table*}

\subsection{Study Procedure}
Each participant took part in a 1-2 hour interview covering three main topics: 1) Their typical design workflow, 2) Past design projects, and 3) Current AI tool usage. We asked participants to prepare three specific projects: their most recent project, a typical project, and the most challenging project in their work experience. For each project, we explored the design task, the procedures they followed, and their overall approach. This included discussing the use of design tools, methods for research and brainstorming, reference materials collected for various design elements, presentation of sketches to directors or clients, and the challenges encountered throughout the process.

\subsection{Findings}
To analyze the data, we organized and summarized the transcribed interview recordings, and one of the authors, with prior experience as a professional concept designer, developed a coding framework to identify key themes for thematic analysis. Two art directors reviewed the coding framework from an animation studio and an art-outsourcing studio, each managing 15 and 40 concept designers, respectively.
Thematic analysis was discussed collaboratively among a team of three people to ensure consensus and validity. This process revealed patterns in concept designers’ research and brainstorming workflows, the purposes of the references they gathered, and the challenges they faced with traditional and AI design tools.

\subsubsection{Challenges during researching}

% RR
Our participants employed a variety of tools during their research. They used search engines like Google\footnote{Google, www.google.com} to gather information and chatbots like ChatGPT\footnote{ChatGPT, https://chatgpt.com/} to explore topics in depth. For initial visual exploration and reference gathering, they relied on online platforms such as Pinterest\footnote{Pinterest, www.pinterest.com}, portfolio websites like Artstation\footnote{Artstation, https://www.artstation.com/}, and image databases like Shutterstock\footnote{Shutterstock, https://www.shutterstock.com/}. Most participants mentioned that this process is straightforward when the briefing is clear, and the themes are familiar, like “\textit{cyberpunk streets}” (P1) or “\textit{Japanese shrines}” (P3). However, when specifications were vague, or the topic was less common—a frequent challenge in environment concept design—they reported greater difficulty in finding relevant information and references. These observations align with findings from previous literature~\cite{80lv2020, bigbadWorld2015, interview2020, iterationandreference2023}. 

This challenge usually arises from two main issues. First, designers often struggle to find search queries and references that align with their design intentions. “\textit{The client asked me to design an internet world for a celebrity, showcasing her popularity. I spent half a day trying keywords like ‘digital world,’ ‘internet world,’ and ‘matrix world’ on Pinterest, but found nothing suitable}” (P1). “\textit{Often, I remember having seen a similar reference before, but now I don't know how to find it}” (P2). Second, traditional search tools often fail to provide sufficient references for unique design topics. One participant noted, “\textit{I was asked to design a Grand Mayan market and a Mayan ballcourt, 80\% based on history. I couldn’t find any relevant design work, and the references on Pinterest were fragmented and lacked useful information}” (P4). Another added, “\textit{We frequently get tasks that require blending different styles and themes, but it's hard to find similar concept art online}” (P3).

\subsubsection{Challenges during brainstorming}
Concept designers often need to create 3–5 design variations per environment, a task that becomes challenging for uncommon designs \cite{bigbadWorld2015}. Most participants noted they typically have only “\textit{half}” (P6)  to “\textit{one}” (P12) day for idea exploration, leaving little time for deeper creative development. “\textit{I need to reserve the entire afternoon for sketching, leaving only the morning for research and exploring different possibilities}” (P4). Designers rely on visual exploration for inspiration, but time constraints and the challenges outlined in the previous section often limit their access to diverse references, restricting creative ideas and exploration. For example, “\textit{I was asked to design an Aztec village with three variations, but the architectural references I found on Pinterest all looked quite similar. With a tight schedule, the final designs I proposed ended up being somewhat alike}” (P9).

Designers spend significant effort in creating design variations, especially for complex design specifications. “\textit{A recent project involved designing a Chinese Steampunk world with realistic and plausible designs. With no existing references, I spent an entire day just sketching one building}” (P12). Generating innovative designs or integrating unique elements into a cohesive vision is another common challenge. “\textit{I often spend a great deal of time contemplating what elements to add to enhance the richness of the scene}” (P5). Also, they often have to try multiple combinations of design elements from references to create a suitable outcome. As one participant explained, “\textit{I often fill an entire A3 canvas with sketches to explore various possible design combinations}” (P2). Furthermore, within the same project, designers are usually tasked with creating multiple scenes within a shared setting, such as “\textit{creating various architectures and their interiors within the same game environment}”(P3).

\subsubsection{Problems with current AI design tools}
To address these challenges in traditional workflows, many studios and designers have started integrating GenAI into their processes~\cite{boucher2024resistance, ko2023large, filmhandbook2024, vimpari2023adapt}. In our study, all participants had experience using AI design tools, with 9 of them already integrating these tools into their workflow. The AI tools used included Midjourney\footnote{MidJourney, https://www.midjourney.com/}, Stable Diffusion\footnote{StableDiffusion, https://stablediffusionweb.com/}, DALL-E\footnote{Dall-E, https://openai.com/index/dall-e-3/}, and more advanced systems like Comfy UI~\cite{comfyui}. However, we found no consistent usage patterns. 6 participants (P1, P3-P4, P8, P11-12) primarily used these tools for image generation, formulating prompts based on concrete ideas and modifying them if the results did not align with their vision. Only 3 participants (P5, P9, P10) used the tools for ideation, providing simple inputs to explore topics. We identified several reasons why current AI design tools are not yet effective ideation tools for concept designers.

Most AI design tools, like Stable Diffusion and Midjourney, rely on text-based prompts that often require complex inputs, such as multiple keywords or lengthy descriptions ~\cite{mahdavi2024ai}. This contrasts with designers' typical workflow, which starts with simple keyword searches on platforms like Pinterest and progresses to image-based exploration. “\textit{As a concept designer, I don’t want to spend time crafting precise prompts}” (P2). Additionally, crafting a suitable prompt often requires a clear idea in advance, making it difficult to use during the initial ideation. “\textit{We usually use this tool to generate images only when we already have a clear idea in mind}” (P3). Many participants noted they struggled to create prompts that generated the desired outcomes. “\textit{I tried modifying the prompt in MidJourney several times, but I still couldn’t get what I wanted}” (P6). Furthermore, most image-generation AI tools struggle to produce diverse results from similar input, limiting their usefulness for breadth idea exploration. “\textit{I have to re-craft the prompt to get something noticeably different}” (P9). “\textit{I feel like everything the AI generates looks pretty much the same every time, similar compositions, styles, and often stereotypical elements}” (P1).

Concept designers require grounded and accurate information to support their designs. However, AI hallucinations pose a significant barrier, discouraging designers from adopting AI tools. “\textit{I usually avoid using AI-generated images as reference pictures because relying on incorrect content could lead to even worse outcomes}” (P6). Additionally, AI design tools often do not provide enough information in the generated outputs. A common issue is the lack of detail, particularly in the structure of objects, making it difficult for designers to identify visual elements for further reference. As one participant noted, “\textit{The content generated by AI is usually only useful to me as a mood reference because the details are often a complete mess}” (P3). Although some tools offer detailed prompts based on simple inputs, they can be challenging to interpret, such as “\textit{AI-generated images often include some interesting elements, but I don’t know what they are. As a result, I can’t incorporate them into my design}” (P1). Moreover, the generated images often do not align accurately with the prompts, “\textit{AI-generated images often include additional elements that are not specified in the prompt}” (P10).

The iterative process is key to achieving a great design for concept designers~\cite{iterationandreference2023}. However, AI design tools often lack the control and flexibility needed to refine output after generation. As one participant noted, “\textit{I only wanted to change the style of one building, but the entire image ended up changing}” (P12). Another shared, “\textit{The generated results often make me question how my changes to the prompt are actually affecting the outcome}” (P3). Moreover, the lack of detailed information accompanying AI-generated images hinders further ideation, “\textit{The AI-generated images contain many visual elements, but without information about them, I don’t know how to modify or adjust them}” (P5).

These factors combined make current AI image-generation tools difficult to use for visual idea exploration and challenging to integrate into a concept designer's workflow.

\begin{figure*}
    \centering
    \includegraphics[width=1\linewidth]{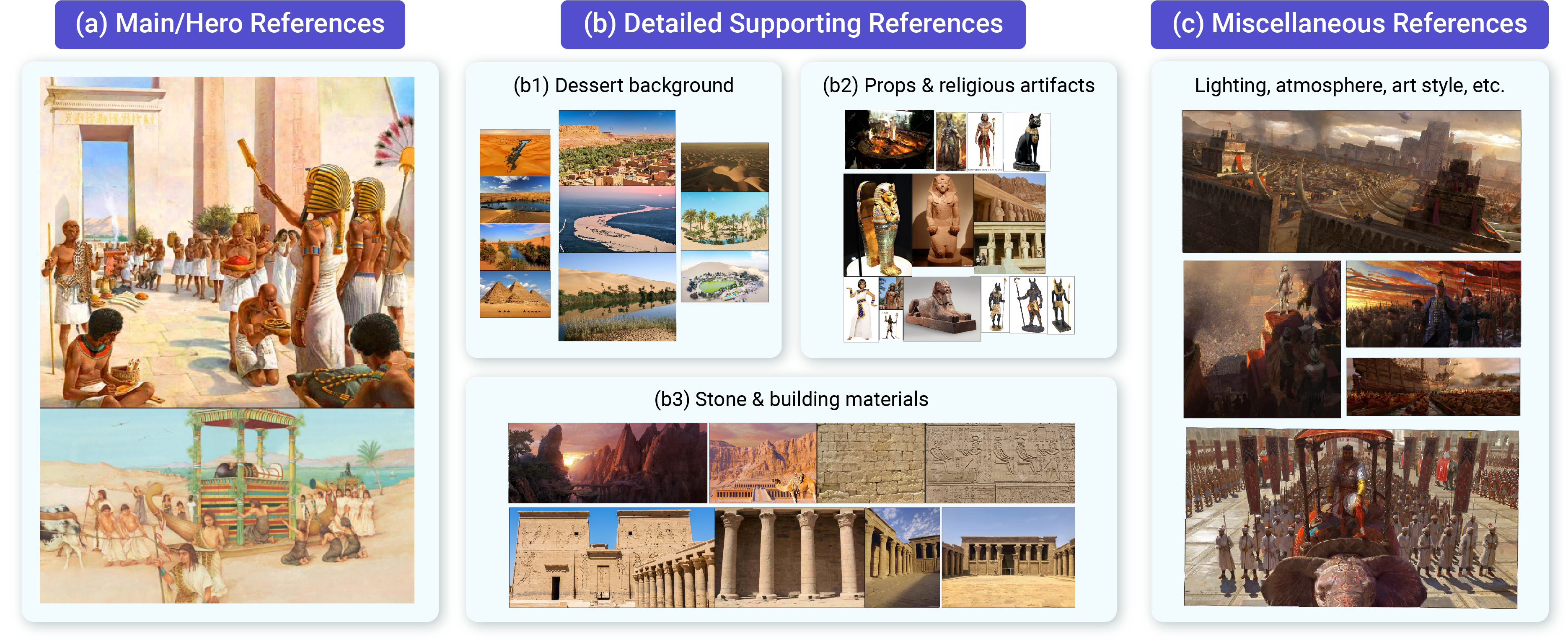}
    \caption{Based on our formative study, concept designers categorize references into three types: (a) Hero (or Main) References: These align closely with the designer's creative vision, conveying the overall story, mood, or design, and are crucial for guiding the project. (b) Detailed Supporting References: These provide specific details, like structure or texture, helping designers accurately implement finer aspects of the design. (c) Miscellaneous References: These cover a range of purposes, including lighting, atmosphere, and color palette, supporting various design elements.}
    \Description{The figure categorizes design references into three types. (a) Main/Hero References depict key visual inspirations that establish the story, mood, or overall design direction, represented by historical Egyptian-themed illustrations. (b) Detailed Supporting References provide specific elements, including (b1) desert backgrounds with landscapes and oases, (b2) props and religious artifacts such as statues and clothing, and (b3) stone and building materials with examples of ancient architecture and textures. (c) Miscellaneous References cover aspects like lighting, atmosphere, and composition, illustrated with cinematic and artistic depictions of grand historical scenes. This classification helps designers structure their reference gathering process effectively.}
    \label{fig:type}
\end{figure*}

\subsubsection{Type of references collected for environment concept design}
Environment concept designers gather diverse reference sets tailored to specific tasks, each serving different purposes. Designers also have unique ways of sourcing and organizing references. To understand these patterns, we analyzed reference frequency and collaborated with designers, identifying the following categories:
\begin{itemize}
    \item \textbf{Hero (or Main) Reference:} These references closely reflect the designer’s creative intent, aligning with the design theme and serving as a guide for establishing the overall mood, shapes, and composition of the design (Figure \ref{fig:type}-a). 
    \item \textbf{Detailed Supporting Reference:} These references provide specific detailed contents that support the design of the project. Typically, photographs provide specific details, such as mechanical structures or architectural features, offering accuracy and reliability for refining intricate design aspects  (Figure \ref{fig:type}-b).
    \item \textbf{Miscellaneous Reference:} Designers often collect references like lighting, atmosphere, art style, color palette, composition, and shot angle to enhance their designs based on project needs. These references, guided by the project's goals or the designer's vision, are not always essential and are categorized as Miscellaneous References. (Figure \ref{fig:type}-c).
\end{itemize}

\subsection{Design Goals}
Based on our findings, we proposed three design goals to better support concept designers during the ideation stage:
\begin{itemize}
    \item \textbf{DG1: Breadth exploration:} To help designers efficiently explore a wide range of ideas and gain a comprehensive understanding of the design topic, the system should support the brainstorming of various ideas using input methods that align with their workflow. This could include allowing users to input natural language instructions, such as task specifications, or directly upload relevant references.
\end{itemize}
\begin{itemize}
    \item \textbf{DG2: Depth exploration:} The system should offer detailed information and references to help designers refine and expand their design solutions while deepening their understanding of both the generated ideas and the design task. Moreover, the provided information should align with the designers' usual reference-gathering practices.
\end{itemize}
\begin{itemize}
    \item \textbf{DG3: Flexible iterative exploration:} The system should allow users to refine design ideas while maintaining control easily. It should support the efficient exploration of variations on the same theme to ensure consistency and creative flow. Designers should be able to narrow or expand the design space as needed, enhancing the creative process.
\end{itemize}

\begin{figure*}[htbp]  
    \centering
    \includegraphics[width=0.8\textwidth]{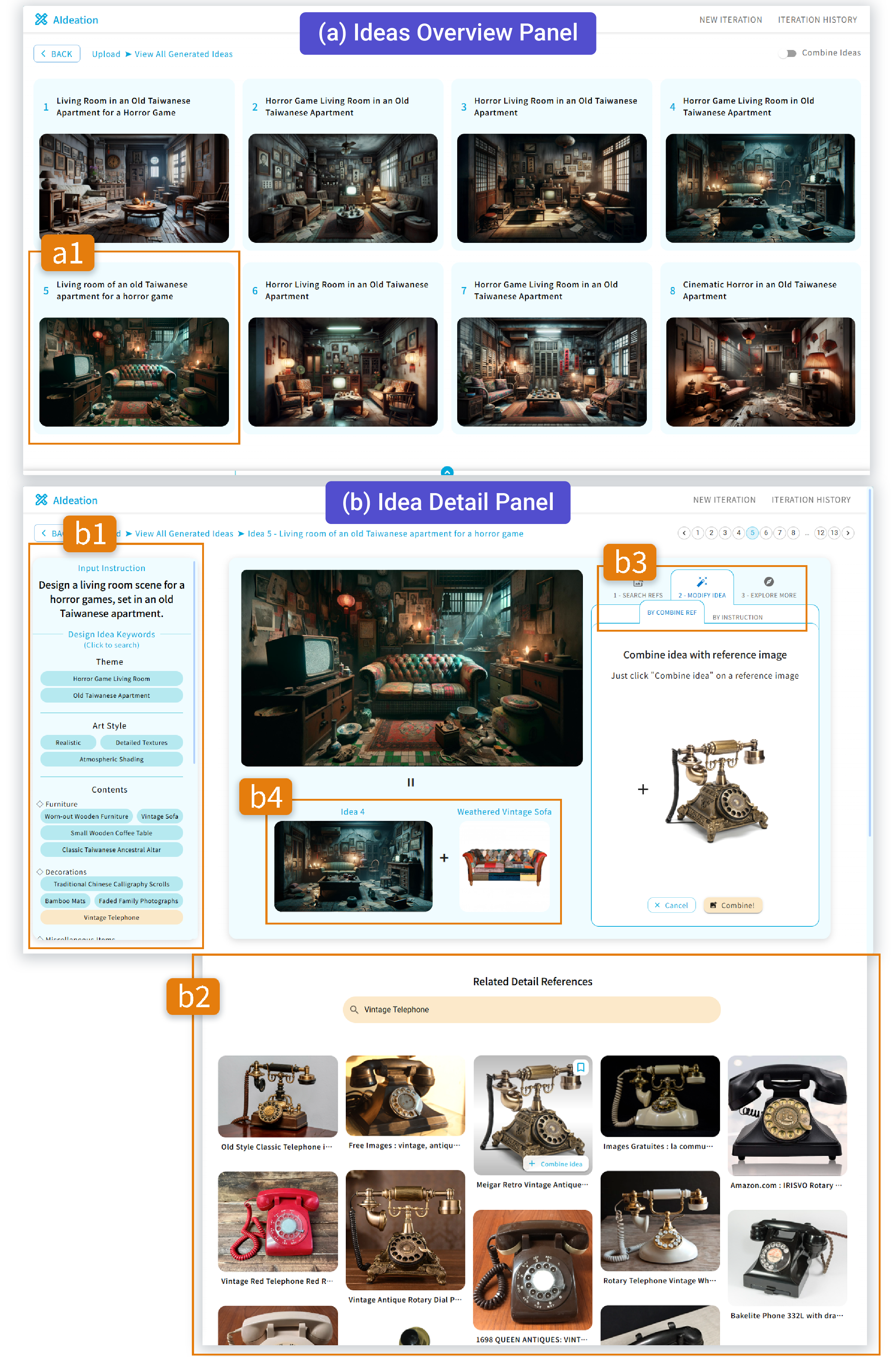} % Changed \linewidth to \textwidth
    \caption{The main interface of AIdeation includes (a) the Ideas Overview Panel, displaying all brainstormed design ideas as images with titles based on user input. Users can select an idea to view in (b) the Idea Detail Panel, which provides detailed information on the selected idea. (b1) The left sidebar lists key elements extracted from the idea, categorized into six groups as keywords. Users can select a keyword to view related search results in (b2). (b3) The right panel allows users to refine the idea by combining it with references or by instruction. (b4) Below the current idea, its origin is shown; in this case, the idea was generated by combining "Idea 4" and a colorful sofa.}
    \Description{The figure presents the AIdeation interface, divided into two main sections. (a) The Ideas Overview Panel displays multiple AI-generated design ideas as images with titles. Users can select an idea (a1) to explore further. (b) The Idea Detail Panel provides more information on the selected idea. (b1) The left sidebar lists extracted key elements, categorized into six keyword groups. Clicking a keyword retrieves related search results (b2), such as antique telephones. (b3) The right panel allows users to refine the idea by combining it with references or modifying it by instruction. (b4) Below the current idea, its origin is shown, demonstrating that this idea was created by merging "Idea 4" with a weathered vintage sofa. The interface supports iterative ideation by integrating brainstorming, reference searching, and refinement.}
    \label{fig:ui}
\end{figure*}

\section{SYSTEM \& IMPLEMENTATION}
We propose AIdeation, a system that integrates multiple generative models to enhance concept designers' early ideation phase. Unlike existing tools, AIdeation combines the strengths of traditional and AI-driven approaches, streamlining the process by unifying research, brainstorming, and design idea refinement into a cohesive, iterative workflow.

\subsection{System Components}
AIdeation's key design aligns closely with the system’s design goals: Breadth exploration through brainstorming, Depth exploration via Research, and Flexible iterative exploration through Refining Design Ideas. To illustrate how AIdeation supports the early ideation phase for concept designers, we present a real-world design task from one of our users, who was tasked with creating game environments for a horror game set in traditional Taiwanese apartments.

\subsubsection{Brainstorming: Supporting breadth exploration}
After receiving the design specification, the designer inputs the instruction into AIdeation: "Design a living room scene for a horror game set in an old Taiwanese apartment." AIdeation generates 8 distinct design ideas, each featuring various elements accurately aligned with the specification. In our design, the ideas are described across six key design elements: Theme, Contents, Art Style, Lighting and Atmosphere, Color Palette, and Shot Angle. These categories are derived from observations in our formative study. We use "Theme" to represent the main reference, as designers normally base their primary searches on the central theme of the design task. "Content" covers detailed references for objects and elements within the scene. The other categories were selected based on the references most frequently used by concept designers. Composition was excluded due to current AI limitations. These categories were later reviewed by the same art directors mentioned in Section 4.

% The ideas are described across 6 categories of design elements: Theme, Art Style, Contents, Lighting and Atmosphere, Color Palette, and Shot Angle. These categories are based on our formative study observations, reflecting how designers collect and prioritize specific design elements in their references.

The ideas are presented as generated images in an ideas overview panel (Figure \ref{fig:ui}-a), offering a clear visual summary of each design and its key components, which can serve as potential hero references. This approach directly addresses challenges identified in the formative study, enabling designers to efficiently grasp the design topic while exploring a diverse range of visuals that align with the design specification and can be incorporated into their creative process.

\subsubsection{Research: Supporting depth exploration}
After selecting a design idea of interest (Figure \ref{fig:ui}-a1), the designer is directed to the idea detail panel (Figure \ref{fig:ui}-b), which provides in-depth information about the chosen idea. The left-side information bar displays key elements of the generated image extracted as keywords (Figure \ref{fig:ui}-b1), organized into 6 categories corresponding to the design idea description. In the "Content" category, elements are further divided into subcategories like "Central Focus" and "Background" due to the volume of information. This structure helps the designer clearly understand the composition of the design and easily identify specific elements in the generated image.

The idea detail panel also allows users to explore supporting references by clicking on relevant keywords. When a keyword is selected, corresponding search results are displayed in the same panel (Figure \ref{fig:ui}-b2), giving access to additional information and detailed references. Combined with diverse outputs from brainstorming, these features provide users with a broader array of ideas and information, facilitating deeper exploration and a more comprehensive understanding of the design topic and generated concepts.

\subsubsection{Refining idea: Supporting flexible iterative exploration}
Following this, AIdeation allows designers to refine the selected design idea using the detailed information provided through a flexible iterative approach—either by combining it with additional references or refining it through specific instructions (Figure \ref{fig:ui}-b3). These options enable users to either expand their exploration or narrow and focus their design scope, depending on their creative needs. 

After identifying a reference of interest based on the selected keyword (Figure \ref{fig:ui}-b2), the designer can combine it with the current idea to generate 5 new design variations. Figure \ref{fig:ui}-b illustrates the result of combining a previous design idea with a selected reference (Figure \ref{fig:ui}-b4). AIdeation adjusts the original design, such as transforming the style of the sofa to match the selected reference, demonstrating how the design scope can be refined. Conversely, if the reference is less related to the original elements, the new design will be more diverse, offering additional creative possibilities.

For the "refine by instruction" feature, once the designer identifies specific elements in the current design, they can use natural language to instruct AIdeation on what to change. These refinements can be based on AIdeation's provided information or the designer’s creative vision. The system then generates 5 new designs that incorporate the user’s instructions, maintaining the essence of the original idea while introducing diversity.

\subsubsection{Next ideation cycle for exploration}
AIdeation enables users to begin the next brainstorming cycle seamlessly based on the current design idea (Figure \ref{fig:ui}-b3, "Explore More"). This feature meets the need for designers to create related tasks based on an existing environment, as noted by participants in the formative study. For example, the designer could input “\textit{design a kitchen based on this}” using the design idea from Figure \ref{fig:ui}-b, efficiently expanding on the current concept.

\begin{figure*}
    \centering
    \includegraphics[width=1\linewidth]{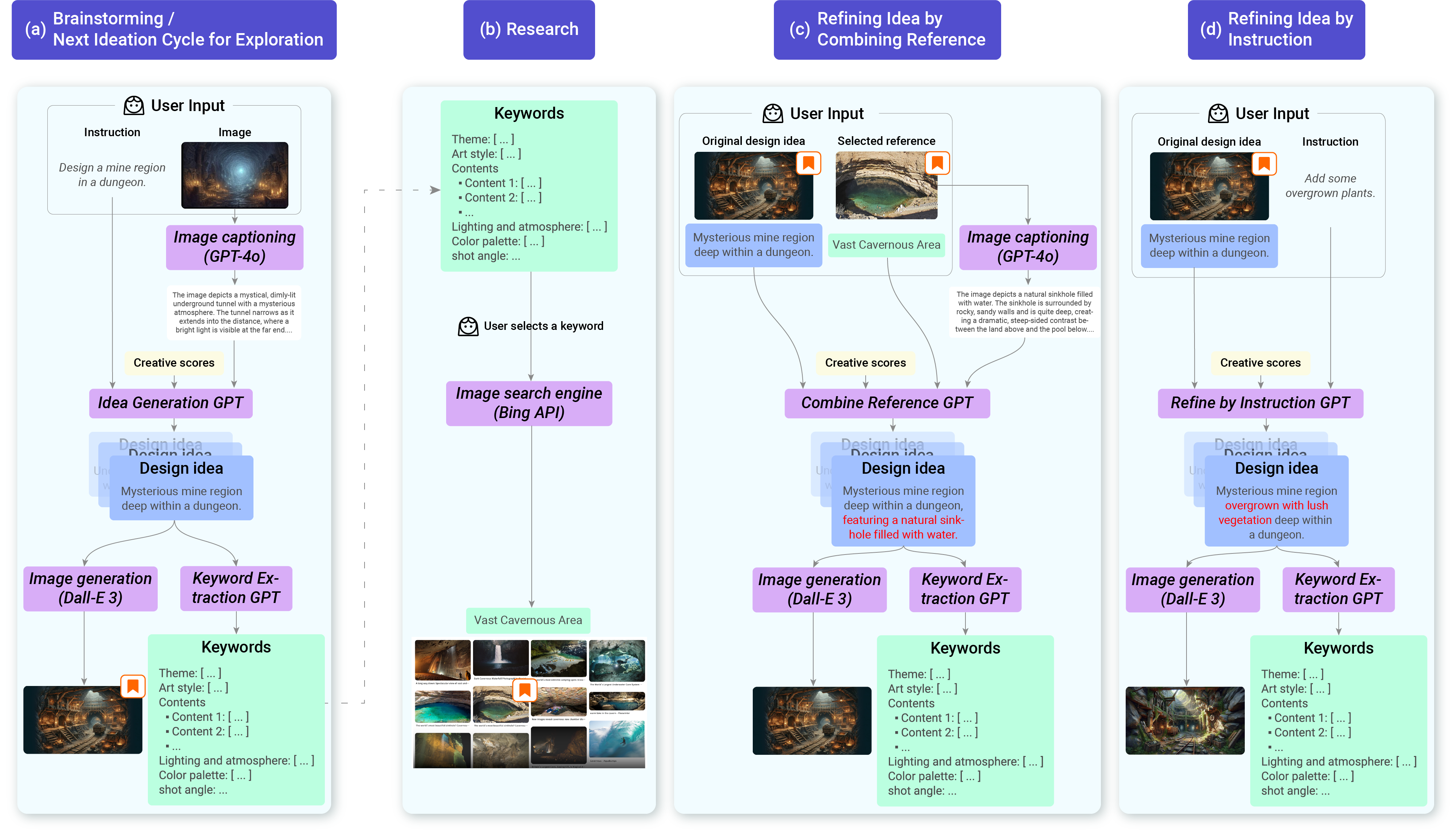}
    \caption{Technical pipeline of AIdeation: (a) The user’s input image is captioned by a vision model and processed by Idea Generation GPT, which integrates instructions and creative score to generate design ideas description. This idea is then converted into keywords, and DALL-E 3 generates an image with the idea description. (b) User-selected keywords initiate a Bing Image Search, returning a set of relevant images. When the user wants to refine the idea, (c) a selected reference is captioned by a vision model and processed by Combine Reference GPT, merging it with the original idea to create modified designs based on the creative score. (d) In contrast, AIdeation also supports refining ideas by instruction. The original idea and user instructions are processed by Refine by Instruction GPT, along with the creative score, to generate additional refined ideas.}
    \Description{The figure illustrates the AIdeation technical pipeline, consisting of four stages. (a) Brainstorming begins with a user-provided input and image, which is captioned by a vision model and processed by Idea Generation GPT to create a design idea. Keywords are extracted, and DALL·E 3 generates a corresponding image. (b) Research allows users to select keywords, triggering a Bing Image Search that retrieves relevant images for further exploration. (c) Refining by Combining Reference integrates a selected reference with the original idea using Combine Reference GPT, producing a modified design with an updated description and newly generated image. (d) Refining by Instruction enables users to modify the idea by providing textual input, which is processed by Refine by Instruction GPT to create an adjusted design. Each step incorporates a creative score to guide idea development iteratively.}
    \label{fig:technical}
\end{figure*}

\subsection{Technical Implementation}
\subsubsection{Brainstorming and research}
AIdeation accepts both textual instructions and image inputs. As shown in Figure \ref{fig:technical}-a, when an image is provided, it is first processed by the GPT-4o Vision model, which generates an image caption. Then, we use Idea Generation GPT—a specially prompted LLM (GPT-4o-2024-05-13 as the base model)—designed for generating environment concept design descriptions (see prompt in Appendix \ref{AppendixC}). The instruction, image caption (if applicable), and creative score are then passed into the LLM. Based on the user’s input, the model generates multiple design ideas in parallel. Each idea was assigned a creative score, ranging from 0 to 1, to reflect the diversity of the outputs. A higher creative score prompts Idea Generation GPT to produce more innovative design descriptions. The output format is detailed in Section 5.1.1.

Each generated idea is processed in two ways: 1) important information is extracted as keywords using a prompted Keyword Extraction GPT (see Appendix \ref{AppendixD}) and displayed in the idea detail panel, and 2) the idea is input into an image generation model to create an image, which is shown in the ideas overview panel. For this work, we used DALL-E 3 as the image generation model due to its ability to interpret natural language prompts, understand complex instructions, and generate corresponding images \footnote{Dall-E, https://openai.com/index/dall-e-3/}. The images are produced at a resolution of 1792x1024, suitable for environment concept design. The entire generation process takes approximately 30 seconds, or around 40 seconds on average when an image is included as input.

In the Idea Detail Panel, when a keyword is selected from the left-side information bar, it is sent to the backend, where the Bing Image Search API \footnote{Bing Image Search API, https://www.microsoft.com/en-us/bing/apis/bing-image-search-api} retrieves a batch of 50 images (Figure \ref{fig:technical}-b). Scrolling to the end of the page triggers an additional batch of images.

\subsubsection{Refining idea and next ideation cycle for exploration}
When a user selects a reference to combine with the current design idea, the reference is processed through the GPT-4o Vision model to generate image captions (Figure \ref{fig:technical}-c). The selected keyword, image caption, creative score (following the same distribution as brainstorming), and the current design description are then input into the Combine Reference GPT (see Appendix \ref{AppendixE}). This GPT modifies the design description by incorporating details from the reference image based on the selected keyword. For example, in Figure \ref{fig:ui}-b4, the keyword "Weathered Vintage Sofa" updates the corresponding section of the original design idea with the sofa's style from the reference image. If the reference is less related to the current design elements, the GPT will make broader adjustments, incorporating the reference while modifying other parts of the description. The level of modification is influenced by the creative score—higher scores result in more significant changes and diverse combinations, offering both control and variety. 

Figure \ref{fig:technical}-d illustrates the technical process of the "Refine by Instruction" feature. Like the brainstorming and reference combination processes, the prompted "Refine by Instruction GPT" (see Appendix \ref{AppendixF}) uses the user’s instruction, creative score, and current design description as inputs. The GPT adjusts the design based on the instruction, with the creative score determining the extent of changes and creativity. The modified design descriptions follow the same format, allowing for later keyword extraction and image generation, just like in the brainstorming process. Both Idea Refinement process takes a similar amount of time as Brainstorming.

For the Next Ideation Cycle for Exploration, the process follows the same structure as the brainstorming phase, with the key difference being that image captioning is replaced by the current design idea description.

\section{SUMMATIVE STUDY}
% 一開始就要講清楚high level的關聯性
% Ideation結果要講清楚是搜集Reference過程

Our summative study examines the effectiveness of a creativity support tool in enhancing designers' early ideation workflows and fostering their creative processes. Rather than evaluating the quality of final design outputs, the study focuses on how AIdeation supports the workflow compared to designers' existing processes. We conducted a within-subject comparative study with 16 professional environment concept designers, focusing on three key aspects:

% To evaluate whether AIdeation enhances designers' early ideation workflow and supports their creative process, we conducted a within-subject comparative study with 16 professional environment concept designers, focusing on three key aspects: 
\begin{itemize}
    \item \textbf{A1:} Support for the Ideation Process
    \item \textbf{A2:} Quality and Efficiency of Ideation 
    \item \textbf{A3:} Workflow Support at Each Stage
\end{itemize}
Given the diversity of concept designers' workflows, we set the baseline to each participant's preferred existing workflow. Participants were free to use any of their current methods, such as image databases, search engines, and AI design tools like Midjourney or DALL-E 3, or a combination of these. For those with no prior experience using AI design tools, we provided access to ChatGPT-4 with DALL-E 3 and a brief tutorial, as AIdeation is also based on DALL-E 3. Participants could then decide whether to incorporate this into their process.

% 設計的時候就寫說設計原因
% 論述強度要加強

\subsection{Study Design}
\subsubsection{Procedure}
The study lasted 2 to 2.5 hours, beginning with a 10-minute briefing. Participants completed a 30-minute design task under each condition, with each task preceded by a 10-minute practice session. They also received a 10-minute tutorial on AIdeation. To minimize bias from prior experience, participants were briefed on key differences between AIdeation and other AI tools, such as its use of natural language inputs instead of prompts and its more limited styling options. After each task, participants took a 10-minute break. Both the conditions and design topics were counterbalanced. After both tasks, participants completed a questionnaire and a 20-30 minute post-study interview. They were compensated approximately 34 USD.

\subsubsection{Task overview}
Participants completed a design task that replicated their ideation workflow using both the baseline method and AIdeation. For each condition, participants were assigned an environment concept design topic and instructed to gather at least three sets of references for interior and exterior designs using the widely used PureRef reference board ~\footnote{Pureref, https://www.pureref.com/}. The design topics were: 1) a Mayan Observatory and Planetarium with an observation hall featuring a dome, and 2) a Tibetan Meditation Research Center with a main research hall. Both tasks required blending traditional and contemporary architectural styles, a common challenge in real-world projects. Participants were asked to follow their ideation process, including research and brainstorming, ensuring that the selected references would suit future designs and client presentations, thereby simulating real-world constraints. Each reference set is needed to represent a distinct ideation result, including at least one main reference supported by detailed references.

\subsubsection{Pilot study and refinements}
Initially, we designed a 40-minute task, which also included a sketching stage after the research and brainstorming stages. We piloted this design with two professional concept designers, but neither was able to complete the tasks, reporting feeling "extremely stressed" as such tasks typically require a full day. Additionally, sketching diverted their focus from exploring the AIdeation system, despite it being the primary focus of the study. Based on these findings, we excluded sketching from the final study design. The revised tasks and design topics were reviewed and validated by three art directors from animation, game, and art outsourcing studios, who confirmed that using references alone to communicate with clients is a common practice, especially under time constraints.

\subsubsection{Measurements}
The questionnaire focused on comparing the support provided by each condition for the ideation process and different workflow stages. Participants rated their preferences across various aspects. For the ideation process, they evaluated breadth and depth of exploration support, flexibility in idea exploration, efficiency in generating diverse ideas, quality of ideas, and creativity enhancement. For example, we asked the participants about their preferences using questions such as, “\textit{Which system allows you to generate a variety of design ideas more efficiently?}” For workflow support, they compared the systems on information collection efficiency, reference-gathering efficiency, the usefulness of collected references, and which system better supported the visual presentation of ideas. Additionally, participants provided feedback on their overall satisfaction and enjoyment while also comparing task difficulty and efficiency between the two conditions. The full questionnaire can be found in Appendix \ref{AppendixA}. All responses were measured using a 7-point Likert scale, where 7 indicated a strong preference for AIdeation, and 1 indicated a preference for their original workflow. For questions related to efficiency, participants were asked to disregard image generation time. This measurement approach aligns with previous findings on preference elicitation, emphasizing the importance of task sensitivity and granularity in capturing meaningful differences between options~\cite{evangelidis2024task}. Specifically, using comparative questions enhances sensitivity to utility differences, while choosing a 7-point scale balances granularity and interpretability for moderate differences in preferences. A one-sample Wilcoxon signed-rank test was performed to evaluate whether responses differed significantly from the neutral midpoint (4). This non-parametric test is appropriate for analyzing the ordinal data collected through the 7-point Likert scale questionnaire, as it does not assume a normal distribution and is well-suited for assessing central tendency differences in ordinal data~\cite{conover1999practical}. By testing whether the median response significantly deviates from the neutral point, this approach effectively determines whether participants exhibited a systematic preference for one condition over the other. This methodology is supported by previous research on the suitability of nonparametric tests for ordinal data and preference-based measures~\cite{roberson1995analysis, capanu2006testing, taheri2013generalization} and is consistent with previous studies employing similar analytical frameworks~\cite{chen2023aircharge}.

% A one-sample Wilcoxon signed-rank test was performed to evaluate whether responses significantly differed from the neutral midpoint (4). This non-parametric test is suitable for analyzing the ordinal data collected through the 7-point Likert scale questionnaire, which evaluates preferences between two conditions. This approach is supported by prior research~\cite{roberson1995analysis, capanu2006testing, taheri2013generalization} and is consistent with previous studies employing similar methodologies~\cite{chen2023aircharge}.

% In the in-depth interview, we first explored participants' attitudes toward AI-generated images, prior experience with AI design tools, typical ideation strategies, and how they approached the task using AIdeation. We then focused on how their experience with AIdeation differed from their original workflow, particularly in terms of the ideation process and overall workflow. Participants also explained the reasoning behind their questionnaire ratings. The interview concluded with discussions on their favorite AIdeation features, suggestions for improvements, and how they might integrate AIdeation into their workflow. 
%RR
In the in-depth interview, we first explored participants' attitudes toward AI-generated images, prior experience with AI design tools, typical ideation strategies, and how they approached the task using AIdeation. We then focused on how their experience with AIdeation differed from their original workflow, particularly in terms of the ideation process and overall workflow. For example, we asked participants to compare their experiences when searching for references using two different approaches. Additionally, we requested that they explain the reasoning behind their questionnaire ratings. For instance, they were asked to elaborate on why they preferred AIdeation for better efficiency. The interview concluded with discussions on their favorite AIdeation features, suggestions for improvements, and how they might integrate AIdeation into their workflow. The detailed interview questions can be found in Appendix \ref{AppendixB}. The interview data were analyzed similarly to the formative study. Three researchers summarized the transcribed recordings, and a former concept designer on the team identified key themes for thematic analysis. The findings were then reviewed and discussed among the researchers to ensure consensus.

\subsubsection{Evaluation approach}
We adopted a self-report approach, aligning with prior research in the HCI and creativity communities~\cite{lubos2024llm,satyanarayan2019critical,palani2022don, son2024genquery}. Participants provided feedback based on their experiences, evaluating AIdeation across key dimensions as a creativity support tool. For the ideation results, participants self-assessed which method provided better support for idea quality and efficiency during the tasks. Given the difficulty of directly comparing outputs between two conditions (collected references), external expert evaluation was not conducted. Instead, participants were asked during interviews to explain their questionnaire ratings in detail, providing qualitative insights into their experiences. To complement this, we later conducted a field study to assess AIdeation's impact on real-world projects, focusing on idea quality, quantity, and creativity.

\subsubsection{Participants}
% RR
We recruited 16 professional environment concept designers from various industries and five studios, including animation (P1, P4, P6, P16-P20), game (P5, P14-P15, P22), art outsourcing (P3, P13), and freelancing (P2, P21). 6 participants (P1-P6) had also participated in the earlier formative study. Participants had between 1 and 12 years of professional concept design experience (M = 4.6, SD = 3.2).

\begin{figure*}
    \centering
    \includegraphics[width=1\linewidth]{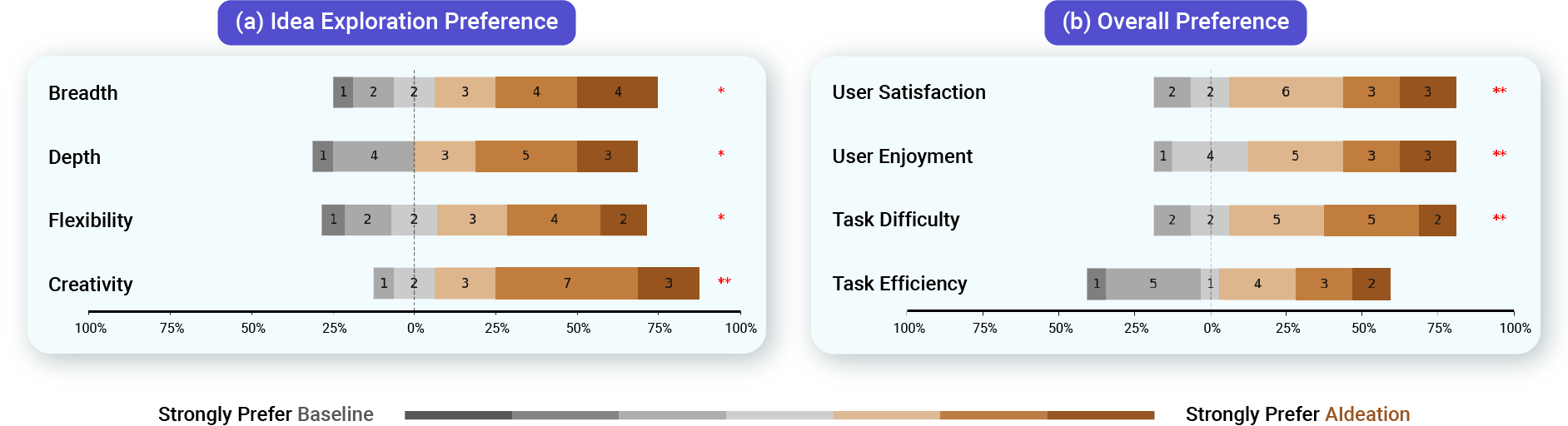}
    \caption{The distribution of user preference for baseline vs. AIdeation: (a) Preference rating on a 7-point Likert scale for idea Exploration; (b) Overall preference for Satisfaction, Enjoyment, Task Difficulty, and Task Efficiency}
    \Description{The figure presents the distribution of user preferences comparing a baseline method with AIdeation. (a) Idea Exploration Preference is rated on a 7-point Likert scale across four aspects: breadth, depth, flexibility, and creativity. AIdeation is preferred in all categories, with significant differences indicated by asterisks. (b) Overall Preference evaluates user satisfaction, enjoyment, task difficulty, and task efficiency. AIdeation scores higher in satisfaction and enjoyment with significant differences, while task difficulty and efficiency also favor AIdeation. The scale ranges from strongly preferring the baseline to strongly preferring AIdeation, showing a clear trend toward AIdeation's advantages in ideation tasks.}
    \label{fig:ideation result}
\end{figure*}

\section{RESULTS \& FINDINGS}
\subsection{A1: Supporting Ideation Process}
In this section, we first address our key aspects using the information collected from the study. Following that, we will report additional notable qualitative findings based on our observations.
\subsubsection{Breadth, depth, and flexibility of idea exploration
}
Figure \ref{fig:ideation result}-a shows that participants preferred AIdeation for breadth (Mean = 5.19, \textit{p} = 0.014), depth (Mean = 5.00, \textit{p} = 0.033), and flexibility (Mean = 4.93, \textit{p} = 0.046) in idea exploration, with 69\% expressing a preference for AIdeation in terms of both breadth and depth. Participants mentioned: “\textit{The randomization provided by AIdeation offers a lot of possibilities}” (P19). and “\textit{The additional references, combined with my existing ideas, really expanded my design space}” (P21). Most participants found AIdeation offered better diversity than their original tools with the same input (P1, P3, P5, P13, P15-P16, P19, P21-P22). “\textit{Compared to MidJourney, AIdeation provided much more diversity, and I didn't even need to think of a prompt}” (P16). “\textit{Each iteration offered significant variation, helping me break out of my usual direction and explore new ideas}” (P4). However, some participants noted limitations in atmospheric or stylistic diversity (P2, P6, P20). “\textit{I feel like when I input 'Mayan Architecture,' the system often gave me stereotypical results}” (P2). “\textit{I always get a similar style from AI images}” (P6).

Regarding the depth of exploration, five participants noted that they could efficiently narrow down their design focus with AIdeation (P3, P18, P20-P22). “\textit{Compared to the design ideas I find on Artstation or Pinterest, which cannot be modified, I can use AIdeation to refine the idea}” (P21). Three participants also mentioned that AIdeation accurately provided detailed information and references that helped in further design work (P3, P5, P18). “\textit{AIdeation offers so much information that would normally take a lot of time to gather from different platforms}” (P18). However, two participants mentioned that AIdeation lacked image-to-image search functionality, which prevented them from obtaining similar images and restricted deeper exploration (P1, P6).

Finally, 56\% of the participants preferred the flexibility of exploration with AIdeation. Most suggested that AIdeation was easy to use and its functionality helped them access diverse design elements (P1, P14-P15, P16-P18, P22). However, three users found it challenging to achieve their desired results with AIdeation due to the lack of detailed control (P1, P2, P6). “\textit{The layout kept changing when I combined references or refined by instructions, but I wanted to keep that layout}” (P2). “\textit{I just wanted to adjust the atmosphere of the image, but I couldn't do that with this system}” (P6).

\subsubsection{Creativity}
Compared to their original workflow, participants significantly preferred AIdeation to enhance creativity (Figure \ref{fig:ideation result} -a, Mean = 5.56, \textit{p} = 0.001), with 81\% expressing a preference for it. Many users noted that AIdeation provided unexpected brainstorming results (P1, P13-P18, P21-P22). Several participants appreciated the keywords provided by AIdeation, mentioning that they could simply refer to the detail information bar for inspiration (P1, P4-P5, P13-P14). Furthermore, combining their original design ideas with the additional reference introduced different ways of thinking (P1, P4, P14, P17, P21-P22). As one participant mentioned, “\textit{The system combined different styles and content in various ways, which gave me a lot of inspiration}” (P21). Another added, “\textit{Combining diverse or uncommon elements sparked new ideas}” (P1).

\subsubsection{Overall satisfaction, task efficiency, and difficulty}
Figure \ref{fig:ideation result}-b shows the distribution of the overall satisfaction, enjoyment, and preferences of participants regarding task difficulty and efficiency. The participants significantly preferred AIdeation in terms of overall satisfaction (Mean = 5.19, \textit{p} = 0.005) and enjoyment (Mean = 5.19, \textit{p} = 0.005), with 75\% and 69\% expressing a preference, respectively. “\textit{The system is really easy to use, compared with other AI tools I used before}” (P16). “\textit{I really enjoy seeing the result of combination, each time I was surprised by unexpected results}” (P1). Results also suggest that participants significantly preferred AIdeation for reducing task difficulty (Mean = 5.19, \textit{p} = 0.005). Participants explained that AIdeation allowed them to work without thinking too much (P3, P16, P18) and provided well-organized and relevant information that would normally take more time to gather using their original workflow (P1, P4-P5, P19), such as “\textit{Those keywords truly helped me quickly find a variety of reference materials}” (P4).

However, the results for task efficiency were mixed. Despite being asked to ignore image generation time, some participants who preferred their baseline workflow noted that waiting for results in AIdeation took longer compared to browsing visuals continuously on Pinterest (P1, P17-P19). Additionally, three participants barely completed their tasks because they spent most of their time trying to achieve an exact match to their idea but were unsuccessful (P2,P6, P18). One participant who preferred the baseline for both satisfaction and efficiency mentioned, “\textit{I kept modifying the prompt but couldn’t get the result I wanted, which was frustrating}” (P6).

In summary, compared to their original workflow, AIdeation improved the creativity of participants and provided better support for both breadth and depth of exploration with improved flexibility. In general, the participants reported greater satisfaction and enjoyment with AIdeation and significantly reduced the difficulty of the task.

\begin{figure*}
    \centering
    \includegraphics[width=1\linewidth]{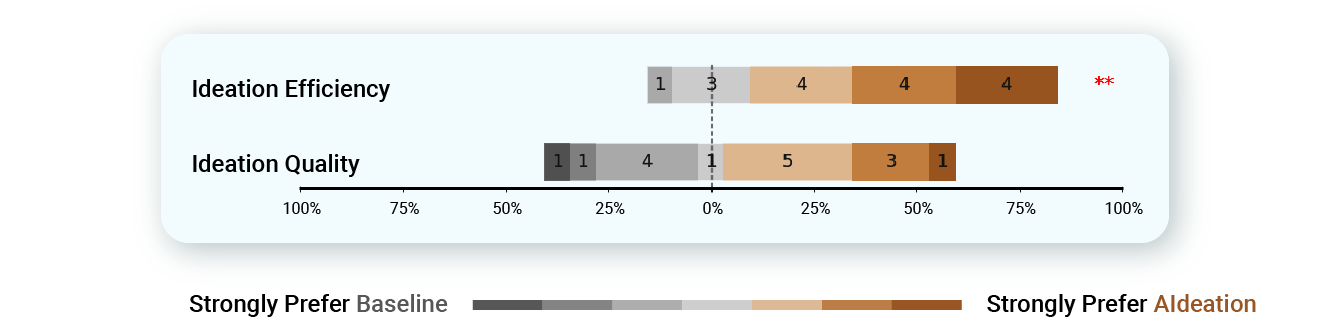}
    \caption{The distribution of user preference for baseline vs. AIdeation:  Preference rating on a 7-point Likert scale for Quality and Efficiency of Ideation}
    \Description{The figure presents user preferences comparing a baseline method with AIdeation on a 7-point Likert scale. Design Ideas Efficiency shows a strong preference for AIdeation, with a significant difference indicated by an asterisk. Design Ideas Quality is more evenly distributed, though AIdeation is still preferred. The scale ranges from strongly preferring the baseline to strongly preferring AIdeation, demonstrating that AIdeation enhances efficiency and is generally favored for idea quality}
    \label{fig:design idea result}
\end{figure*}

\subsection{A2: Quality and Efficiency of Ideation}
Figure \ref{fig:design idea result} shows the distribution of the participants' preferences for various aspects of the ideation process. The results indicate that the participants significantly preferred AIdeation to efficiently generate a variety of design ideas (Mean = 5.44, \textit{p} = 0.003) while maintaining a quality similar to their original workflow (Mean = 4.31, \textit{p} = 0.41). 75\% of participants preferred AIdeation for its efficiency in assisting with design idea generation. 

Most participants noted that AIdeation quickly offered a wide range of diverse design elements that aligned with the topic (P1, P13-P16, P18, P21- P22). “\textit{In my original workflow, I need to first think of keywords to search on Pinterest. Even if I find an interesting design, it's hard to extract key information from the image. AIdeation, however, provides plenty of ideas with well-organized information}” (P22). “\textit{In MidJourney, I have to come up with the idea first, but AIdeation already presents many ideas}” (P14).

Half of the participants noted that AIdeation significantly reduced the time needed to combine two design elements (P1, P3-P5, P13, P18, P21-P22). “\textit{Before, I had to manually combine two design elements, and if the result wasn’t ideal, the time was wasted. With AIdeation, I can instantly combine elements with some level of control, saving me a lot of time}” (P13).
Additionally, most users took advantage of AIdeation's "explore more" functionality to quickly generate interior design ideas based on their selected exterior design ideas. Most suggested that this feature significantly helped them obtain diverse interior results without starting the process from scratch (P1, P3, P15-P16, P18, P20, P22). “\textit{The ability to generate images in this narrative-driven way is truly a groundbreaking innovation}” (P18).  “\textit{This will be very useful if I need to design multiple rooms within the same building}” (P13).

Regarding ideation quality, some participants noted that their original workflow with Pinterest often yielded more diverse and unexpected design elements, leading to fresh ideas. In contrast, AIdeation tended to offer more relevant design elements (P2, P6). Other participants who preferred the baseline or remained neutral noted that they could achieve similar or better quality using their original workflow (P6, P13, P18, P20). “\textit{The AI-generated images are very different from what I have in mind. If it were a real photo, the quality would be better, and the details would be more accurate}” (P6).

In summary, the participants considered AIdeation to have helped them generate more diverse design ideas while maintaining quality similar to their original workflow. 

\begin{figure*}
    \centering
    \includegraphics[width=1\linewidth]{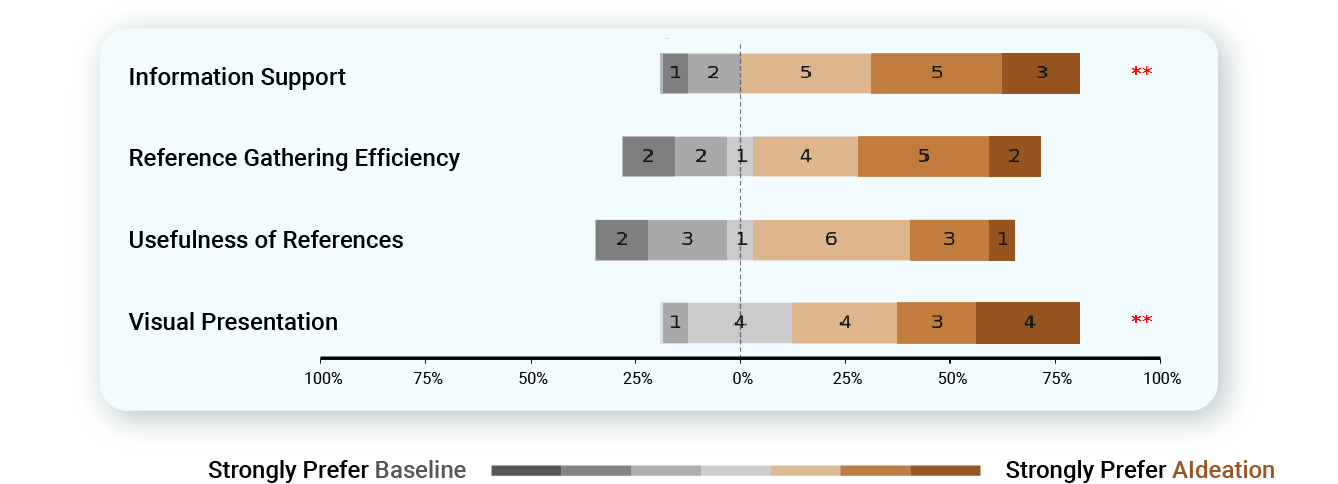}
    \caption{The distribution of user preference for baseline vs. AIdeation:  Preference rating on a 7-point Likert scale for Workflow Support}
    \Description{The figure compares user preferences for baseline vs. AIdeation on a 7-point Likert scale. AIdeation is significantly preferred for information support and visual presentation. Reference gathering efficiency and usefulness of references also favor AIdeation but with a more balanced distribution. The scale ranges from strongly preferring the baseline to strongly preferring AIdeation, showing AIdeation improves reference gathering and workflow support.}
    \label{fig:workflow result}
\end{figure*}

\subsection{A3: Workflow Support at Each Stage}
\subsubsection{Support for research, reference gathering and visual presentation}
As shown in Figure \ref{fig:workflow result}, participants significantly preferred AIdeation for efficiently gathering information for the design task (Mean = 5.25, \textit{p} = 0.009), with 81\% expressing a preference. Many found that AIdeation helped them quickly understand the design topic (P1-P3, P15-P17, P19, P21-P22). One participant noted, “\textit{Normally, we spend hours researching complex topics without existing references, but AIdeation provided a whole package of concepts and information instantly, saving a lot of time}” (P3). Participants mentioned that the keywords and detailed supporting references provided by AIdeation for each design idea helped them quickly obtain the information needed to develop their designs. “\textit{Even if the AI-generated image didn't fit my needs, I could use the keywords provided by the system to find a lot of useful information}” (P5). “\textit{I don’t need to come up with keywords or read through a lot of text when using AIdeation}” (P13).

The participants also significantly preferred AIdeation to better visually present the design idea (Figure \ref{fig:workflow result}, Mean = 5.31, \textit{p} = 0.004), with 69\% expressing preference. “\textit{Usually, I have to sketch multiple design versions because no reference fits my needs. But with AIdeation, I can select elements from several generated images and directly present my idea to the client}” (P3).
However, there was no significant preference regarding the efficiency of gathering references or the usefulness of references with AIdeation. Participants who preferred the baseline stated that “\textit{Even though AIdeation provides accurate references, the overall quantity is much less compared to my original workflow (Pinterest)}” (P1).
% “\textit{I usually prefer photorealistic references, but the generated images always have an ‘AI style’.}” (P6)

\subsubsection{Integrating AIdeation into workflow}
At the end of the study, we asked the participants if and how they would use AIdeation in their real-world projects. Most of the participants indicated that they would use AIdeation immediately after receiving design specifications to explore different concepts (P1-P6, P13-P16, P18-P22). Several noted that AIdeation is more efficient and user-friendly then other AI tools (P1, P13-P14, P18, P21-P22). Three participants found it particularly useful for initiating new designs based on existing concepts (P3, P5, P15). Many suggested using AIdeation's output to communicate more effectively with directors or clients, thereby significantly increasing efficiency (P3-P5, P13, P16-P17, P21-P22). Additionally, two users mentioned potential applications for photobashing with AIdeation's outputs (P15, P20).

In summary, the participants found AIdeation to be more efficient in collecting relevant information and visually presenting the ideation results. While some preferred traditional methods for sourcing references, most considered AIdeation's suggestions valuable for ideation. Additionally, it showed the potential to streamline workflows and enhance client communication.

\subsection{Qualitative Findings on AIdeation Usage}
\subsubsection{Impact of intuitive vs. Analytical usage on AIdeation}
Participants who intuitively engaged in AIdeation and freely experimented achieved better results than those who overanalyzed the process. Those who quickly iterated without perfecting the prompts generated more diverse and creative outputs, while participants who spent excessive time refining the inputs produced fewer results and found the tool less efficient.
For example, P5 embraced a simple and iterative approach, generating 16 hero references in 4 ideation cycles with 10 refinements, aligning well with her creative vision. In contrast, P2 spent considerable time crafting inputs and struggled with unsatisfactory outputs, completing 8 cycles with 4 refinements but only 6 hero references, ultimately perceiving the tool as less effective.

\subsubsection{Controllability of AIdeation}
AIdeation received polarized opinions regarding its controllability. On the positive side, many participants were impressed with the degree to which the tool understood their intentions, especially when combining references and refining with instruction. (P3-P5, P13, P15, P20) “\textit{For me, ease of use is the most important factor for an AI tool. AIdeation met that goal and was able to capture the key points I wanted}” (P15). “\textit{Combining references allows me to control specific parts and choose what to merge, which is extremely helpful}” (P19). However, some participants had contrasting experiences. “\textit{I wanted to keep the material but change the layout from square to round, and after trying several times, it still didn’t work}” (P18). Interestingly, some participants appreciated the lower level of controllability. "\textit{Each generated idea had significant variation, and I could extract different elements from them}" (P16).

% move from discussion, need to be shorten
\subsubsection{Expectations toward AIdeation and their effect}
Users' expectations toward AIdeation strongly influenced their experience. When seen mainly as an image generation tool—a common perception in our study—users often produced less diverse results, focusing on precise prompts and modifications (see Section 7.4.1). This mindset, common among users familiar with other AI design tools, emphasizes control and exact results. As one participant noted, “\textit{With AI, I expect a complete design. Unlike Pinterest, where I look for elements, AI images seem finished, so I feel the need to tweak them for clients}” (P18). This approach can limit opportunities, as users may overlook valuable design elements. In contrast, treating AIdeation like browsing Pinterest encourages the discovery and integration of new ideas, enriching the creative process.

% > Line 865: should it say, “The system is really [easy] to use…”?
% > It would be useful for the authors to clarify whether the field study took place after the ideation study and whether any further development or refinement to the AIdeation system took place in-between. The field study set-up lacks some important details, such as how the participants were briefed or understood how to use the system (some of which depends on whether they were same participants from the previous activities or not?) nor does it describe what or how data was captured (e.g., interviews, surveys, diary study?).
% > 
% > The findings from the field study are brief and not currently well organised, e.g., S1 and S2 presented as distinct sections, then S3 and S4 folded into one section that also covers some detail from S1 and S2. It is not clear whether the field study offers anything new to the paper as a whole.

\section{FIELD STUDY}
After the summative study, we conducted a week-long field study across four studios to evaluate how AIdeation supported production projects in terms of quality, creativity, and efficiency, particularly with external assessment and validation by directors and clients. 
We also examined how AIdeation integrated into the design process, influenced the creative workflow, and addressed challenges or limitations encountered in real-world project settings.
% To evaluate AIdeation's practical application in concept design, we conducted a week-long field study across 4 studios. This approach allowed us to observe how AIdeation integrates into real-world workflows and assess its impact on the creative process. Our primary focus was on examining how AIdeation enhances the design process, facilitates the generation of diverse, high-quality solutions, and identifies the challenges or limitations it encounters.

\subsection{Participants: Studios and Designers}
We recruited 8 participants across 4 studios (S1-S4) from the summative study, as their familiarity with AIdeation made them well-suited to integrate AIdeation into their commercial projects. To control for selection bias, these participants had an average satisfaction score of 5.25, closely aligned with the overall mean score of 5.19 from the summative study.

S1 (P1, P16-P18) is a visual effects (VFX) studio focusing on designing environments for animation, films, TV shows, and advertisements; S2 (P3, P13) is an art outsourcing studio; S3 (P22) is an AAA game studio; and S4 (P5) is a game studio known for creating side-scrolling Metroidvania games, a subgenre of action-adventure and platformer games that are known for their non-linear exploration and progression. The professional concept design experience of the participants ranged from 2 to 11 years (mean = 4.5, SD = 2.9), similar to that experienced with an average of 4.6 years from the summative study. 
%Since participants from each studio completed one project, and we collected data from four studios, we refer to these studios as S1 through S4 in this study.

\begin{table*}[t]
\centering
\footnotesize
\begin{tabular}{|c|c|c|c|c|c|c|c|c|}
\hline
Studio ID  & Field & Task Type & AIdeation Usage \% & Other Tools Used & Env. & Ideas Gen. & Cycles & Ideas Used\\
\hline
1  & Animation, Films, TV shows &Environment Design & 40\% & Pinterest, Midjourney & 3 & 93 & 14 & 14\\
\hline
2  & Art Outsourcing &Matte Painting & 100\% & None & 1 & 105 & 10 & 5\\
\hline
3 & Game & Visual Development & 80\% & Pinterest & 3 & 652 & 45 & 28 \\
\hline
4  & Game & Environment Design & 90\% & Pinterest & 5 & 242 & 29 & 23 \\
\hline
\end{tabular}
\caption{Summary of the 4 studios' usage of AIdeation: studio's field, project tasks, usage of AIdeation among all tools used, number of environment design tasks, total ideas generated, total ideation cycles, and the number of ideas generated by AIdeation that were used in the final output of the designers.}
\Description{This table summarizes the usage of AIdeation across four studios, providing details about their fields, project tasks, and tool utilization. The table highlights the percentage of AIdeation usage among all tools, other tools used, the number of environments designed, total ideas generated, ideation cycles completed, and the number of ideas selected for final outputs.
 Studio 1:  Working in the animation, film, and TV show industry, this studio focused on environment design tasks. They used AIdeation for 40\% of their workflow alongside Pinterest and Midjourney, generating 93 ideas across 14 cycles and finalizing 14 ideas.  Studio 2:  Specializing in matte painting under art outsourcing, this studio relied exclusively on AIdeation, achieving 100\% usage. They generated 105 ideas in 10 cycles, with 5 ideas selected for their final output.  Studio 3:  A game studio focusing on visual development tasks, this studio utilized AIdeation for 80\% of its workflow alongside Pinterest. They completed 45 cycles, generating 652 ideas, and selected 28 for the final output.  Studio 4:  Another game studio working on environment design, this studio used AIdeation for 90\% of its workflow, also integrating Pinterest. They generated 242 ideas across 29 cycles, with 23 ideas selected for their final designs. Overall, the four studios collectively generated 1,092 ideas across 98 cycles, selecting 60 ideas that contributed to the creation of 12 environments. The data highlights the varying degrees of AIdeation adoption and its integration with other tools across different industries and tasks.}
\label{tab:studio_production}
\end{table*}

\subsection{Study Procedure and Evaluation}
We deployed the same AIdeation system from the summative study on AWS (Amazon Web Services) EC2, assigning each studio its own EC2 instance. 
We asked designers to incorporate AIdeation into their current projects, emphasizing its use during early ideation alongside other design tools.
We used a combination of diary studies and interviews to explore participants’ experiences with AIdeation. Participants documented their workflows, including how they used the tool, the ideas they selected, their iteration processes, and the progression from initial input to final results.

We then conducted 30-minute online interviews in which participants assessed how AIdeation impacted their design outcomes and efficiency. The participants estimated the time typically required for the project based on previous experience and compared it with the actual time spent using AIdeation. Participants also reported on directors' and clients' feedback on the results. Additionally, they shared how AIdeation supported their projects, highlighted new insights gained during its use in real-world projects, and identified areas for improvement. 
% The detailed questions are provided in Appendix X. 
Although the studios shared the final production results with the authors as part of the field study, they withheld permission for us to publish them due to NDAs and unreleased games.

\begin{table*}[t]
\centering
\footnotesize
\begin{tabular}{|c|c|c|c|c|}
\hline
Studio ID & Actual Time (AIdeation) & Estimated Time (Original) & Time Difference & Time Difference (\%) \\
\hline
1 & 1.5 Working Days & 2 Working Days & -0.5 Working Days & -25\% \\
\hline
2 & 5 Hours & 4 Hours & +1 Hour & +25\% \\
\hline
3 & 2 Working Days & 5 Working Days & -3 Working Days & -60\% \\
\hline
4 & 6 Working Days & 14 Working Days & -8 Working Days & -57\% \\
\hline
\end{tabular}

\caption{Comparison of concept design time using AIdeation (actual) vs. the same designs using existing workflow (estimated).}
\Description{Comparison of Estimated Time Using Original Workflow vs Actual Time Using AIdeation. This table provides a comparison of the time efficiency between traditional workflows and workflows incorporating AIdeation across four different studios. The table highlights both the original time estimates (without AIdeation) and the actual time taken using AIdeation for completing their respective design tasks. The table also calculates the time saved (or extended) and expresses the difference as a percentage. Studio ID: A unique identifier for each studio, linking it to the projects discussed in the study. Estimated Time (Original): The time each studio estimated they would spend on the design task using their traditional workflow (without AIdeation). Time is expressed either in working days or hours, depending on the scope of the task. Actual Time (AIdeation): The actual time spent on the same task using AIdeation. This column shows how long the process took with the AI tool, also expressed in working days or hours. Time Difference: This column indicates the difference in time between the estimated time using the traditional workflow and the actual time spent using AIdeation. A negative value indicates time saved, while a positive value shows time overrun. Time Difference (\%): The percentage change in time compared to the original workflow. A negative percentage represents a reduction in time (i.e., improved efficiency), while a positive percentage represents an increase in time (i.e., decreased efficiency). Breakdown of Studio Performance: Studio 1: Estimated Time: 2 working days Actual Time with AIdeation: 1.5 working days Time Difference: -0.5 working days Time Difference (\%): -25\% Result: Studio 1 saved 0.5 working days, resulting in a 25\% improvement in efficiency when using AIdeation. Studio 2: Estimated Time: 4 hours Actual Time with AIdeation: 5 hours Time Difference: +1 hour Time Difference (\%): +25\% Result: Studio 2 experienced a slight decrease in efficiency, taking 1 extra hour (25\% longer) to complete their task using AIdeation. Studio 3: Estimated Time: 5 working days Actual Time with AIdeation: 2 working days Time Difference: -3 working days Time Difference (\%): -60\% Result: Studio 3 saw a significant improvement, reducing their working time by 3 days, equating to a 60\% improvement in efficiency. Studio 4: Estimated Time: 14 working days Actual Time with AIdeation: 6 working days Time Difference: -8 working days Time Difference (\%): -57\% Result: Studio 4 also reported a major time reduction, saving 8 working days, which translates to a 57\% improvement in efficiency. Summary of Findings: Studio 1 and Studio 3 reported moderate and significant improvements in time efficiency, respectively, reducing their work time by 25\% and 60\% through the use of AIdeation. Studio 4 experienced the largest time savings, cutting down their work by 8 days, which resulted in a 57\% reduction. However, Studio 2 experienced a slight decrease in efficiency, taking 1 additional hour (25\% longer) than estimated, likely due to the challenges mentioned in their case study.}
\label{tab:time_comparison}
\end{table*}

\subsection{Results and Findings}
% 從Overview開始做
% 先講accross 4個的全面的結果
% 做了幾個場景，用了多久，生出多少Idea
% 可以摘要Table 1, 給個Summary
% 要講一下重點
% 這裡是給Quantitative Overview

% Table 2 的也要做一個Summary
% 還有report client / director的feedback
% \subsubsection{Integration of AIdeation into Workflow}
Table \ref{tab:studio_production} presents background information on the project each studio was working on and how they utilized AIdeation in their current project (where cycles are defined as starting a new input or using the "explore more" feature). The four studios collectively generated 1,092 ideas across 98 cycles, ultimately selecting 60 ideas, which contributed to the design of 12 environments in total.

% The following sections offer an in-depth look at two case studies (Figure 10), followed by a broader analysis of AIdeation’s performance, its support for workflow, and its impact on ideation efficiency and quality.

\begin{figure*}
    \centering
    \includegraphics[width=1\linewidth]{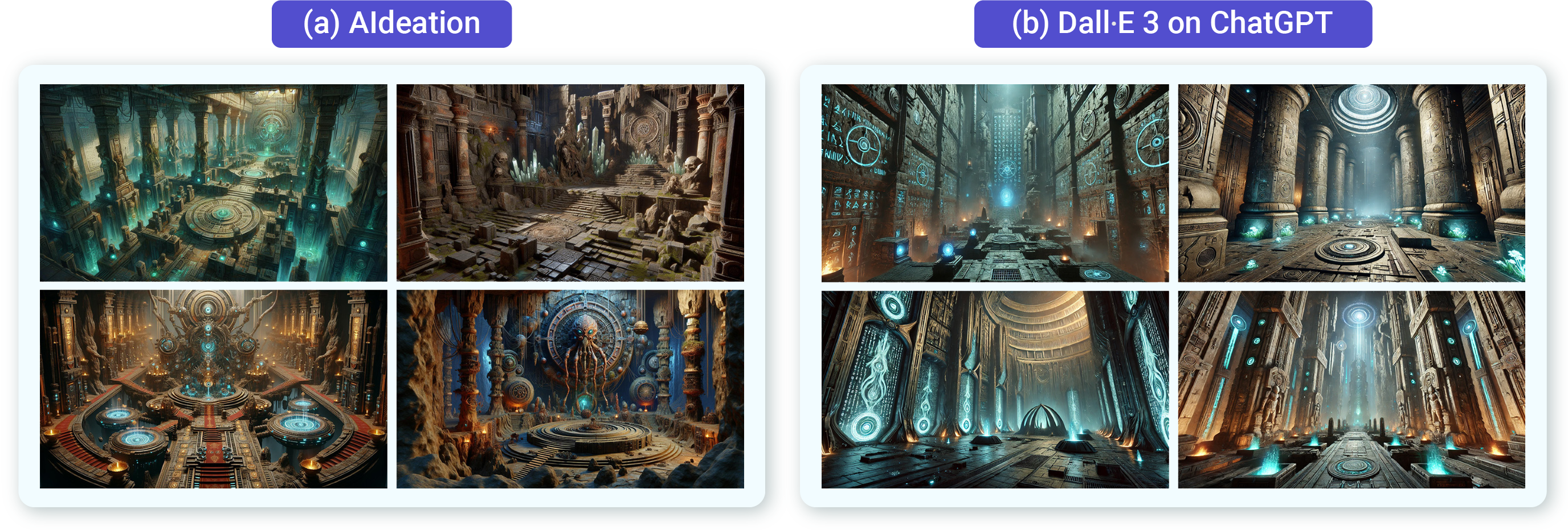}
    \caption{A comparison between the initial outputs from (a) AIdeation and (b) DALL-E 3 on ChatGPT, using the same input provided by Field Study Participant S3, revealed notable differences. The participant observed that AIdeation produced designs with significantly greater diversity and richness compared to those generated by DALL-E 3 on ChatGPT.}
    \Description{The figure compares design outputs generated by (a) AIdeation and (b) DALL·E 3 on ChatGPT using the same input from a field study participant. AIdeation produces more diverse and visually rich designs, incorporating intricate structures and varied compositions. In contrast, DALL·E 3 on ChatGPT generates more uniform designs with consistent lighting and patterns. The participant noted AIdeation’s greater variety in architectural complexity and artistic expression.}
    \label{fig:design tool}
\end{figure*}

\subsubsection{Design efficiency, quality, and creativity}
Table \ref{tab:time_comparison} compares the estimated time for their original workflows with the actual time spent using AIdeation. Both S3 and S4 reported significant efficiency improvements, with time spent reduced from 5 days to 2 days and 14 days to 6 days, respectively. Both participants emphasized that AIdeation helped them identify a design direction, particularly when they were unsure how to begin working from the provided design specifications. “\textit{(AIdeation) Can quickly provide multiple directions for our team to explore and develop}”(S4). However, S2 experienced a slight decrease in efficiency, they stated that “\textit{The client didn't seem to favor the artistic style generated by AIdeation, and the image generation process was somewhat time-consuming}.”

S1, S3, and S4 reported that AIdeation significantly enhanced the quality of their final designs. All three noted that AIdeation enriched their designs with a variety of elements. Figure \ref{fig:design tool} presents a comparison between the results generated by AIdeation and DALL-E 3 in ChatGPT, using the same input provided by a participant in the field study (S3). According to the participant, AIdeation results demonstrated significantly greater diversity and richness in design. The S1 project leader stated: “\textit{The generated images might not always have the exact level of detail I needed, but I was able to extract many useful design elements}.” The participant from S4 shared that when directors saw the scenes she created using AIdeation, they were “\textit{strikingly impressed}.”

All the studios agreed that AIdeation boosted the creativity of their final design. “\textit{I couldn’t stop exploring new concepts. Every iteration sparked an eagerness to try something I hadn’t thought of before}” (S3). “\textit{There were often some great unexpected outcomes that we ended up incorporating into our designs}” (S2).

Some participants also discovered uses beyond our expectations. For instance, participants from S1 used AIdeation to generate unique patterns, which they found difficult to source online, and incorporated them into their designs. The participant from S4 suggested that sometimes, simply uploading an image without providing any prompts could still generate quite good ideas.

\subsubsection{Challenges}
Participants also reported some issues. Users from S1 and S2 mentioned the styling and aesthetics problems. 
“\textit{The generated scenes were consistently symmetrical. This limited their ability to present the outputs effectively to clients, as the composition and layout lacked variation}” (S1).

Participants desired greater controllability, noting that AIdeation lacked a gradual generation feature. “\textit{Sometimes I just wanted to remove one element from the idea or adjust the composition, but the whole image changed}” (S4). “\textit{AIdeation tended to generate overly complex designs in the initial cycle, I needed to include instructions to simplify the designs}” (S3). A user from S1 suggested adding the Inpainting functionality for more detailed control.

% In summary, AIdeation presented both benefits and challenges in creative workflows. It boosted efficiency and creativity, especially in complex projects, but may also generated overly complex designs and had limitations in composition and styling. Users appreciated its ability to quickly generate diverse ideas but also emphasized the need for more control and faster generation times. Despite these issues, AIdeation proved helpful in early-stage ideation, with potential for improvement in control and customization to enhance its effectiveness across different projects.

\subsection{Case Studies}
We selected two cases to discuss both negative and positive feedback in depth: S2 showed limitations of AIdeation and opporunities for improvement, while S4 demonstrated significant improvements in quality, efficiency, and creativity.

\subsubsection{Case study \#1 - Large mountainous environment scenes (S2)}
Two environment concept designers were tasked with creating a mountain scene featuring a narrow path winding upward to a massive rocky summit. The final design would be used as a poster and web page background. For this project, the designers had previously used MidJourney's output to communicate with their clients and switched to using AIdeation exclusively for the field study. Starting with the client’s specifications and styling keywords, they first picked 25 ideas from 10 cycles and ultimately selected 5 images for the client to confirm the design direction. The client chose 2 images as the main reference points, which the designers then used to draft the final result. Figure \ref{fig:field study}-a shows examples of generated images and selected ones. 

% 遇到的問題是畫面客戶比較難想像成品 主要是畫面排列不甚自然 
% 在景深表現上空間感比較難拉出遙遠的空間感

While the designers reported that AIdeation improved the ideation process for them, their client strongly preferred the aesthetic qualities and depth rendering of MidJourney over Dall-E, which was used by AIdeation. 
In terms of working time, the designers exceeded their initial 4-hour estimate by one hour, unlike other projects that reported significant time savings. A key factor was that this project had clear and detailed design specifications from the client, making ideation a smaller part of the task. Most of the time was spent generating images that accurately aligned with the client’s design requirements and aesthetic preferences.
%The client also struggled to visualize the final product from AIdeation's outputs, which lacked strong spatial representation. 
%AIdeation's longer generation time also contributed to them.
%The designers attributed the inefficiency to two factors: 1) the project focused more on matte painting, emphasizing aesthetics over design complexity, and 2) their lack of familiarity with AIdeation. In this case, the client's design specifications were already clear and detailed, so the designers primarily needed the tool to generate images that accurately aligned with the design requirements. 

\subsubsection{Case study \#2 - Metroidvania game environment design (S4)}
The designer used AIdeation to create three key game scenes and the backgrounds for two secondary scenes, which will be directly incorporated into the final game. We selected the most interesting case from the key scenes: a steel bridge scene. The designer started with a rough concept from another game scene and was tasked with designing a steel bridge in the same style. The bridge's structure needed to be both complex and coherent. Before using AIdeation, she had spent two days experimenting with various approaches but could not create a satisfactory design.

\begin{figure*}
    \centering
    \includegraphics[width=1\linewidth]{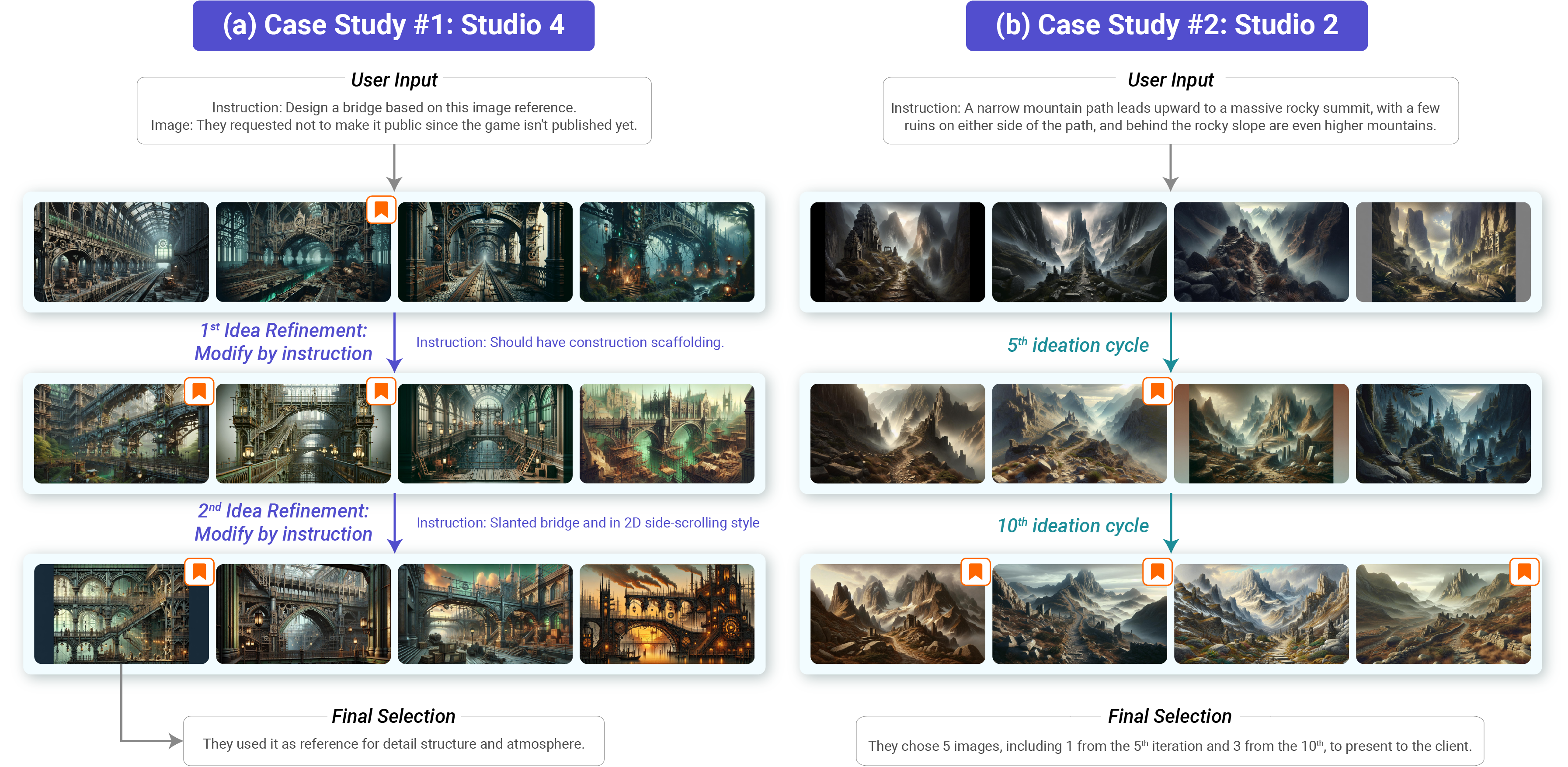}
    \caption{The workflow and results of using AIdeation on real-world projects from two designers in two studios (S4 and S2) are as follows: (a) The designer (S4) was tasked with creating a functional steel bridge based on an existing concept design. After just two idea refinements, the designer achieved the desired result, supporting the detailed structure design that she had already spent two days conceptualizing; (b) The designer was tasked with creating a mountain scene matte painting for a webpage background. After 10 cycles of ideation, they selected five results to present to the client. However, the entire process took an hour longer than their original workflow estimate.}
    \Description{The figure presents two case studies of designers using AIdeation in professional workflows. (a) In Studio 4, a designer was tasked with creating a functional steel bridge based on an existing concept. After two refinement cycles modifying structure and style, they finalized a design that complemented their previous two days of work. (b) In Studio 2, a designer iteratively refined a mountain scene matte painting for a webpage background. After 10 ideation cycles, they selected five images to present to the client. While AIdeation supported detailed exploration, the process took an hour longer than their original workflow estimate.}
    \label{fig:field study}
\end{figure*}

With AIdeation, she input the reference concept design with brief instructions. After the initial generation and only two iterations of "refining by instruction", she obtained the design elements she needed to realize her creative vision (Figure \ref{fig:field study}-b). For other tasks, she followed the same process, completing her designs and the art director was “\textit{strikingly impressed}.” The designer reported: “\textit{With a tight deadline for an upcoming game update and complex design tasks, I was prepared to work overtime for two weeks. Thanks to AIdeation providing multiple design solutions and many design elements, I managed to work overtime for just one week}” (S4).

In this case, the designer was tasked with creating complex scenes featuring intricate structures while adhering to a specific style, a process that typically demands extensive research, brainstorming, and finding suitable references. %AIdeation directly provided diverse design solutions and references, supporting the logical coherence of the designs. This enabled the designer to quickly go through multiple ideation cycles. Additionally, AIdeation's iterative exploration feature effectively helped the designer refine their ideas precisely.

% 這邊再放一個延續的work
% Studio 4 後面用了 1.5個月， 用AIdeation跑了69個cycle, 生了590個Idea, 做完了22個場景。因為這次死線是固定的，所以估計不出節省的時間，但品質是大幅提升。在使用AIdeation的這1.5個月，設計出來的場景讓主管覺得很震撼。品質超過沒有使用AIdeation前的作品。

% We are happy to share that after the field study, studio 1 and 4 keep using AIdeation. studio 1 後面用了 0.5個月，用AIdeation跑了35個cycle,生了221個Idea,做了6個場景。 用AIdeation大概節省了40 \% 的時間。其中用AIdeation生了很多特殊圖樣和雕紋大幅提升效率（原本要自己手繪）和作品品質. Studio 4 後面用了 1.5個月， 用AIdeation跑了69個cycle, 生了590個Idea, 做完了22個場景。因為這次死線是固定的，所以估計不出節省的時間，但品質是大幅提升。在使用AIdeation的這1.5個月，設計出來的場景讓主管覺得很震撼。品質超過沒有使用AIdeation前的作品。

\subsection{Continued Usage in Production beyond the Field Study}
We are excited to share that after the end of the field study, Studios 1 and 4 have continued to use AIdeation in production to date. For example, in one of Studio 1's projects, it iterated 35 cycles and 221 ideas to create 6 scenes in 2 weeks, saving approximately 40\% of the time while significantly enhancing quality, especially with intricate patterns and textures. In one of Studio 4's projects, it iterated 69 cycles and 590 ideas to create 22 scenes in 6 weeks. While time savings couldn't be measured due to a fixed deadline, the quality of their outputs greatly exceeded previous work, “\textit{The team leader was amazed by the quality of my work over the past 1.5 months}.” remarked by the participant from S4.

%In summary, AIdeation presented both benefits and challenges in creative workflows. It boosted efficiency and creativity, especially in complex projects, but may also generate overly complex designs with limitations in composition and styling. Users appreciated its ability to quickly generate diverse ideas but also emphasized the need for more control and faster generation times. Despite these issues, AIdeation proved helpful in early-stage ideation, with potential for improvement in control and customization to enhance its effectiveness across different projects.

\section{DISCUSSION, LIMITATIONS, AND FUTURE WORK}

% 2. AI expose Stucture，透明度，很容易讀（可以generalize），不知道背後在幹嘛，很難control，我們organize讓user能夠立刻了解這些key attributes。其他AI工具也可以參考這樣的做法。（expose這些attribute，讓使用者可以更精准控制） 1. 提供系統化資訊很快了解2 提供界面很快修改

\subsection{Addressing Barriers to Adoption: Transparency, Accuracy, and User Perceptions in AI Design Tools}
A significant proportion of designers and artists exhibit resistance to the adoption of GenAI tools ~\cite{kawakami2024impact, jiang2023ai}. Although concerns such as copyright and other factors discussed previously play a role, another critical reason for this reluctance is the lack of transparency in these systems ~\cite{zhang2024confrontation, shi2023understanding}. Without a clear understanding of the underlying actions of the system, users struggle to control and communicate with it effectively, ultimately reducing acceptance and adoption~\cite{auernhammer2020human, Usmani2023Human-Centered}. 
% 2nd review
To address this, prior work has explored enhancing interpretability and user control through multi-modal feedback and visualization. XCreation ~\cite{yan2023xcreation} integrates an entity-relation graph to visually map picture elements and their relationships, making generative structures more transparent. In product design, PhotoDreamer ~\cite{zhang2024protodreamer} allows designers to prototype with physical materials while AI interprets their inputs, providing clear feedback on how designs evolve. And AutoSpark~\cite{chen2024autospark} enables fine-grained comparisons to improve text-image relevance. 

On the other hand, AIdeation is specifically designed to meet the needs of concept designers by breaking down brainstorming results into visuals and categorized keywords, helping designers quickly grasp key attributes. Building on this understanding, AIdeation enables designers to fine-tune elements precisely. At each step, it eliminates the traditional need for designers to spend excessive time interpreting generated images or manually crafting and modifying complex prompts, while still preserving high-level control over design directions. As one participant noted: “\textit{Compared to other image-generation tools I've used before, I can clearly see what to do next, making it much more efficient to achieve the desired outcome}” (P14). By enhancing AI transparency and control of creative directions, AI design tools would improve engagement, foster human-AI collaboration, and improve user satisfaction, as proposed by human-centered AI design principles~\cite{shneiderman2022human}.

Hallucination is another critical concern in GenAI, 
Hegazy et al.\cite{hegazy2023evolution} identified potential issues with using GenAI in architectural design, such as a lack of consideration for structural feasibility and inconsistencies in generated outcomes. Similarly, concept designers rely heavily on factual, real-world information\cite{maleki2024ai, monteith2024artificial}, distinguishing them from other design disciplines. Both formative and summative studies revealed that designers hesitate to adopt AI tools due to fears of receiving inaccurate output, compounded by a general preference to avoid over-reliance on others' designs. As one participant (P2) explained: “\textit{I mainly use photos as references and avoid concept art since, despite looking good, it may lack thorough, factual research. AI-generated images have the same problem}.” 
While prior work in architectural design explores pre-trained models and ControlNet~\cite{zhang2023adding} to improve accuracy~\cite{chen2024enhancing}, these methods are unsuitable for concept design due to its broader scope.
To mitigate this problem, AIdeation integrates non-AI image search to provide real-world reference images, supporting the design elements of its generated ideas and aligning with designers' existing workflow for reference gathering. This approach significantly increased designers' willingness to engage with the tool. As another participant (P15) noted, “\textit{Although I still don't like AI-generated images, the keywords and references are very useful}.” These findings, coupled with our observations in Section 7.4.3, highlight the substantial impact of user attitudes and expectations on their experience with AI systems, a conclusion supported by recent research~\cite{kang2024impact}.

% Hegazy et al ~\cite{hegazy2023evolution} stated the possible problems using GenAI in architectural design, such as Lack of consideration for structural design feasibility and Inconsistency of outcomes.

These issues also extend to other domains. While GenAI is powerful, designers need to identify and address the root causes of possible negative attitudes toward it. A user-centered approach is helpful in identifying the root causes, making it possible to design strategies to specifically address each of users' concerns, incorporating both GenAI and traditional approaches as needed. 
%Ensuring systems are trustworthy, transparent, and aligned with workflows is critical. Addressing these challenges can lead to greater adoption and improved user satisfaction in future systems.

% 3. 這個領域有高比例反AI，1. 覺得資訊是錯誤的 （我們的設計就是用真實世界的search reference來support）User 對 AI 的態度差異。要注意User的negative thought。工具好用但不是每個user都這樣。要給什麼建議，針對對AI負面的user (資訊是不正確的) 用真實的東西去Back AI產出的Idea
% \textbf{對ai負面態度的原因很多類型，針對不同原因去設計}

\subsection{Implication for GenAI in Iterative Ideation}
% 4. 不同階段user會要不同的control。拆成不同Phase，creative process convergence divergence。過去的做法是會用slider。我們是拆成brainstorming和refinement (concept designer workflow)。不同creative process，背後概念都是這樣。怎麼去support 這個progression
Unlike existing AI tools commonly used by concept designers, which typically follow a linear, one-step solution, AIdeation adopts a nonlinear and iterative approach that aligns more closely with designers' ideation processes. This design philosophy is similar to frameworks such as OptiMuse ~\cite{OptiMuse} and DesignGPT ~\cite{ding2023designgpt}, recognizing iteration as a fundamental aspect of the design process ~\cite{adams1999cognitive}, and many prior work has incorporated this principles ~\cite{hou2024c2ideas, han2024teams}.
At different stages of the design process, designers may require varying levels of divergent and convergent thinking, along with cognitive processes that balance exploring both breadth and depth. ~\cite{tversky2011creativity, goldschmidt2016linkographic}. Tools such as RoomDreaming used sliders to control the diversity of visual outputs ~\cite{wang2024roomdreaming}, while GenQuery employs visual search and image combination techniques to dynamically shift focus ~\cite{son2024genquery}. In contrast, AIdeation organizes the functionality into modular components, where designers can switch between based on their current needs, providing the flexibility to adapt to different phases of the creative process. 
These concepts apply to most creative processes. Future work could explore how GenAI can support different stages of ideation across various creative domains while allowing users to seamlessly switch between them.

During interviews, many designers highlighted that AIdeation was significantly easier to control and communicate with compared to other AI tools they had used. One participant noted, “\textit{I feel that AIdeation can effectively understand how I wish to modify the current idea}” (P6). This observation highlights the importance of systems that understand user intentions and support clear and effective communication. ~\cite{verganti2020innovation, shneiderman2022human}. Previous work, such as IntentTuner, has proposed frameworks to integrate human intentions into fine-tuning general image generation systems ~\cite{zeng2024intenttuner}. In contrast, AIdeation uses domain-specific knowledge to guide each AI module, ensuring that it aligns with the different phases of concept design. This approach improves communication between the tool and designers.

The principles behind AIdeation can guide the future development of AI-assisted design tools. One promising direction is exploring how GenAI can better support collaboration, enabling directors and designers to co-create in shared workflows. Such systems could act as a communication bridge,  integrating team inputs and supporting both broad exploration and focused refinement. This aligns with the frameworks of Han et al., which highlights AI’s role in enhancing team creativity ~\cite{han2024teams}.

\subsection{Integrating GenAI into the Design Workflow with a Human-Centered AI Approach}
% 1. Revisit Contribution and Research Goal
% We are integrating AI into a group that already using GenAI
% Objective: Restate the research goals and contributions of AIdeation.
% Highlight how AIdeation fills gaps in existing tools by combining breadth and depth in idea exploration. ()
% Note its role in advancing HCI knowledge by supporting divergent and convergent thinking in iterative workflows.
% Notice: Emphasize novelty and address specific challenges (e.g., inefficiency in traditional workflows, limited control in current AI tools).

% 1. 從最broad開始講，大家都用AI tool， 但沒效率。各個domain都有這樣的work。
% Highlight human centered approach, 可以更符合使用者需求, 有兩間還在用
%The findings from both studies indicate that AIdeation effectively integrates multiple GenAI models to support the ideation process and address the complex workflows of concept designers. 

While GenAI tools are increasingly used by designers across various domains, research shows they often fail to align with user-centered design principles. These shortcomings often result in user reluctance and inefficiencies~\cite{vimpari2023adapt, zhang2024confrontation, mahdavi2024ai}. Aligned with established principles of human-centered AI design ~\cite{shneiderman2022human, xu2023transitioning, auernhammer2020human}, AIdeation provides a solution that prioritizes the needs and workflows of concept designers.

Previous research in various design domains has demonstrated the use of GenAI to simplify nuanced tasks, enabling designers to rapidly explore various visual concepts~\cite{wang2024roomdreaming, davis2024fashioning, oh2024lumimood}. Furthermore, studies have demonstrated the effectiveness of AI multi-agent collaboration in managing complex tasks~\cite{talebirad2023multi, de2024llmr}. 
AIdeation, on the other hand, deconstructs complex workflows into modular tasks, combining both suitable AI modules and non-AI tools for each phase and integrating them into a cohesive workflow for concept designers. This approach eliminates labor-intensive steps while retaining essential creative decisions, allowing users to focus on the core creative aspects of their work. In this context, GenAI functions as a tool to augment human capabilities ~\cite{chen2023next}. As one participant remarked, “\textit{Using AIdeation felt like being an art director, with multiple design assistants gathering information and proposing ideas}” (P13).

A similar approach can be generalized to other design domains that involve multiple phases of ideation, prototyping, and refinement, such as fashion, graphic, architectural, and industrial design~\cite{camburn2017design, carlgren2016framing}. Although many design fields have already integrated AI tools into their workflows~\cite{anantrasirichai2022artificial}, these tools often do not align with domain-specific needs, which presents a significant opportunity for HCI researchers to bridge this gap.
Instead of relying on one-size-fits-all AI solutions, researchers should use domain expertise to integrate the right tools, AI or otherwise, into workflows and ensure designers retain control over core creative decisions. This approach results in systems that better meet user needs and outperform traditional or purely AI-driven solutions.

\subsection{Limitations and Future Work}
\subsubsection{Limitations of the study}
% Discuss study duration, self report method
% Reflect on generalizability to other contexts or domains.
% Notice: Frame limitations constructively, linking them to potential future studies.
Due to the difficulty of including the entire ideation process in our summative study and the challenge of directly comparing the results of the ideation between conditions, we relied mainly on self-reported data, which is a limitation of this work. While a follow-up field study evaluated real-world design outputs with input from designers, directors, and clients, it lacked quantitative measures and had less control compared to lab studies. Future research could explore longer summative sessions focused on narrower tasks, like designing a single prop, which is simpler than broader tasks like environment design.

\subsubsection{Controllability}
% Objective: Discuss the trade-offs between user control and automation in AIdeation.
% Address how increased control could either support or hinder creative exploration.
% Notice: Use participant feedback to support arguments about user needs for control.
Although AIdeation emphasizes idea exploration, participants noted its limitations in controlling specific details of generated results. Features like "combine with the reference" and "refine by instruction" provide high-level control but lack the ability to adjust elements such as lighting, atmosphere, camera angles, and composition while preserving other elements. These aspects remain challenging and are active areas of AI research. As one participant (P4) remarked, “\textit{The system covers 70-80\% for client communication, but control over lighting, atmosphere, and camera angles is needed for the final 20\%}.” As AI technology continues to advance, such controllability features could be integrated into AIdeation. Future iterations of AIdeation could integrate such detailed controls to better support designers' focus and refinement during the convergence phase of their work.

\subsubsection{Customization and personalization}
% Customization and Personalization
% Objective: Reflect on how AIdeation can be personalized to suit diverse workflows.
% Explore adaptive AI features, such as tailoring outputs to user preferences or expertise levels.
% Notice: Highlight ethical implications, like avoiding bias and ensuring inclusivity.
Many users noted the limited diversity in art styles, atmosphere, and camera angles, largely due to the constraints of the image generation model used in AIdeation. Different models have distinct strengths; for instance, users appreciated MidJourney for its aesthetic quality, while Stable Diffusion, fine-tuned with LoRA ~\cite{hu2021lora}, offers more style variety and specialized designs. Future updates could let users select specific styles or atmospheres, choose fine-tuned models, or allow the system to automatically pick the most suitable model based on input. Another option could be to generate multiple outputs from different models to better match the design task.

Beyond image generation tuning, AIdeation can be personalized to fit the design field, the designer's specialization, and work habits, similar to the ideas proposed by Long et al.~\cite{long2024not}. The system could adapt to various design domains by modifying the prompts or highlighting specific design elements to better suit individual users. For instance, designers could select a focus, such as environments, props, or characters, and AIdeation would generate customized output accordingly. Although the system currently lacks the ability to retain context from previous sessions, future updates could include memory features and personalized recommendations. Furthermore, incorporating self-adaptive capabilities, where the system adjusts its behavior based on user preferences or current work stage, could further improve its effectiveness, as suggested in previous research ~\cite{macias2013self}.

\section{CONCLUSION}
AIdeation, as an idea exploration tool, offers a unique blend of traditional and AI-driven approaches, providing broad exploration through diverse references while generating new outputs based on user input. This enables designers to efficiently gather inspiration and produce a wide range of unique ideas. In addition, AIdeation streamlines the entire workflow by integrating research, brainstorming, reference gathering, and design refinement into one cohesive process, with the added flexibility of iterating design ideas easily. Our study with 16 professional concept designers revealed that AIdeation improved both the breadth and depth of idea exploration, significantly enhanced creativity, and increased efficiency in generating more diverse ideas of comparable quality to their original workflows. A field study conducted in four design studios further demonstrated that AI-driven ideation could potentially improve project efficiency, particularly for complex design tasks.

%Demonstrate ideation’s potential in aproving // to say how many days they save. Write a very strong result

%

%% The acknowledgments section is defined using the "acks" environment (and NOT an unnumbered section). This ensures the proper identification of the section in the article metadata, and the consistent spelling of the heading.
\begin{acks}
This work was supported by the National Science and Technology Council, Taiwan (NTSC 112-2221-E-002-185-MY3) and the Center of Data Intelligence: Technologies, Applications, and Systems at National Taiwan University (113L900901, 113L900902, 113L900903), funded through the Featured Areas Research Center Program under the Higher Education Sprout Project by the Ministry of Education (MOE) of Taiwan. We also acknowledge support from National Taiwan University, Moonshine Studio, Winking Studios, and Rayark Games. Finally, we extend our gratitude to all participants and reviewers for their valuable feedback.
\end{acks}

\bibliographystyle{ACM-Reference-Format}

\bibliography{main}


\begin{thebibliography}{125}

%%% the Web address.

\ifx \showCODEN    \undefined \def \showCODEN     #1{\unskip}     \fi
\ifx \showDOI      \undefined \def \showDOI       #1{#1}\fi
\ifx \showISBNx    \undefined \def \showISBNx     #1{\unskip}     \fi
\ifx \showISBNxiii \undefined \def \showISBNxiii  #1{\unskip}     \fi
\ifx \showISSN     \undefined \def \showISSN      #1{\unskip}     \fi
\ifx \showLCCN     \undefined \def \showLCCN      #1{\unskip}     \fi
\ifx \shownote     \undefined \def \shownote      #1{#1}          \fi
\ifx \showarticletitle \undefined \def \showarticletitle #1{#1}   \fi
\ifx \showURL      \undefined \def \showURL       {\relax}        \fi
% invisible to TeX
\providecommand\bibfield[2]{#2}
\providecommand\bibinfo[2]{#2}
\providecommand\natexlab[1]{#1}
\providecommand\showeprint[2][]{arXiv:#2}

\bibitem[3dtotal Publishing(2018)]%
        {conceptart2018}
\bibfield{author}{\bibinfo{person}{3dtotal Publishing}.}
  \bibinfo{year}{2018}\natexlab{}.
\newblock \bibinfo{booktitle}{\emph{The Ultimate Concept Art Career Guide}}.
\newblock \bibinfo{publisher}{3dtotal Publishing}.
\newblock


\bibitem[Adams and Atman(1999)]%
        {adams1999cognitive}
\bibfield{author}{\bibinfo{person}{Robin~S Adams} {and}
  \bibinfo{person}{Cynthia~J Atman}.} \bibinfo{year}{1999}\natexlab{}.
\newblock \showarticletitle{Cognitive processes in iterative design behavior}.
  In \bibinfo{booktitle}{\emph{FIE'99 Frontiers in Education. 29th Annual
  Frontiers in Education Conference. Designing the Future of Science and
  Engineering Education. Conference Proceedings (IEEE Cat. No. 99CH37011}},
  Vol.~\bibinfo{volume}{1}. IEEE, \bibinfo{pages}{11A6--13}.
\newblock


\bibitem[adobe(2020)]%
        {adobe2020}
\bibfield{author}{\bibinfo{person}{adobe}.} \bibinfo{year}{2020}\natexlab{}.
\newblock \bibinfo{title}{Bring an idea to life with concept art.}
\newblock
  \bibinfo{howpublished}{\url{https://www.adobe.com/mena_en/creativecloud/illustration/discover/concept-art.html}}.
\newblock


\bibitem[Almeda et~al\mbox{.}(2024)]%
        {almeda2024prompting}
\bibfield{author}{\bibinfo{person}{Shm~Garanganao Almeda}, \bibinfo{person}{JD
  Zamfirescu-Pereira}, \bibinfo{person}{Kyu~Won Kim}, \bibinfo{person}{Pradeep
  Mani~Rathnam}, {and} \bibinfo{person}{Bjoern Hartmann}.}
  \bibinfo{year}{2024}\natexlab{}.
\newblock \showarticletitle{Prompting for Discovery: Flexible Sense-Making for
  AI Art-Making with Dreamsheets}. In \bibinfo{booktitle}{\emph{Proceedings of
  the CHI Conference on Human Factors in Computing Systems}}.
  \bibinfo{pages}{1--17}.
\newblock


\bibitem[Anantrasirichai and Bull(2020)]%
        {Anantrasirichai2020Artificial}
\bibfield{author}{\bibinfo{person}{N. Anantrasirichai} {and}
  \bibinfo{person}{D. Bull}.} \bibinfo{year}{2020}\natexlab{}.
\newblock \showarticletitle{Artificial intelligence in the creative industries:
  a review}.
\newblock \bibinfo{journal}{\emph{Artificial Intelligence Review}}
  \bibinfo{volume}{55} (\bibinfo{year}{2020}), \bibinfo{pages}{589--656}.
\newblock
\urldef\tempurl%
\url{https://doi.org/10.1007/s10462-021-10039-7}
\showDOI{\tempurl}


\bibitem[Anantrasirichai and Bull(2022)]%
        {anantrasirichai2022artificial}
\bibfield{author}{\bibinfo{person}{Nantheera Anantrasirichai} {and}
  \bibinfo{person}{David Bull}.} \bibinfo{year}{2022}\natexlab{}.
\newblock \showarticletitle{Artificial intelligence in the creative industries:
  a review}.
\newblock \bibinfo{journal}{\emph{Artificial intelligence review}}
  \bibinfo{volume}{55}, \bibinfo{number}{1} (\bibinfo{year}{2022}),
  \bibinfo{pages}{589--656}.
\newblock


\bibitem[A.T.~Purcell(1992)]%
        {rw21}
\bibfield{author}{\bibinfo{person}{J.S.~Gero A.T.~Purcell}.}
  \bibinfo{year}{1992}\natexlab{}.
\newblock \showarticletitle{Effects of examples on the results of a design
  activity}.
\newblock \bibinfo{journal}{\emph{Knowledge-Based Systems, Volume 5, Issue 1}}
  (\bibinfo{date}{09} \bibinfo{year}{1992}).
\newblock
\urldef\tempurl%
\url{https://doi.org/10.1016/0950-7051(92)90026-C.}
\showDOI{\tempurl}


\bibitem[Auernhammer(2020)]%
        {auernhammer2020human}
\bibfield{author}{\bibinfo{person}{Jan Auernhammer}.}
  \bibinfo{year}{2020}\natexlab{}.
\newblock \showarticletitle{Human-centered AI: The role of Human-centered
  Design Research in the development of AI}.
\newblock  (\bibinfo{year}{2020}).
\newblock


\bibitem[Boston et~al\mbox{.}(1998)]%
        {boston1998design}
\bibfield{author}{\bibinfo{person}{OP Boston}, \bibinfo{person}{AW Court},
  \bibinfo{person}{SJ Culley}, {and} \bibinfo{person}{CA McMahon}.}
  \bibinfo{year}{1998}\natexlab{}.
\newblock \showarticletitle{Design information issues in new product
  development}. In \bibinfo{booktitle}{\emph{The design productivity debate}}.
  Springer, \bibinfo{pages}{231--254}.
\newblock


\bibitem[Boucher et~al\mbox{.}(2024)]%
        {boucher2024resistance}
\bibfield{author}{\bibinfo{person}{Josiah~D Boucher}, \bibinfo{person}{Gillian
  Smith}, {and} \bibinfo{person}{Yunus~Do{\u{g}}an Telliel}.}
  \bibinfo{year}{2024}\natexlab{}.
\newblock \showarticletitle{Is Resistance Futile?: Early Career Game
  Developers, Generative AI, and Ethical Skepticism}. In
  \bibinfo{booktitle}{\emph{Proceedings of the CHI Conference on Human Factors
  in Computing Systems}}. \bibinfo{pages}{1--13}.
\newblock


\bibitem[Brade et~al\mbox{.}(2023)]%
        {rw23}
\bibfield{author}{\bibinfo{person}{Stephen Brade}, \bibinfo{person}{Bryan
  Wang}, \bibinfo{person}{Mauricio Sousa}, \bibinfo{person}{Sageev Oore}, {and}
  \bibinfo{person}{Tovi Grossman}.} \bibinfo{year}{2023}\natexlab{}.
\newblock \showarticletitle{Promptify: Text-to-Image Generation through
  Interactive Prompt Exploration with Large Language Models}.
\newblock \bibinfo{journal}{\emph{Association for Computing Machinery}}
  (\bibinfo{date}{10} \bibinfo{year}{2023}).
\newblock
\urldef\tempurl%
\url{https://doi.org/10.1145/3586183.3606725}
\showDOI{\tempurl}


\bibitem[Cai et~al\mbox{.}(2023)]%
        {rw22}
\bibfield{author}{\bibinfo{person}{Alice Cai}, \bibinfo{person}{Steven~R Rick},
  \bibinfo{person}{Jennifer~L Heyman}, \bibinfo{person}{Yanxia Zhang},
  \bibinfo{person}{Alexandre Filipowicz}, \bibinfo{person}{Matthew Hong},
  \bibinfo{person}{Matt Klenk}, {and} \bibinfo{person}{Thomas Malone}.}
  \bibinfo{year}{2023}\natexlab{}.
\newblock \showarticletitle{DesignAID: Using Generative AI and Semantic
  Diversity for Design Inspiration}.
\newblock \bibinfo{journal}{\emph{Association for Computing Machinery}}
  (\bibinfo{date}{11} \bibinfo{year}{2023}).
\newblock
\urldef\tempurl%
\url{https://doi.org/10.1145/3582269.3615596}
\showDOI{\tempurl}


\bibitem[Camburn et~al\mbox{.}(2017)]%
        {camburn2017design}
\bibfield{author}{\bibinfo{person}{Bradley Camburn}, \bibinfo{person}{Vimal
  Viswanathan}, \bibinfo{person}{Julie Linsey}, \bibinfo{person}{David
  Anderson}, \bibinfo{person}{Daniel Jensen}, \bibinfo{person}{Richard
  Crawford}, \bibinfo{person}{Kevin Otto}, {and} \bibinfo{person}{Kristin
  Wood}.} \bibinfo{year}{2017}\natexlab{}.
\newblock \showarticletitle{Design prototyping methods: state of the art in
  strategies, techniques, and guidelines}.
\newblock \bibinfo{journal}{\emph{Design Science}}  \bibinfo{volume}{3}
  (\bibinfo{year}{2017}), \bibinfo{pages}{e13}.
\newblock


\bibitem[Capanu et~al\mbox{.}(2006)]%
        {capanu2006testing}
\bibfield{author}{\bibinfo{person}{Marinela Capanu}, \bibinfo{person}{Gregory~A
  Jones}, {and} \bibinfo{person}{Ronald~H Randles}.}
  \bibinfo{year}{2006}\natexlab{}.
\newblock \showarticletitle{Testing for preference using a sum of Wilcoxon
  signed rank statistics}.
\newblock \bibinfo{journal}{\emph{Computational statistics \& data analysis}}
  \bibinfo{volume}{51}, \bibinfo{number}{2} (\bibinfo{year}{2006}),
  \bibinfo{pages}{793--796}.
\newblock


\bibitem[Carlgren et~al\mbox{.}(2016)]%
        {carlgren2016framing}
\bibfield{author}{\bibinfo{person}{Lisa Carlgren}, \bibinfo{person}{Ingo
  Rauth}, {and} \bibinfo{person}{Maria Elmquist}.}
  \bibinfo{year}{2016}\natexlab{}.
\newblock \showarticletitle{Framing design thinking: The concept in idea and
  enactment}.
\newblock \bibinfo{journal}{\emph{Creativity and innovation management}}
  \bibinfo{volume}{25}, \bibinfo{number}{1} (\bibinfo{year}{2016}),
  \bibinfo{pages}{38--57}.
\newblock


\bibitem[Chen et~al\mbox{.}(2024b)]%
        {chen2024enhancing}
\bibfield{author}{\bibinfo{person}{Feiran Chen}, \bibinfo{person}{Mengran Mai},
  \bibinfo{person}{Xinyi Huang}, {and} \bibinfo{person}{Yinghan Li}.}
  \bibinfo{year}{2024}\natexlab{b}.
\newblock \showarticletitle{Enhancing the Sustainability of AI Technology in
  Architectural Design: Improving the Matching Accuracy of Chinese-Style
  Buildings}.
\newblock \bibinfo{journal}{\emph{Sustainability}} \bibinfo{volume}{16},
  \bibinfo{number}{19} (\bibinfo{year}{2024}), \bibinfo{pages}{8414}.
\newblock


\bibitem[Chen et~al\mbox{.}(2024a)]%
        {chen2024autospark}
\bibfield{author}{\bibinfo{person}{Liuqing Chen}, \bibinfo{person}{Qianzhi
  Jing}, \bibinfo{person}{Yixin Tsang}, \bibinfo{person}{Qianyi Wang},
  \bibinfo{person}{Ruocong Liu}, \bibinfo{person}{Duowei Xia},
  \bibinfo{person}{Yunzhan Zhou}, {and} \bibinfo{person}{Lingyun Sun}.}
  \bibinfo{year}{2024}\natexlab{a}.
\newblock \showarticletitle{AutoSpark: Supporting Automobile Appearance Design
  Ideation with Kansei Engineering and Generative AI}. In
  \bibinfo{booktitle}{\emph{Proceedings of the 37th Annual ACM Symposium on
  User Interface Software and Technology}}. \bibinfo{pages}{1--19}.
\newblock


\bibitem[Chen et~al\mbox{.}(2023b)]%
        {chen2023aircharge}
\bibfield{author}{\bibinfo{person}{Po-Yu Chen}, \bibinfo{person}{Ching-Yi
  Tsai}, \bibinfo{person}{Wei-Hsin Wang}, \bibinfo{person}{Chao-Jung Lai},
  \bibinfo{person}{Chia-An Fan}, \bibinfo{person}{Shih~Chin Lin},
  \bibinfo{person}{Chia-Chen Chi}, {and} \bibinfo{person}{Mike~Y Chen}.}
  \bibinfo{year}{2023}\natexlab{b}.
\newblock \showarticletitle{AirCharge: Amplifying Ungrounded Impact Force by
  Accumulating Air Propulsion Momentum}. In
  \bibinfo{booktitle}{\emph{Proceedings of the 36th Annual ACM Symposium on
  User Interface Software and Technology}}. \bibinfo{pages}{1--11}.
\newblock


\bibitem[Chen et~al\mbox{.}(2023a)]%
        {chen2023next}
\bibfield{author}{\bibinfo{person}{Xiang'Anthony' Chen}, \bibinfo{person}{Jeff
  Burke}, \bibinfo{person}{Ruofei Du}, \bibinfo{person}{Matthew~K Hong},
  \bibinfo{person}{Jennifer Jacobs}, \bibinfo{person}{Philippe Laban},
  \bibinfo{person}{Dingzeyu Li}, \bibinfo{person}{Nanyun Peng},
  \bibinfo{person}{Karl~DD Willis}, \bibinfo{person}{Chien-Sheng Wu},
  {et~al\mbox{.}}} \bibinfo{year}{2023}\natexlab{a}.
\newblock \showarticletitle{Next steps for human-centered generative ai: A
  technical perspective}.
\newblock \bibinfo{journal}{\emph{arXiv preprint arXiv:2306.15774}}
  (\bibinfo{year}{2023}).
\newblock


\bibitem[Chen~Chen(2024)]%
        {rw20}
\bibfield{author}{\bibinfo{person}{Thibault Groueix Vladimir G. Kim
  Nadir~Weibel Chen~Chen, Cuong~Nguyen}.} \bibinfo{year}{2024}\natexlab{}.
\newblock \showarticletitle{MemoVis: A GenAI-Powered Tool for Creating
  Companion Reference Images for 3D Design Feedback}.
\newblock \bibinfo{journal}{\emph{ACM Transactions on Computer-Human
  Interaction, 2024}} (\bibinfo{date}{09} \bibinfo{year}{2024}).
\newblock
\urldef\tempurl%
\url{https://doi.org/10.48550/arXiv.2409.06082}
\showDOI{\tempurl}


\bibitem[Cheshire(2021)]%
        {artstation2021}
\bibfield{author}{\bibinfo{person}{Bob Cheshire}.}
  \bibinfo{year}{2021}\natexlab{}.
\newblock \bibinfo{title}{Concept Art: Interview with Bob Cheshire}.
\newblock
  \bibinfo{howpublished}{\url{https://magazine.artstation.com/2021/09/concept-art-interview-with-bob-cheshire/}}.
\newblock


\bibitem[Chiou et~al\mbox{.}(2023)]%
        {DesigningChiou2023}
\bibfield{author}{\bibinfo{person}{Li-Yuan Chiou}, \bibinfo{person}{Peng-Kai
  Hung}, \bibinfo{person}{Rung-Huei Liang}, {and} \bibinfo{person}{Chun-Teng
  Wang}.} \bibinfo{year}{2023}\natexlab{}.
\newblock \showarticletitle{Designing with AI: An Exploration of Co-Ideation
  with Image Generators}.
\newblock  (\bibinfo{year}{2023}), \bibinfo{pages}{1941–1954}.
\newblock
\showISBNx{9781450398930}
\urldef\tempurl%
\url{https://doi.org/10.1145/3563657.3596001}
\showDOI{\tempurl}


\bibitem[Choi et~al\mbox{.}(2024)]%
        {choi2024creativeconnect}
\bibfield{author}{\bibinfo{person}{DaEun Choi}, \bibinfo{person}{Sumin Hong},
  \bibinfo{person}{Jeongeon Park}, \bibinfo{person}{John Joon~Young Chung},
  {and} \bibinfo{person}{Juho Kim}.} \bibinfo{year}{2024}\natexlab{}.
\newblock \showarticletitle{CreativeConnect: Supporting Reference Recombination
  for Graphic Design Ideation with Generative AI}. In
  \bibinfo{booktitle}{\emph{Proceedings of the CHI Conference on Human Factors
  in Computing Systems}}. \bibinfo{pages}{1--25}.
\newblock


\bibitem[Claudia~Eckert(2019)]%
        {rw11}
\bibfield{author}{\bibinfo{person}{Martin~Stacey Claudia~Eckert}.}
  \bibinfo{year}{2019}\natexlab{}.
\newblock \showarticletitle{Sources of inspiration: a language of design}.
\newblock \bibinfo{journal}{\emph{Design Studies, Volume 21, Issue 5, 2000,
  Pages 523-538, ISSN 0142-694X}} (\bibinfo{date}{06} \bibinfo{year}{2019}).
\newblock
\urldef\tempurl%
\url{https://doi.org/10.1016/S0142-694X(00)00022-3.}
\showDOI{\tempurl}


\bibitem[Conover(1999)]%
        {conover1999practical}
\bibfield{author}{\bibinfo{person}{William~Jay Conover}.}
  \bibinfo{year}{1999}\natexlab{}.
\newblock \bibinfo{booktitle}{\emph{Practical nonparametric statistics}}.
  Vol.~\bibinfo{volume}{350}.
\newblock \bibinfo{publisher}{john wiley \& sons}.
\newblock


\bibitem[Contributors(2023)]%
        {comfyui}
\bibfield{author}{\bibinfo{person}{ComfyUI Contributors}.}
  \bibinfo{year}{2023}\natexlab{}.
\newblock \bibinfo{title}{ComfyUI: A powerful and modular Stable Diffusion GUI
  and backend}.
\newblock
  \bibinfo{howpublished}{\url{https://github.com/comfyanonymous/ComfyUI}}.
\newblock


\bibitem[David G.~Jansson(2024)]%
        {rw4}
\bibfield{author}{\bibinfo{person}{Steven M.~Smi David G.~Jansson}.}
  \bibinfo{year}{2024}\natexlab{}.
\newblock \showarticletitle{Design fixation}.
\newblock \bibinfo{journal}{\emph{Design Studies, Volume 12, Issue 1}}
  (\bibinfo{date}{11} \bibinfo{year}{2024}).
\newblock
\urldef\tempurl%
\url{https://doi.org/10.1016/0142-694X(91)90003-F}
\showDOI{\tempurl}


\bibitem[Davis et~al\mbox{.}(2024)]%
        {davis2024fashioning}
\bibfield{author}{\bibinfo{person}{Richard~Lee Davis}, \bibinfo{person}{Thiemo
  Wambsganss}, \bibinfo{person}{Wei Jiang}, \bibinfo{person}{Kevin~Gonyop Kim},
  \bibinfo{person}{Tanja K{\"a}ser}, {and} \bibinfo{person}{Pierre
  Dillenbourg}.} \bibinfo{year}{2024}\natexlab{}.
\newblock \showarticletitle{Fashioning Creative Expertise with Generative AI:
  Graphical Interfaces for Design Space Exploration Better Support Ideation
  Than Text Prompts}. In \bibinfo{booktitle}{\emph{Proceedings of the CHI
  Conference on Human Factors in Computing Systems}}. \bibinfo{pages}{1--26}.
\newblock


\bibitem[De~La~Torre et~al\mbox{.}(2024)]%
        {de2024llmr}
\bibfield{author}{\bibinfo{person}{Fernanda De~La~Torre},
  \bibinfo{person}{Cathy~Mengying Fang}, \bibinfo{person}{Han Huang},
  \bibinfo{person}{Andrzej Banburski-Fahey}, \bibinfo{person}{Judith
  Amores~Fernandez}, {and} \bibinfo{person}{Jaron Lanier}.}
  \bibinfo{year}{2024}\natexlab{}.
\newblock \showarticletitle{Llmr: Real-time prompting of interactive worlds
  using large language models}. In \bibinfo{booktitle}{\emph{Proceedings of the
  CHI Conference on Human Factors in Computing Systems}}.
  \bibinfo{pages}{1--22}.
\newblock


\bibitem[del Mar~Navarro(2022)]%
        {Navarro2022Research}
\bibfield{author}{\bibinfo{person}{María del Mar~Navarro}.}
  \bibinfo{year}{2022}\natexlab{}.
\newblock \showarticletitle{Research for Designers: A Guide to Methods and
  Practice}.
\newblock \bibinfo{journal}{\emph{Design and Culture}}  \bibinfo{volume}{15}
  (\bibinfo{year}{2022}), \bibinfo{pages}{130 -- 133}.
\newblock
\urldef\tempurl%
\url{https://doi.org/10.1080/17547075.2022.2138144}
\showDOI{\tempurl}


\bibitem[Ding et~al\mbox{.}(2023)]%
        {ding2023designgpt}
\bibfield{author}{\bibinfo{person}{Shiying Ding}, \bibinfo{person}{Xinyi Chen},
  \bibinfo{person}{Yan Fang}, \bibinfo{person}{Wenrui Liu},
  \bibinfo{person}{Yiwu Qiu}, {and} \bibinfo{person}{Chunlei Chai}.}
  \bibinfo{year}{2023}\natexlab{}.
\newblock \showarticletitle{DesignGPT: Multi-Agent Collaboration in Design}. In
  \bibinfo{booktitle}{\emph{2023 16th International Symposium on Computational
  Intelligence and Design (ISCID)}}. IEEE, \bibinfo{pages}{204--208}.
\newblock


\bibitem[Dow et~al\mbox{.}(2005)]%
        {dow2005wizard}
\bibfield{author}{\bibinfo{person}{Steven Dow}, \bibinfo{person}{Blair
  MacIntyre}, \bibinfo{person}{Jaemin Lee}, \bibinfo{person}{Christopher
  Oezbek}, \bibinfo{person}{Jay~David Bolter}, {and} \bibinfo{person}{Maribeth
  Gandy}.} \bibinfo{year}{2005}\natexlab{}.
\newblock \showarticletitle{Wizard of Oz support throughout an iterative design
  process}.
\newblock \bibinfo{journal}{\emph{IEEE Pervasive Computing}}
  \bibinfo{volume}{4}, \bibinfo{number}{4} (\bibinfo{year}{2005}),
  \bibinfo{pages}{18--26}.
\newblock


\bibitem[Eckert and Stacey(2000a)]%
        {rw18}
\bibfield{author}{\bibinfo{person}{Claudia Eckert} {and}
  \bibinfo{person}{Martin. Stacey}.} \bibinfo{year}{2000}\natexlab{a}.
\newblock \showarticletitle{Sources of Inspiration: A Language of Design.}
\newblock \bibinfo{journal}{\emph{Design Studies. 21..}} (\bibinfo{date}{09}
  \bibinfo{year}{2000}).
\newblock
\urldef\tempurl%
\url{https://doi.org/10.1016/S0142-694X(00)00022-3.}
\showDOI{\tempurl}


\bibitem[Eckert and Stacey(2000b)]%
        {eckert2000sources}
\bibfield{author}{\bibinfo{person}{Claudia Eckert} {and}
  \bibinfo{person}{Martin Stacey}.} \bibinfo{year}{2000}\natexlab{b}.
\newblock \showarticletitle{Sources of inspiration: a language of design}.
\newblock \bibinfo{journal}{\emph{Design studies}} \bibinfo{volume}{21},
  \bibinfo{number}{5} (\bibinfo{year}{2000}), \bibinfo{pages}{523--538}.
\newblock


\bibitem[Ehsan et~al\mbox{.}(2021)]%
        {Ehsan2021Expanding}
\bibfield{author}{\bibinfo{person}{Upol Ehsan}, \bibinfo{person}{Q. Liao},
  \bibinfo{person}{Michael~J. Muller}, \bibinfo{person}{Mark~O. Riedl}, {and}
  \bibinfo{person}{Justin~D. Weisz}.} \bibinfo{year}{2021}\natexlab{}.
\newblock \showarticletitle{Expanding Explainability: Towards Social
  Transparency in AI systems}.
\newblock \bibinfo{journal}{\emph{Proceedings of the 2021 CHI Conference on
  Human Factors in Computing Systems}} (\bibinfo{year}{2021}).
\newblock
\urldef\tempurl%
\url{https://doi.org/10.1145/3411764.3445188}
\showDOI{\tempurl}


\bibitem[Epstein et~al\mbox{.}(2023)]%
        {Epstein2023Art}
\bibfield{author}{\bibinfo{person}{Ziv Epstein}, \bibinfo{person}{Aaron
  Hertzmann}, \bibinfo{person}{L. Herman}, \bibinfo{person}{Robert Mahari},
  \bibinfo{person}{M. Frank}, \bibinfo{person}{Matthew Groh},
  \bibinfo{person}{Hope Schroeder}, \bibinfo{person}{Amy Smith},
  \bibinfo{person}{Memo Akten}, \bibinfo{person}{Jessica Fjeld},
  \bibinfo{person}{H. Farid}, \bibinfo{person}{Neil Leach}, \bibinfo{person}{A.
  Pentland}, {and} \bibinfo{person}{Olga Russakovsky}.}
  \bibinfo{year}{2023}\natexlab{}.
\newblock \showarticletitle{Art and the science of generative AI}.
\newblock \bibinfo{journal}{\emph{Science}}  \bibinfo{volume}{380}
  (\bibinfo{year}{2023}), \bibinfo{pages}{1110 -- 1111}.
\newblock
\urldef\tempurl%
\url{https://doi.org/10.1126/science.adh4451}
\showDOI{\tempurl}


\bibitem[et~al.(2024)]%
        {pilot}
\bibfield{author}{\bibinfo{person}{Zelun Tony~Zhang et al.}}
  \bibinfo{year}{2024}\natexlab{}.
\newblock \showarticletitle{Beyond Recommendations: From Backward to Forward AI
  Support of Pilots’ Decision-Making Process}.
\newblock \bibinfo{journal}{\emph{ACM Hum.-Comput. Interact}}
  (\bibinfo{year}{2024}).
\newblock


\bibitem[Evangelidis(2024)]%
        {evangelidis2024task}
\bibfield{author}{\bibinfo{person}{Ioannis Evangelidis}.}
  \bibinfo{year}{2024}\natexlab{}.
\newblock \showarticletitle{Task sensitivity and noise: How mechanical
  properties of preference elicitation tasks account for differences in
  preferences across tasks.}
\newblock \bibinfo{journal}{\emph{Decision}} \bibinfo{volume}{11},
  \bibinfo{number}{2} (\bibinfo{year}{2024}), \bibinfo{pages}{283}.
\newblock


\bibitem[Fullerton(2018)]%
        {gameworkshop2018}
\bibfield{author}{\bibinfo{person}{Tracy Fullerton}.}
  \bibinfo{year}{2018}\natexlab{}.
\newblock \bibinfo{booktitle}{\emph{Game Design Workshop: A Playcentric
  Approach to Creating Innovative Games, Fourth Edition}}.
\newblock \bibinfo{publisher}{A K Peters/CRC Press}.
\newblock


\bibitem[Gabriela~Goldschmidt(2019)]%
        {rw12}
\bibfield{author}{\bibinfo{person}{Anat Litan~Sever Gabriela~Goldschmidt}.}
  \bibinfo{year}{2019}\natexlab{}.
\newblock \showarticletitle{Inspiring design ideas with texts}.
\newblock \bibinfo{journal}{\emph{Design Studies,Volume 32, Issue 2,Pages
  139-155,ISSN 0142-694X}} (\bibinfo{date}{06} \bibinfo{year}{2019}).
\newblock
\urldef\tempurl%
\url{https://doi.org/10.1016/j.destud.2010.09.006.}
\showDOI{\tempurl}


\bibitem[Gallet(2023)]%
        {iterationandreference2023}
\bibfield{author}{\bibinfo{person}{Mari Gallet}.}
  \bibinfo{year}{2023}\natexlab{}.
\newblock \bibinfo{title}{Iteration and Reference in Concept Art}.
\newblock
  \bibinfo{howpublished}{\url{https://discover.therookies.co/2023/11/10/iteration-and-reference-in-concept-art/}}.
\newblock


\bibitem[Goldschmidt(2016)]%
        {goldschmidt2016linkographic}
\bibfield{author}{\bibinfo{person}{Gabriela Goldschmidt}.}
  \bibinfo{year}{2016}\natexlab{}.
\newblock \showarticletitle{Linkographic evidence for concurrent divergent and
  convergent thinking in creative design}.
\newblock \bibinfo{journal}{\emph{Creativity research journal}}
  \bibinfo{volume}{28}, \bibinfo{number}{2} (\bibinfo{year}{2016}),
  \bibinfo{pages}{115--122}.
\newblock


\bibitem[Han et~al\mbox{.}(2024)]%
        {han2024teams}
\bibfield{author}{\bibinfo{person}{Yuanning Han}, \bibinfo{person}{Ziyi Qiu},
  \bibinfo{person}{Jiale Cheng}, {and} \bibinfo{person}{RAY LC}.}
  \bibinfo{year}{2024}\natexlab{}.
\newblock \showarticletitle{When Teams Embrace AI: Human Collaboration
  Strategies in Generative Prompting in a Creative Design Task}. In
  \bibinfo{booktitle}{\emph{Proceedings of the CHI Conference on Human Factors
  in Computing Systems}}. \bibinfo{pages}{1--14}.
\newblock


\bibitem[Hegazy and Saleh(2023)]%
        {hegazy2023evolution}
\bibfield{author}{\bibinfo{person}{Muhammad Hegazy} {and}
  \bibinfo{person}{Ahmed Saleh}.} \bibinfo{year}{2023}\natexlab{}.
\newblock \showarticletitle{Evolution of AI role in architectural design:
  between parametric exploration and machine hallucination}.
\newblock \bibinfo{journal}{\emph{MSA Engineering Journal}}
  \bibinfo{volume}{2}, \bibinfo{number}{2} (\bibinfo{year}{2023}),
  \bibinfo{pages}{262--288}.
\newblock


\bibitem[Hois et~al\mbox{.}(2019)]%
        {Hois2019How}
\bibfield{author}{\bibinfo{person}{J. Hois}, \bibinfo{person}{Dimitra
  Theofanou-Fuelbier}, {and} \bibinfo{person}{Alischa~Janine Junk}.}
  \bibinfo{year}{2019}\natexlab{}.
\newblock \showarticletitle{How to Achieve Explainability and Transparency in
  Human AI Interaction}.
\newblock  (\bibinfo{year}{2019}), \bibinfo{pages}{177--183}.
\newblock
\urldef\tempurl%
\url{https://doi.org/10.1007/978-3-030-23528-4_25}
\showDOI{\tempurl}


\bibitem[Hou et~al\mbox{.}(2024)]%
        {hou2024c2ideas}
\bibfield{author}{\bibinfo{person}{Yihan Hou}, \bibinfo{person}{Manling Yang},
  \bibinfo{person}{Hao Cui}, \bibinfo{person}{Lei Wang}, \bibinfo{person}{Jie
  Xu}, {and} \bibinfo{person}{Wei Zeng}.} \bibinfo{year}{2024}\natexlab{}.
\newblock \showarticletitle{C2Ideas: Supporting Creative Interior Color Design
  Ideation with a Large Language Model}. In
  \bibinfo{booktitle}{\emph{Proceedings of the CHI Conference on Human Factors
  in Computing Systems}}. \bibinfo{pages}{1--18}.
\newblock


\bibitem[Hsuan-Yi~Wu(2016)]%
        {Imagination}
\bibfield{author}{\bibinfo{person}{Vic~Callaghan Hsuan-Yi~Wu}.}
  \bibinfo{year}{2016}\natexlab{}.
\newblock \showarticletitle{From Imagination to Innovation: A Creative
  Development Process}.
\newblock  (\bibinfo{year}{2016}).
\newblock


\bibitem[Hu et~al\mbox{.}(2021)]%
        {hu2021lora}
\bibfield{author}{\bibinfo{person}{Edward~J. Hu}, \bibinfo{person}{Yelong
  Shen}, \bibinfo{person}{Phillip Wallis}, \bibinfo{person}{Zeyuan Allen-Zhu},
  \bibinfo{person}{Yuanzhi Li}, \bibinfo{person}{Shean Wang}, {and}
  \bibinfo{person}{Weizhu Chen}.} \bibinfo{year}{2021}\natexlab{}.
\newblock \showarticletitle{LoRA: Low-Rank Adaptation of Large Language
  Models}. In \bibinfo{booktitle}{\emph{International Conference on Learning
  Representations}}.
\newblock
\urldef\tempurl%
\url{https://arxiv.org/abs/2106.09685}
\showURL{%
\tempurl}


\bibitem[Hua(2019)]%
        {rw14}
\bibfield{author}{\bibinfo{person}{Min. Hua}.} \bibinfo{year}{2019}\natexlab{}.
\newblock \showarticletitle{The Roles of Sketching in Supporting Creative
  Design}.
\newblock \bibinfo{journal}{\emph{The Design Journal}} (\bibinfo{date}{06}
  \bibinfo{year}{2019}).
\newblock
\urldef\tempurl%
\url{https://doi.org/10.1080/14606925.2019.1655187}
\showDOI{\tempurl}


\bibitem[Huang et~al\mbox{.}(2024)]%
        {PlantoGraphy2024}
\bibfield{author}{\bibinfo{person}{Rong Huang}, \bibinfo{person}{Haichuan Lin},
  \bibinfo{person}{Chuanzhang Chen}, \bibinfo{person}{Kang Zhang}, {and}
  \bibinfo{person}{Wei Zeng}.} \bibinfo{year}{2024}\natexlab{}.
\newblock \showarticletitle{PlantoGraphy: Incorporating Iterative Design
  Process into Generative Artificial Intelligence for Landscape Rendering}.
\newblock , Article \bibinfo{articleno}{168} (\bibinfo{year}{2024}),
  \bibinfo{numpages}{19}~pages.
\newblock
\showISBNx{9798400703300}
\urldef\tempurl%
\url{https://doi.org/10.1145/3613904.3642824}
\showDOI{\tempurl}


\bibitem[Huang et~al\mbox{.}(2022)]%
        {navipath2023}
\bibfield{author}{\bibinfo{person}{Ting-Hao~'Kenneth' Huang},
  \bibinfo{person}{Boyang Li}, \bibinfo{person}{Ying Xu}, \bibinfo{person}{Alan
  Black}, {and} \bibinfo{person}{Jeffrey~P. Bigham}.}
  \bibinfo{year}{2022}\natexlab{}.
\newblock \showarticletitle{NaviPath: Enhancing User Navigation and Interaction
  in Complex Information Spaces}. In \bibinfo{booktitle}{\emph{Proceedings of
  the CHI Conference on Human Factors in Computing Systems}}.
  \bibinfo{pages}{39--50}.
\newblock


\bibitem[Javaid(2021)]%
        {rw6}
\bibfield{author}{\bibinfo{person}{james~paul. Javaid, Syed \&~Pandarakalam}.}
  \bibinfo{year}{2021}\natexlab{}.
\newblock \showarticletitle{The Association of Creativity with Divergent and
  Convergent Thinking}.
\newblock \bibinfo{journal}{\emph{Psychiatria Danubina. 33. 133-139}}
  (\bibinfo{date}{06} \bibinfo{year}{2021}).
\newblock
\urldef\tempurl%
\url{https://doi.org/10.24869/psyd.2021.133}
\showDOI{\tempurl}


\bibitem[Jeon et~al\mbox{.}(2021)]%
        {rw29}
\bibfield{author}{\bibinfo{person}{Youngseung Jeon}, \bibinfo{person}{Seungwan
  Jin}, \bibinfo{person}{Patrick~C. Shih}, {and} \bibinfo{person}{Kyungsik
  Han}.} \bibinfo{year}{2021}\natexlab{}.
\newblock \showarticletitle{FashionQ: An AI-Driven Creativity Support Tool for
  Facilitating Ideation in Fashion Design}.
\newblock \bibinfo{journal}{\emph{Association for Computing Machinery}}
  (\bibinfo{date}{05} \bibinfo{year}{2021}).
\newblock
\urldef\tempurl%
\url{https://doi.org/10.1145/3411764.3445093}
\showDOI{\tempurl}


\bibitem[Jiang et~al\mbox{.}(2023)]%
        {jiang2023ai}
\bibfield{author}{\bibinfo{person}{Harry~H Jiang}, \bibinfo{person}{Lauren
  Brown}, \bibinfo{person}{Jessica Cheng}, \bibinfo{person}{Mehtab Khan},
  \bibinfo{person}{Abhishek Gupta}, \bibinfo{person}{Deja Workman},
  \bibinfo{person}{Alex Hanna}, \bibinfo{person}{Johnathan Flowers}, {and}
  \bibinfo{person}{Timnit Gebru}.} \bibinfo{year}{2023}\natexlab{}.
\newblock \showarticletitle{AI Art and its Impact on Artists}. In
  \bibinfo{booktitle}{\emph{Proceedings of the 2023 AAAI/ACM Conference on AI,
  Ethics, and Society}}. \bibinfo{pages}{363--374}.
\newblock


\bibitem[Johnson(1997)]%
        {johnson1997analysis}
\bibfield{author}{\bibinfo{person}{Eric~William Johnson}.}
  \bibinfo{year}{1997}\natexlab{}.
\newblock \bibinfo{booktitle}{\emph{Analysis and refinement of iterative design
  processes}}.
\newblock \bibinfo{publisher}{University of Notre Dame}.
\newblock


\bibitem[Kang et~al\mbox{.}(2024)]%
        {kang2024impact}
\bibfield{author}{\bibinfo{person}{Sangbum Kang}, \bibinfo{person}{Yongjoo
  Choi}, {and} \bibinfo{person}{Boyoung Kim}.} \bibinfo{year}{2024}\natexlab{}.
\newblock \showarticletitle{Impact of motivation factors for using generative
  AI services on continuous use intention: Mediating trust and acceptance
  attitude}.
\newblock \bibinfo{journal}{\emph{Social Sciences}} \bibinfo{volume}{13},
  \bibinfo{number}{9} (\bibinfo{year}{2024}), \bibinfo{pages}{475}.
\newblock


\bibitem[Kawakami and Venkatagiri(2024)]%
        {kawakami2024impact}
\bibfield{author}{\bibinfo{person}{Reishiro Kawakami} {and}
  \bibinfo{person}{Sukrit Venkatagiri}.} \bibinfo{year}{2024}\natexlab{}.
\newblock \showarticletitle{The Impact of Generative AI on Artists}. In
  \bibinfo{booktitle}{\emph{Proceedings of the 16th Conference on Creativity \&
  Cognition}}. \bibinfo{pages}{79--82}.
\newblock


\bibitem[Khang~Le(2005)]%
        {skillful2005}
\bibfield{author}{\bibinfo{person}{Felix~Yoon Khang~Le, Mike~Yamada}.}
  \bibinfo{year}{2005}\natexlab{}.
\newblock \bibinfo{booktitle}{\emph{The Skillful Huntsman: Visual Development
  of a Grimm Tale at Art Center College of Design}}.
\newblock \bibinfo{publisher}{Design Studio Press}.
\newblock


\bibitem[Kim(2018)]%
        {rw8}
\bibfield{author}{\bibinfo{person}{Pierce~R.A. Kim, K.H.}}
  \bibinfo{year}{2018}\natexlab{}.
\newblock \showarticletitle{Convergent Versus Divergent Thinking}.
\newblock \bibinfo{journal}{\emph{Encyclopedia of Creativity, Invention,
  Innovation and Entrepreneurship. Springer, New York, NY}} (\bibinfo{date}{06}
  \bibinfo{year}{2018}).
\newblock
\urldef\tempurl%
\url{https://doi.org/10.1007/978-1-4614-3858-8_22}
\showDOI{\tempurl}


\bibitem[Kim(2024)]%
        {Kim2024Establishing}
\bibfield{author}{\bibinfo{person}{Sunnie S.~Y. Kim}.}
  \bibinfo{year}{2024}\natexlab{}.
\newblock \showarticletitle{Establishing Appropriate Trust in AI through
  Transparency and Explainability}.
\newblock \bibinfo{journal}{\emph{Extended Abstracts of the CHI Conference on
  Human Factors in Computing Systems}} (\bibinfo{year}{2024}).
\newblock
\urldef\tempurl%
\url{https://doi.org/10.1145/3613905.3638184}
\showDOI{\tempurl}


\bibitem[Knearem et~al\mbox{.}(2023)]%
        {Knearem2023Exploring}
\bibfield{author}{\bibinfo{person}{Tiffany Knearem}, \bibinfo{person}{M.
  Khwaja}, \bibinfo{person}{Yuling Gao}, \bibinfo{person}{Frank Bentley}, {and}
  \bibinfo{person}{Clara Kliman-Silver}.} \bibinfo{year}{2023}\natexlab{}.
\newblock \showarticletitle{Exploring the future of design tooling: The role of
  artificial intelligence in tools for user experience professionals}.
\newblock \bibinfo{journal}{\emph{Extended Abstracts of the 2023 CHI Conference
  on Human Factors in Computing Systems}} (\bibinfo{year}{2023}).
\newblock
\urldef\tempurl%
\url{https://doi.org/10.1145/3544549.3573874}
\showDOI{\tempurl}


\bibitem[Ko et~al\mbox{.}(2023)]%
        {ko2023large}
\bibfield{author}{\bibinfo{person}{Hyung-Kwon Ko}, \bibinfo{person}{Gwanmo
  Park}, \bibinfo{person}{Hyeon Jeon}, \bibinfo{person}{Jaemin Jo},
  \bibinfo{person}{Juho Kim}, {and} \bibinfo{person}{Jinwook Seo}.}
  \bibinfo{year}{2023}\natexlab{}.
\newblock \showarticletitle{Large-scale text-to-image generation models for
  visual artists’ creative works}. In \bibinfo{booktitle}{\emph{Proceedings
  of the 28th international conference on intelligent user interfaces}}.
  \bibinfo{pages}{919--933}.
\newblock


\bibitem[Li et~al\mbox{.}(2022)]%
        {li2022analyzing}
\bibfield{author}{\bibinfo{person}{Yuan Li}, \bibinfo{person}{Yinglong Zhang},
  {and} \bibinfo{person}{Robert Capra}.} \bibinfo{year}{2022}\natexlab{}.
\newblock \showarticletitle{Analyzing information resources that support the
  creative process}. In \bibinfo{booktitle}{\emph{Proceedings of the 2022
  Conference on Human Information Interaction and Retrieval}}.
  \bibinfo{pages}{180--190}.
\newblock


\bibitem[Lilly(2015)]%
        {bigbadWorld2015}
\bibfield{author}{\bibinfo{person}{Eliott~J. Lilly}.}
  \bibinfo{year}{2015}\natexlab{}.
\newblock \bibinfo{booktitle}{\emph{Big Bad World of Concept Art for Video
  Games: An Insider's Guide for Students}}.
\newblock \bibinfo{publisher}{Design Studio Press}.
\newblock


\bibitem[Lindley(2013)]%
        {rw16}
\bibfield{author}{\bibinfo{person}{Xiang \& Helmes John \& Morris Richard \&
  Meek~Sam. Lindley, Siân \&~Cao}.} \bibinfo{year}{2013}\natexlab{}.
\newblock \showarticletitle{Towards a Tool for Design Ideation: Insights from
  Use of SketchStorm}.
\newblock \bibinfo{journal}{\emph{HCI 2013 - 27th International British
  Computer Society Human Computer Interaction Conference: The Internet of
  Things}} (\bibinfo{date}{09} \bibinfo{year}{2013}).
\newblock
\urldef\tempurl%
\url{https://doi.org/10.14236/ewic/HCI2013.13}
\showDOI{\tempurl}


\bibitem[Linsey et~al\mbox{.}(2011)]%
        {linsey2011experimental}
\bibfield{author}{\bibinfo{person}{Julie~S Linsey}, \bibinfo{person}{Emily~F
  Clauss}, \bibinfo{person}{Tolga Kurtoglu}, \bibinfo{person}{Jeremy~T Murphy},
  \bibinfo{person}{Kristin~L Wood}, {and} \bibinfo{person}{Arthur~B Markman}.}
  \bibinfo{year}{2011}\natexlab{}.
\newblock \showarticletitle{An experimental study of group idea generation
  techniques: understanding the roles of idea representation and viewing
  methods}.
\newblock  (\bibinfo{year}{2011}).
\newblock


\bibitem[Long et~al\mbox{.}(2024)]%
        {long2024not}
\bibfield{author}{\bibinfo{person}{Tao Long}, \bibinfo{person}{Katy~Ilonka
  Gero}, {and} \bibinfo{person}{Lydia~B Chilton}.}
  \bibinfo{year}{2024}\natexlab{}.
\newblock \showarticletitle{Not Just Novelty: A Longitudinal Study on Utility
  and Customization of an AI Workflow}. In
  \bibinfo{booktitle}{\emph{Proceedings of the 2024 ACM Designing Interactive
  Systems Conference}}. \bibinfo{pages}{782--803}.
\newblock


\bibitem[Lubart(2016)]%
        {rw9}
\bibfield{author}{\bibinfo{person}{Todd. Lubart}.}
  \bibinfo{year}{2016}\natexlab{}.
\newblock \showarticletitle{CREATIVITY AND CONVERGENT THINKING: REFLECTIONS,
  CONNECTIONS AND PRACTICAL CONSIDERATIONS}.
\newblock \bibinfo{journal}{\emph{RUDN Journal of Psychology and Pedagogics.
  7-15.}} (\bibinfo{date}{06} \bibinfo{year}{2016}).
\newblock
\urldef\tempurl%
\url{https://doi.org/10.22363/2313-1683-2016-4-7-15.}
\showDOI{\tempurl}


\bibitem[Lubos et~al\mbox{.}(2024)]%
        {lubos2024llm}
\bibfield{author}{\bibinfo{person}{Sebastian Lubos}, \bibinfo{person}{Thi
  Ngoc~Trang Tran}, \bibinfo{person}{Alexander Felfernig},
  \bibinfo{person}{Seda Polat~Erdeniz}, {and} \bibinfo{person}{Viet-Man Le}.}
  \bibinfo{year}{2024}\natexlab{}.
\newblock \showarticletitle{LLM-generated Explanations for Recommender
  Systems}. In \bibinfo{booktitle}{\emph{Adjunct Proceedings of the 32nd ACM
  Conference on User Modeling, Adaptation and Personalization}}.
  \bibinfo{pages}{276--285}.
\newblock


\bibitem[Mac{\'\i}as-Escriv{\'a} et~al\mbox{.}(2013)]%
        {macias2013self}
\bibfield{author}{\bibinfo{person}{Frank~D Mac{\'\i}as-Escriv{\'a}},
  \bibinfo{person}{Rodolfo Haber}, \bibinfo{person}{Raul Del~Toro}, {and}
  \bibinfo{person}{Vicente Hernandez}.} \bibinfo{year}{2013}\natexlab{}.
\newblock \showarticletitle{Self-adaptive systems: A survey of current
  approaches, research challenges and applications}.
\newblock \bibinfo{journal}{\emph{Expert Systems with Applications}}
  \bibinfo{volume}{40}, \bibinfo{number}{18} (\bibinfo{year}{2013}),
  \bibinfo{pages}{7267--7279}.
\newblock


\bibitem[Mahdavi~Goloujeh et~al\mbox{.}(2024)]%
        {mahdavi2024ai}
\bibfield{author}{\bibinfo{person}{Atefeh Mahdavi~Goloujeh},
  \bibinfo{person}{Anne Sullivan}, {and} \bibinfo{person}{Brian Magerko}.}
  \bibinfo{year}{2024}\natexlab{}.
\newblock \showarticletitle{Is It AI or Is It Me? Understanding Users’ Prompt
  Journey with Text-to-Image Generative AI Tools}. In
  \bibinfo{booktitle}{\emph{Proceedings of the CHI Conference on Human Factors
  in Computing Systems}}. \bibinfo{pages}{1--13}.
\newblock


\bibitem[Maleki et~al\mbox{.}(2024)]%
        {maleki2024ai}
\bibfield{author}{\bibinfo{person}{Negar Maleki}, \bibinfo{person}{Balaji
  Padmanabhan}, {and} \bibinfo{person}{Kaushik Dutta}.}
  \bibinfo{year}{2024}\natexlab{}.
\newblock \showarticletitle{AI hallucinations: a misnomer worth clarifying}. In
  \bibinfo{booktitle}{\emph{2024 IEEE Conference on Artificial Intelligence
  (CAI)}}. IEEE, \bibinfo{pages}{133--138}.
\newblock


\bibitem[Marc(2023)]%
        {randomguidebook2023}
\bibfield{author}{\bibinfo{person}{Adrian Marc}.}
  \bibinfo{year}{2023}\natexlab{}.
\newblock \bibinfo{booktitle}{\emph{The Random Guidebook of Concept Designers :
  Tips and Tricks} (\bibinfo{edition}{1st} ed.)}.
\newblock \bibinfo{publisher}{JOLUA}.
\newblock
\showISBNx{9730378061}


\bibitem[Mccormack et~al\mbox{.}(2020)]%
        {Mccormack2020Design}
\bibfield{author}{\bibinfo{person}{J. Mccormack}, \bibinfo{person}{P.
  Hutchings}, \bibinfo{person}{Toby Gifford}, \bibinfo{person}{M. Yee-King},
  \bibinfo{person}{M.~T. Llano}, {and} \bibinfo{person}{M. d'Inverno}.}
  \bibinfo{year}{2020}\natexlab{}.
\newblock \showarticletitle{Design Considerations for Real-Time Collaboration
  with Creative Artificial Intelligence}.
\newblock \bibinfo{journal}{\emph{Organised Sound}}  \bibinfo{volume}{25}
  (\bibinfo{year}{2020}), \bibinfo{pages}{41 -- 52}.
\newblock
\urldef\tempurl%
\url{https://doi.org/10.1017/S1355771819000451}
\showDOI{\tempurl}


\bibitem[Michael~Rabiger(2020)]%
        {directing2020}
\bibfield{author}{\bibinfo{person}{Mick Hurbis-Cherrier Michael~Rabiger}.}
  \bibinfo{year}{2020}\natexlab{}.
\newblock \bibinfo{booktitle}{\emph{Directing: Film Techniques and
  Aesthetics}}.
\newblock \bibinfo{publisher}{Routledge}.
\newblock


\bibitem[Monteith et~al\mbox{.}(2024)]%
        {monteith2024artificial}
\bibfield{author}{\bibinfo{person}{Scott Monteith}, \bibinfo{person}{Tasha
  Glenn}, \bibinfo{person}{John~R Geddes}, \bibinfo{person}{Peter~C Whybrow},
  \bibinfo{person}{Eric Achtyes}, {and} \bibinfo{person}{Michael Bauer}.}
  \bibinfo{year}{2024}\natexlab{}.
\newblock \showarticletitle{Artificial intelligence and increasing
  misinformation}.
\newblock \bibinfo{journal}{\emph{The British Journal of Psychiatry}}
  \bibinfo{volume}{224}, \bibinfo{number}{2} (\bibinfo{year}{2024}),
  \bibinfo{pages}{33--35}.
\newblock


\bibitem[M{\"u}ller et~al\mbox{.}(2013)]%
        {muller2013muse}
\bibfield{author}{\bibinfo{person}{Oliver M{\"u}ller}, \bibinfo{person}{Stefan
  Debortoli}, {and} \bibinfo{person}{Stefan Seidel}.}
  \bibinfo{year}{2013}\natexlab{}.
\newblock \showarticletitle{MUSE: implementation of a design theory for systems
  that support convergent and divergent thinking}. In
  \bibinfo{booktitle}{\emph{Design Science at the Intersection of Physical and
  Virtual Design: 8th International Conference, DESRIST 2013, Helsinki,
  Finland, June 11-12, 2013. Proceedings 8}}. Springer,
  \bibinfo{pages}{438--445}.
\newblock


\bibitem[Newland et~al\mbox{.}(1987)]%
        {newland1987understanding}
\bibfield{author}{\bibinfo{person}{Paul Newland}, \bibinfo{person}{James~A
  Powell}, {and} \bibinfo{person}{Chris Creed}.}
  \bibinfo{year}{1987}\natexlab{}.
\newblock \showarticletitle{Understanding architectural designers' selective
  information handling}.
\newblock \bibinfo{journal}{\emph{Design Studies}} \bibinfo{volume}{8},
  \bibinfo{number}{1} (\bibinfo{year}{1987}), \bibinfo{pages}{2--16}.
\newblock


\bibitem[Oh et~al\mbox{.}(2024)]%
        {oh2024lumimood}
\bibfield{author}{\bibinfo{person}{Jeongseok Oh}, \bibinfo{person}{Seungju
  Kim}, {and} \bibinfo{person}{Seungjun Kim}.} \bibinfo{year}{2024}\natexlab{}.
\newblock \showarticletitle{LumiMood: A Creativity Support Tool for Designing
  the Mood of a 3D Scene}. In \bibinfo{booktitle}{\emph{Proceedings of the CHI
  Conference on Human Factors in Computing Systems}}. \bibinfo{pages}{1--21}.
\newblock


\bibitem[Palani et~al\mbox{.}(2022)]%
        {palani2022don}
\bibfield{author}{\bibinfo{person}{Srishti Palani}, \bibinfo{person}{David
  Ledo}, \bibinfo{person}{George Fitzmaurice}, {and} \bibinfo{person}{Fraser
  Anderson}.} \bibinfo{year}{2022}\natexlab{}.
\newblock \showarticletitle{” I don’t want to feel like I’m working in a
  1960s factory”: The Practitioner Perspective on Creativity Support Tool
  Adoption}. In \bibinfo{booktitle}{\emph{Proceedings of the 2022 CHI
  Conference on Human Factors in Computing Systems}}. \bibinfo{pages}{1--18}.
\newblock


\bibitem[Park and Hannafin(1993)]%
        {park1993empirically}
\bibfield{author}{\bibinfo{person}{Innwoo Park} {and}
  \bibinfo{person}{Michael~J Hannafin}.} \bibinfo{year}{1993}\natexlab{}.
\newblock \showarticletitle{Empirically-based guidelines for the design of
  interactive multimedia}.
\newblock \bibinfo{journal}{\emph{Educational Technology Research and
  Development}} \bibinfo{volume}{41}, \bibinfo{number}{3}
  (\bibinfo{year}{1993}), \bibinfo{pages}{63--85}.
\newblock


\bibitem[Peng et~al\mbox{.}(2024)]%
        {DesignPrompt2024}
\bibfield{author}{\bibinfo{person}{Xiaohan Peng}, \bibinfo{person}{Janin Koch},
  {and} \bibinfo{person}{Wendy~E. Mackay}.} \bibinfo{year}{2024}\natexlab{}.
\newblock \showarticletitle{DesignPrompt: Using Multimodal Interaction for
  Design Exploration with Generative AI}.
\newblock  (\bibinfo{year}{2024}), \bibinfo{pages}{804–818}.
\newblock
\showISBNx{9798400705830}
\urldef\tempurl%
\url{https://doi.org/10.1145/3643834.3661588}
\showDOI{\tempurl}


\bibitem[Qin(2023)]%
        {qin2023does}
\bibfield{author}{\bibinfo{person}{Jiayang Qin}.}
  \bibinfo{year}{2023}\natexlab{}.
\newblock \showarticletitle{How does Text-to-image AI Affect Indie Game
  Designers and Artists?}
\newblock \bibinfo{journal}{\emph{Journal of Innovation and Development}}
  \bibinfo{volume}{5}, \bibinfo{number}{3} (\bibinfo{year}{2023}),
  \bibinfo{pages}{107--111}.
\newblock


\bibitem[R{\"a}ssa(2018)]%
        {rassa2018concept}
\bibfield{author}{\bibinfo{person}{Julia R{\"a}ssa}.}
  \bibinfo{year}{2018}\natexlab{}.
\newblock \showarticletitle{Concept art creation methodologies: Visual
  Development of “Rock Boy”}.
\newblock  (\bibinfo{year}{2018}).
\newblock


\bibitem[Rezwana and Maher(2023)]%
        {rezwana2023designing}
\bibfield{author}{\bibinfo{person}{Jeba Rezwana} {and}
  \bibinfo{person}{Mary~Lou Maher}.} \bibinfo{year}{2023}\natexlab{}.
\newblock \showarticletitle{Designing creative AI partners with COFI: A
  framework for modeling interaction in human-AI co-creative systems}.
\newblock \bibinfo{journal}{\emph{ACM Transactions on Computer-Human
  Interaction}} \bibinfo{volume}{30}, \bibinfo{number}{5}
  (\bibinfo{year}{2023}), \bibinfo{pages}{1--28}.
\newblock


\bibitem[Rick et~al\mbox{.}(2023)]%
        {Rick2023Supermind}
\bibfield{author}{\bibinfo{person}{Steven~R Rick}, \bibinfo{person}{Gianni
  Giacomelli}, \bibinfo{person}{Haoran Wen}, \bibinfo{person}{Robert~J.
  Laubacher}, \bibinfo{person}{Nancy Taubenslag}, \bibinfo{person}{Jennifer~L.
  Heyman}, \bibinfo{person}{Max~Sina Knicker}, \bibinfo{person}{Younes Jeddi},
  \bibinfo{person}{Hendrik Maier}, \bibinfo{person}{Stephen Dwyer},
  \bibinfo{person}{Pranav Ragupathy}, {and} \bibinfo{person}{Thomas Malone}.}
  \bibinfo{year}{2023}\natexlab{}.
\newblock \showarticletitle{Supermind Ideator: Exploring generative AI to
  support creative problem-solving}.
\newblock \bibinfo{journal}{\emph{ArXiv}}  \bibinfo{volume}{abs/2311.01937}
  (\bibinfo{year}{2023}).
\newblock
\urldef\tempurl%
\url{https://doi.org/10.48550/arXiv.2311.01937}
\showDOI{\tempurl}


\bibitem[Roberson et~al\mbox{.}(1995)]%
        {roberson1995analysis}
\bibfield{author}{\bibinfo{person}{Paula~K Roberson}, \bibinfo{person}{SJ
  Shema}, \bibinfo{person}{DJ Mundfrom}, {and} \bibinfo{person}{TM Holmes}.}
  \bibinfo{year}{1995}\natexlab{}.
\newblock \showarticletitle{Analysis of paired Likert data: how to evaluate
  change and preference questions.}
\newblock \bibinfo{journal}{\emph{Family medicine}} \bibinfo{volume}{27},
  \bibinfo{number}{10} (\bibinfo{year}{1995}), \bibinfo{pages}{671--675}.
\newblock


\bibitem[Rogers(2014)]%
        {levelup2014}
\bibfield{author}{\bibinfo{person}{Scott Rogers}.}
  \bibinfo{year}{2014}\natexlab{}.
\newblock \bibinfo{booktitle}{\emph{Level Up! The Guide to Great Video Game
  Design}}.
\newblock \bibinfo{publisher}{Wiley}.
\newblock


\bibitem[Satyanarayan et~al\mbox{.}(2019)]%
        {satyanarayan2019critical}
\bibfield{author}{\bibinfo{person}{Arvind Satyanarayan},
  \bibinfo{person}{Bongshin Lee}, \bibinfo{person}{Donghao Ren},
  \bibinfo{person}{Jeffrey Heer}, \bibinfo{person}{John Stasko},
  \bibinfo{person}{John Thompson}, \bibinfo{person}{Matthew Brehmer}, {and}
  \bibinfo{person}{Zhicheng Liu}.} \bibinfo{year}{2019}\natexlab{}.
\newblock \showarticletitle{Critical reflections on visualization authoring
  systems}.
\newblock \bibinfo{journal}{\emph{IEEE transactions on visualization and
  computer graphics}} \bibinfo{volume}{26}, \bibinfo{number}{1}
  (\bibinfo{year}{2019}), \bibinfo{pages}{461--471}.
\newblock


\bibitem[Sawant(2024)]%
        {filmhandbook2024}
\bibfield{author}{\bibinfo{person}{Ojas~D. Sawant}.}
  \bibinfo{year}{2024}\natexlab{}.
\newblock \bibinfo{booktitle}{\emph{Visual Storytelling with Generative AI: A
  Practical Handbook for modern Filmmakers and Content Creators}}.
\newblock \bibinfo{publisher}{Independently published}.
\newblock


\bibitem[Scharowski et~al\mbox{.}(2023)]%
        {Scharowski2023Exploring}
\bibfield{author}{\bibinfo{person}{Nicolas Scharowski}, \bibinfo{person}{S.
  Perrig}, \bibinfo{person}{Melanie Svab}, \bibinfo{person}{K. Opwis}, {and}
  \bibinfo{person}{Florian Brühlmann}.} \bibinfo{year}{2023}\natexlab{}.
\newblock \showarticletitle{Exploring the effects of human-centered AI
  explanations on trust and reliance}.
\newblock \bibinfo{journal}{\emph{Frontiers Comput. Sci.}}  \bibinfo{volume}{5}
  (\bibinfo{year}{2023}).
\newblock
\urldef\tempurl%
\url{https://doi.org/10.3389/fcomp.2023.1151150}
\showDOI{\tempurl}


\bibitem[Schell(2008)]%
        {artofgame2008}
\bibfield{author}{\bibinfo{person}{Jesse Schell}.}
  \bibinfo{year}{2008}\natexlab{}.
\newblock \bibinfo{booktitle}{\emph{The Art of Game Design: A Book of Lenses}}.
\newblock \bibinfo{publisher}{CRC Press}.
\newblock


\bibitem[Shi et~al\mbox{.}(2023)]%
        {shi2023understanding}
\bibfield{author}{\bibinfo{person}{Jingyu Shi}, \bibinfo{person}{Rahul Jain},
  \bibinfo{person}{Runlin Duan}, {and} \bibinfo{person}{Karthik Ramani}.}
  \bibinfo{year}{2023}\natexlab{}.
\newblock \showarticletitle{Understanding Generative AI in Art: An Interview
  Study with Artists on G-AI from an HCI Perspective}.
\newblock \bibinfo{journal}{\emph{arXiv preprint arXiv:2310.13149}}
  (\bibinfo{year}{2023}).
\newblock


\bibitem[Shneiderman(2022)]%
        {shneiderman2022human}
\bibfield{author}{\bibinfo{person}{Ben Shneiderman}.}
  \bibinfo{year}{2022}\natexlab{}.
\newblock \bibinfo{booktitle}{\emph{Human-centered AI}}.
\newblock \bibinfo{publisher}{Oxford University Press}.
\newblock


\bibitem[Singh et~al\mbox{.}(2023)]%
        {Singh2023Artificial}
\bibfield{author}{\bibinfo{person}{Hardeep Singh}, \bibinfo{person}{Kamaljeet
  Kaur}, {and} \bibinfo{person}{Preet~Pinder Singh}.}
  \bibinfo{year}{2023}\natexlab{}.
\newblock \showarticletitle{Artificial Intelligence as a facilitator for Film
  Production Process}.
\newblock \bibinfo{journal}{\emph{2023 International Conference on Artificial
  Intelligence and Smart Communication (AISC)}} (\bibinfo{year}{2023}),
  \bibinfo{pages}{969--972}.
\newblock
\urldef\tempurl%
\url{https://doi.org/10.1109/AISC56616.2023.10085082}
\showDOI{\tempurl}


\bibitem[Son et~al\mbox{.}(2024)]%
        {son2024genquery}
\bibfield{author}{\bibinfo{person}{Kihoon Son}, \bibinfo{person}{DaEun Choi},
  \bibinfo{person}{Tae~Soo Kim}, \bibinfo{person}{Young-Ho Kim}, {and}
  \bibinfo{person}{Juho Kim}.} \bibinfo{year}{2024}\natexlab{}.
\newblock \showarticletitle{GenQuery: Supporting Expressive Visual Search with
  Generative Models}. In \bibinfo{booktitle}{\emph{Proceedings of the CHI
  Conference on Human Factors in Computing Systems}}. \bibinfo{pages}{1--19}.
\newblock


\bibitem[Stamps(1999)]%
        {Stamps1999Demographic}
\bibfield{author}{\bibinfo{person}{A. Stamps}.}
  \bibinfo{year}{1999}\natexlab{}.
\newblock \showarticletitle{Demographic Effects in Environmental Aesthetics: A
  Meta-Analysis}.
\newblock \bibinfo{journal}{\emph{Journal of Planning Literature}}
  \bibinfo{volume}{14} (\bibinfo{year}{1999}), \bibinfo{pages}{155 -- 175}.
\newblock
\urldef\tempurl%
\url{https://doi.org/10.1177/08854129922092630}
\showDOI{\tempurl}


\bibitem[Steven~Ascher(2019)]%
        {filmmaker2019}
\bibfield{author}{\bibinfo{person}{Edward~Pincus Steven~Ascher}.}
  \bibinfo{year}{2019}\natexlab{}.
\newblock \bibinfo{booktitle}{\emph{The Filmmakers Handbook: A Comprehensive
  Guide for the Digital Age}}.
\newblock \bibinfo{publisher}{PLUME BOOKS}.
\newblock


\bibitem[Stigliani and Ravasi(2018)]%
        {Stigliani2018The}
\bibfield{author}{\bibinfo{person}{I. Stigliani} {and} \bibinfo{person}{Davide
  Ravasi}.} \bibinfo{year}{2018}\natexlab{}.
\newblock \showarticletitle{The Shaping of Form: Exploring Designers’ Use of
  Aesthetic Knowledge}.
\newblock \bibinfo{journal}{\emph{Organization Studies}}  \bibinfo{volume}{39}
  (\bibinfo{year}{2018}), \bibinfo{pages}{747 -- 784}.
\newblock
\urldef\tempurl%
\url{https://doi.org/10.1177/0170840618759813}
\showDOI{\tempurl}


\bibitem[Taheri and Hesamian(2013)]%
        {taheri2013generalization}
\bibfield{author}{\bibinfo{person}{SM Taheri} {and} \bibinfo{person}{Gholamreza
  Hesamian}.} \bibinfo{year}{2013}\natexlab{}.
\newblock \showarticletitle{A generalization of the Wilcoxon signed-rank test
  and its applications}.
\newblock \bibinfo{journal}{\emph{Statistical Papers}}  \bibinfo{volume}{54}
  (\bibinfo{year}{2013}), \bibinfo{pages}{457--470}.
\newblock


\bibitem[Talebirad and Nadiri(2023)]%
        {talebirad2023multi}
\bibfield{author}{\bibinfo{person}{Yashar Talebirad} {and}
  \bibinfo{person}{Amirhossein Nadiri}.} \bibinfo{year}{2023}\natexlab{}.
\newblock \showarticletitle{Multi-agent collaboration: Harnessing the power of
  intelligent llm agents}.
\newblock \bibinfo{journal}{\emph{arXiv preprint arXiv:2306.03314}}
  (\bibinfo{year}{2023}).
\newblock


\bibitem[Tokarev(2020)]%
        {80lv2020}
\bibfield{author}{\bibinfo{person}{Kirill Tokarev}.}
  \bibinfo{year}{2020}\natexlab{}.
\newblock \bibinfo{title}{Working as a Concept Artist in the Fast-Paced Film
  Industry}.
\newblock
  \bibinfo{howpublished}{\url{https://80.lv/articles/working-as-a-concept-artist-in-the-fast-paced-film-industry/}}.
\newblock


\bibitem[Tseng et~al\mbox{.}(2024)]%
        {Tseng2024KeyframerEA}
\bibfield{author}{\bibinfo{person}{Tiffany Tseng}, \bibinfo{person}{Ruijia
  Cheng}, {and} \bibinfo{person}{Jeffrey Nichols}.}
  \bibinfo{year}{2024}\natexlab{}.
\newblock \showarticletitle{Keyframer: Empowering Animation Design using Large
  Language Models}.
\newblock \bibinfo{journal}{\emph{ArXiv}}  \bibinfo{volume}{abs/2402.06071}
  (\bibinfo{year}{2024}).
\newblock
\urldef\tempurl%
\url{https://api.semanticscholar.org/CorpusID:267616691}
\showURL{%
\tempurl}


\bibitem[Tversky and Chou(2011)]%
        {tversky2011creativity}
\bibfield{author}{\bibinfo{person}{Barbara Tversky} {and}
  \bibinfo{person}{Juliet~Y Chou}.} \bibinfo{year}{2011}\natexlab{}.
\newblock \showarticletitle{Creativity: depth and breadth}. In
  \bibinfo{booktitle}{\emph{Design creativity 2010}}. Springer,
  \bibinfo{pages}{209--214}.
\newblock


\bibitem[Usmani et~al\mbox{.}(2023)]%
        {Usmani2023Human-Centered}
\bibfield{author}{\bibinfo{person}{Usman~Ahmad Usmani}, \bibinfo{person}{A.
  Happonen}, {and} \bibinfo{person}{J. Watada}.}
  \bibinfo{year}{2023}\natexlab{}.
\newblock \showarticletitle{Human-Centered Artificial Intelligence: Designing
  for User Empowerment and Ethical Considerations}.
\newblock \bibinfo{journal}{\emph{2023 5th International Congress on
  Human-Computer Interaction, Optimization and Robotic Applications (HORA)}}
  (\bibinfo{year}{2023}), \bibinfo{pages}{01--05}.
\newblock
\urldef\tempurl%
\url{https://doi.org/10.1109/HORA58378.2023.10156761}
\showDOI{\tempurl}


\bibitem[Uusitalo et~al\mbox{.}(2024)]%
        {claytoplay2024}
\bibfield{author}{\bibinfo{person}{Severi Uusitalo}, \bibinfo{person}{Antti
  Salovaara}, \bibinfo{person}{Tero Jokela}, {and} \bibinfo{person}{Marja
  Salmimaa}.} \bibinfo{year}{2024}\natexlab{}.
\newblock \showarticletitle{”Clay to Play With”: Generative AI Tools in UX
  and Industrial Design Practice}.
\newblock  (\bibinfo{year}{2024}), \bibinfo{pages}{1566–1578}.
\newblock
\showISBNx{9798400705830}
\urldef\tempurl%
\url{https://doi.org/10.1145/3643834.3661624}
\showDOI{\tempurl}


\bibitem[Valluri(2020)]%
        {interview2020}
\bibfield{author}{\bibinfo{person}{Siddhartha Valluri}.}
  \bibinfo{year}{2020}\natexlab{}.
\newblock \bibinfo{title}{Game Concept Artist Interview}.
\newblock
  \bibinfo{howpublished}{\url{https://theinterviewportal.com/2020/08/16/game-concept-artist-interview/}}.
\newblock


\bibitem[Venigandla et~al\mbox{.}(2024)]%
        {Venigandla2024Hybrid}
\bibfield{author}{\bibinfo{person}{Kamala Venigandla}, \bibinfo{person}{Navya
  Vemuri}, {and} \bibinfo{person}{Naveen Vemuri}.}
  \bibinfo{year}{2024}\natexlab{}.
\newblock \showarticletitle{Hybrid Intelligence Systems Combining Human
  Expertise and AI/RPA for Complex Problem Solving}.
\newblock \bibinfo{journal}{\emph{International Journal of Innovative Science
  and Research Technology (IJISRT)}} (\bibinfo{year}{2024}).
\newblock
\urldef\tempurl%
\url{https://doi.org/10.38124/ijisrt/ijisrt24mar2039}
\showDOI{\tempurl}


\bibitem[Verganti et~al\mbox{.}(2020)]%
        {verganti2020innovation}
\bibfield{author}{\bibinfo{person}{Roberto Verganti}, \bibinfo{person}{Luca
  Vendraminelli}, {and} \bibinfo{person}{Marco Iansiti}.}
  \bibinfo{year}{2020}\natexlab{}.
\newblock \showarticletitle{Innovation and design in the age of artificial
  intelligence}.
\newblock \bibinfo{journal}{\emph{Journal of product innovation management}}
  \bibinfo{volume}{37}, \bibinfo{number}{3} (\bibinfo{year}{2020}),
  \bibinfo{pages}{212--227}.
\newblock


\bibitem[Vimpari et~al\mbox{.}(2023)]%
        {vimpari2023adapt}
\bibfield{author}{\bibinfo{person}{Veera Vimpari}, \bibinfo{person}{Annakaisa
  Kultima}, \bibinfo{person}{Perttu H{\"a}m{\"a}l{\"a}inen}, {and}
  \bibinfo{person}{Christian Guckelsberger}.} \bibinfo{year}{2023}\natexlab{}.
\newblock \showarticletitle{“An Adapt-or-Die Type of Situation”:
  Perception, Adoption, and Use of Text-to-Image-Generation AI by Game Industry
  Professionals}.
\newblock \bibinfo{journal}{\emph{Proceedings of the ACM on Human-Computer
  Interaction}} \bibinfo{volume}{7}, \bibinfo{number}{CHI PLAY}
  (\bibinfo{year}{2023}), \bibinfo{pages}{131--164}.
\newblock


\bibitem[Wang et~al\mbox{.}(2024b)]%
        {wang2024roomdreaming}
\bibfield{author}{\bibinfo{person}{Shun-Yu Wang}, \bibinfo{person}{Wei-Chung
  Su}, \bibinfo{person}{Serena Chen}, \bibinfo{person}{Ching-Yi Tsai},
  \bibinfo{person}{Marta Misztal}, \bibinfo{person}{Katherine~M Cheng},
  \bibinfo{person}{Alwena Lin}, \bibinfo{person}{Yu Chen}, {and}
  \bibinfo{person}{Mike~Y Chen}.} \bibinfo{year}{2024}\natexlab{b}.
\newblock \showarticletitle{RoomDreaming: Generative-AI Approach to
  Facilitating Iterative, Preliminary Interior Design Exploration}. In
  \bibinfo{booktitle}{\emph{Proceedings of the CHI Conference on Human Factors
  in Computing Systems}}. \bibinfo{pages}{1--20}.
\newblock


\bibitem[Wang et~al\mbox{.}(2023)]%
        {wang2023reprompt}
\bibfield{author}{\bibinfo{person}{Yunlong Wang}, \bibinfo{person}{Shuyuan
  Shen}, {and} \bibinfo{person}{Brian~Y Lim}.} \bibinfo{year}{2023}\natexlab{}.
\newblock \showarticletitle{Reprompt: Automatic prompt editing to refine
  ai-generative art towards precise expressions}. In
  \bibinfo{booktitle}{\emph{Proceedings of the 2023 CHI conference on human
  factors in computing systems}}. \bibinfo{pages}{1--29}.
\newblock


\bibitem[Wang et~al\mbox{.}(2024a)]%
        {PromptCharm2024}
\bibfield{author}{\bibinfo{person}{Zhijie Wang}, \bibinfo{person}{Yuheng
  Huang}, \bibinfo{person}{Da Song}, \bibinfo{person}{Lei Ma}, {and}
  \bibinfo{person}{Tianyi Zhang}.} \bibinfo{year}{2024}\natexlab{a}.
\newblock \showarticletitle{PromptCharm: Text-to-Image Generation through
  Multi-modal Prompting and Refinement}.
\newblock , Article \bibinfo{articleno}{185} (\bibinfo{year}{2024}),
  \bibinfo{numpages}{21}~pages.
\newblock
\showISBNx{9798400703300}
\urldef\tempurl%
\url{https://doi.org/10.1145/3613904.3642803}
\showDOI{\tempurl}


\bibitem[Weaver(2022)]%
        {rw15}
\bibfield{author}{\bibinfo{person}{Jacob \& Merzdorf Hillary \& Dorozhkin Denis
  \& Douglas Kerrie \& Linsey~Julie. Weaver, Morgan \&~Buck}.}
  \bibinfo{year}{2022}\natexlab{}.
\newblock \showarticletitle{Investigating Priming Effects of Sketch Evaluation
  Instructions on Idea Generation Productivity}.
\newblock \bibinfo{journal}{\emph{ASME 2022 International Design Engineering
  Technical Conferences and Computers and Information in Engineering
  Conference}} (\bibinfo{date}{11} \bibinfo{year}{2022}).
\newblock
\urldef\tempurl%
\url{https://doi.org/10.1115/DETC2022-91313}
\showDOI{\tempurl}


\bibitem[WeiliZhu(2019)]%
        {rw10}
\bibfield{author}{\bibinfo{person}{WeiliJiang MengPei~andYanjieSu WeiliZhu,
  SiyuanShang}.} \bibinfo{year}{2019}\natexlab{}.
\newblock \showarticletitle{Convergent thinking moderates the relationship
  between divergent thinking and scientific creativity.}
\newblock \bibinfo{journal}{\emph{Creativity Research Journal 31, 3 (2019),
  320–328.}} (\bibinfo{date}{06} \bibinfo{year}{2019}).
\newblock
\urldef\tempurl%
\url{https://doi.org/10.1080/10400419.2019.1641685}
\showDOI{\tempurl}


\bibitem[Winkler~Pettersson(2010)]%
        {rw17}
\bibfield{author}{\bibinfo{person}{Kjellin A. Lind M. \& Seipel~S.
  Winkler~Pettersson, L.}} \bibinfo{year}{2010}\natexlab{}.
\newblock \showarticletitle{On the role of visual references in collaborative
  visualization.}
\newblock \bibinfo{journal}{\emph{Information Visualization, 9(2), 98–114.}}
  (\bibinfo{date}{09} \bibinfo{year}{2010}).
\newblock
\urldef\tempurl%
\url{https://doi.org/10.1057/ivs.2009.2}
\showDOI{\tempurl}


\bibitem[Xie(2023)]%
        {rw5}
\bibfield{author}{\bibinfo{person}{Xueying Xie}.}
  \bibinfo{year}{2023}\natexlab{}.
\newblock \showarticletitle{The cognitive process of creative design: A
  perspective of divergent thinking}.
\newblock \bibinfo{journal}{\emph{Thinking Skills and Creativity, Volume 48}}
  (\bibinfo{date}{06} \bibinfo{year}{2023}).
\newblock
\urldef\tempurl%
\url{https://doi.org/10.1016/j.tsc.2023.101266.}
\showDOI{\tempurl}


\bibitem[Xu et~al\mbox{.}(2023)]%
        {xu2023transitioning}
\bibfield{author}{\bibinfo{person}{Wei Xu}, \bibinfo{person}{Marvin~J Dainoff},
  \bibinfo{person}{Liezhong Ge}, {and} \bibinfo{person}{Zaifeng Gao}.}
  \bibinfo{year}{2023}\natexlab{}.
\newblock \showarticletitle{Transitioning to human interaction with AI systems:
  New challenges and opportunities for HCI professionals to enable
  human-centered AI}.
\newblock \bibinfo{journal}{\emph{International Journal of Human--Computer
  Interaction}} \bibinfo{volume}{39}, \bibinfo{number}{3}
  (\bibinfo{year}{2023}), \bibinfo{pages}{494--518}.
\newblock


\bibitem[Yan et~al\mbox{.}(2023)]%
        {yan2023xcreation}
\bibfield{author}{\bibinfo{person}{Zihan Yan}, \bibinfo{person}{Chunxu Yang},
  \bibinfo{person}{Qihao Liang}, {and} \bibinfo{person}{Xiang'Anthony' Chen}.}
  \bibinfo{year}{2023}\natexlab{}.
\newblock \showarticletitle{XCreation: A Graph-based Crossmodal Generative
  Creativity Support Tool}. In \bibinfo{booktitle}{\emph{Proceedings of the
  36th Annual ACM Symposium on User Interface Software and Technology}}.
  \bibinfo{pages}{1--15}.
\newblock


\bibitem[Zeng et~al\mbox{.}(2024)]%
        {zeng2024intenttuner}
\bibfield{author}{\bibinfo{person}{Xingchen Zeng}, \bibinfo{person}{Ziyao Gao},
  \bibinfo{person}{Yilin Ye}, {and} \bibinfo{person}{Wei Zeng}.}
  \bibinfo{year}{2024}\natexlab{}.
\newblock \showarticletitle{IntentTuner: An Interactive Framework for
  Integrating Human Intentions in Fine-tuning Text-to-Image Generative Models}.
  In \bibinfo{booktitle}{\emph{Proceedings of the CHI Conference on Human
  Factors in Computing Systems}}. \bibinfo{pages}{1--18}.
\newblock


\bibitem[Zhang et~al\mbox{.}(2024)]%
        {zhang2024protodreamer}
\bibfield{author}{\bibinfo{person}{Hongbo Zhang}, \bibinfo{person}{Pei Chen},
  \bibinfo{person}{Xuelong Xie}, \bibinfo{person}{Chaoyi Lin},
  \bibinfo{person}{Lianyan Liu}, \bibinfo{person}{Zhuoshu Li},
  \bibinfo{person}{Weitao You}, {and} \bibinfo{person}{Lingyun Sun}.}
  \bibinfo{year}{2024}\natexlab{}.
\newblock \showarticletitle{ProtoDreamer: A Mixed-prototype Tool Combining
  Physical Model and Generative AI to Support Conceptual Design}. In
  \bibinfo{booktitle}{\emph{Proceedings of the 37th Annual ACM Symposium on
  User Interface Software and Technology}}. \bibinfo{pages}{1--18}.
\newblock


\bibitem[Zhang et~al\mbox{.}(2023)]%
        {zhang2023adding}
\bibfield{author}{\bibinfo{person}{Lvmin Zhang}, \bibinfo{person}{Anyi Rao},
  {and} \bibinfo{person}{Maneesh Agrawala}.} \bibinfo{year}{2023}\natexlab{}.
\newblock \showarticletitle{Adding conditional control to text-to-image
  diffusion models}. In \bibinfo{booktitle}{\emph{Proceedings of the IEEE/CVF
  International Conference on Computer Vision}}. \bibinfo{pages}{3836--3847}.
\newblock


\bibitem[Zhang and Li(2024)]%
        {zhang2024confrontation}
\bibfield{author}{\bibinfo{person}{Shuning Zhang} {and}
  \bibinfo{person}{Shixuan Li}.} \bibinfo{year}{2024}\natexlab{}.
\newblock \showarticletitle{" Confrontation or Acceptance": Understanding
  Novice Visual Artists' Perception towards AI-assisted Art Creation}.
\newblock \bibinfo{journal}{\emph{arXiv preprint arXiv:2410.14925}}
  (\bibinfo{year}{2024}).
\newblock


\bibitem[Zhou et~al\mbox{.}(2024a)]%
        {OptiMuse}
\bibfield{author}{\bibinfo{person}{Jiayi Zhou}, \bibinfo{person}{Renzhong Li},
  \bibinfo{person}{Junxiu Tang}, \bibinfo{person}{Tan Tang},
  \bibinfo{person}{Haotian Li}, \bibinfo{person}{Weiwei Cui}, {and}
  \bibinfo{person}{Yingcai Wu}.} \bibinfo{year}{2024}\natexlab{a}.
\newblock \showarticletitle{Understanding Nonlinear Collaboration between Human
  and AI Agents: A Co-design Framework for Creative Design}. In
  \bibinfo{booktitle}{\emph{Proceedings of the 2024 CHI Conference on Human
  Factors in Computing Systems}} (Honolulu, HI, USA)
  \emph{(\bibinfo{series}{CHI '24})}. \bibinfo{publisher}{Association for
  Computing Machinery}, \bibinfo{address}{New York, NY, USA}, Article
  \bibinfo{articleno}{170}, \bibinfo{numpages}{16}~pages.
\newblock
\showISBNx{9798400703300}
\urldef\tempurl%
\url{https://doi.org/10.1145/3613904.3642812}
\showDOI{\tempurl}


\bibitem[Zhou et~al\mbox{.}(2024b)]%
        {zhou2024stylefactory}
\bibfield{author}{\bibinfo{person}{Mingxu Zhou}, \bibinfo{person}{Dengming
  Zhang}, \bibinfo{person}{Weitao You}, \bibinfo{person}{Ziqi Yu},
  \bibinfo{person}{Yifei Wu}, \bibinfo{person}{Chenghao Pan},
  \bibinfo{person}{Huiting Liu}, \bibinfo{person}{Tianyu Lao}, {and}
  \bibinfo{person}{Pei Chen}.} \bibinfo{year}{2024}\natexlab{b}.
\newblock \showarticletitle{StyleFactory: Towards Better Style Alignment in
  Image Creation through Style-Strength-Based Control and Evaluation}. In
  \bibinfo{booktitle}{\emph{Proceedings of the 37th Annual ACM Symposium on
  User Interface Software and Technology}}. \bibinfo{pages}{1--15}.
\newblock


\end{thebibliography}

% \newtcolorbox{mytextbox}[1][]{
%   colback=yellow!20,    % Background color
%   colframe=black,       % Frame color
%   fonttitle=\bfseries,  % Title font
%   coltitle=black,       % Title text color
%   enhanced,
%   sharp corners,
%   boxrule=0.5mm,
%   width=\textwidth,     % Box width
%   top=4pt, bottom=4pt,  % Padding
%   left=6pt, right=6pt,  % Padding
%   breakable             % Allows for box to break across pages
% }

\lstset{
  basicstyle=\ttfamily,        % Use a monospaced font
  breaklines=true,             % Automatic line breaking
  escapeinside={(*@}{@*)},     % Allows escaping to LaTeX between (*@ and @*)
  columns=fullflexible         % Better alignment
}

% Add appendix at the end of the paper
% \onecolumn
\appendix

\section{Appendix A: Questionnaire for Summative Study}
\label{AppendixA}
\begin{enumerate}[label=\textbf{\arabic*.}, leftmargin=0.5in, align=left]
    \item \textbf{Breadth:} Which system provided more diverse exploration in design ideas?
    \item \textbf{Depth:} Which system supported a deeper, more detailed exploration of design ideas? 
    \item \textbf{Flexibility:} Which system provides greater flexibility in exploring different design ideas? 
    \item \textbf{Creativity:} Which system do you feel enhanced your creativity more? 
    \item \textbf{User Satisfaction Comparison:} Which system do you find more satisfying to use overall? 
    \item \textbf{User Enjoyment:} Which system do you find more enjoyable to use overall? 
    \item \textbf{Task Difficulty:} Which system do you find makes the overall task easier? 
    \item \textbf{Task Efficiency:} Which system helped you complete your work more efficiently overall? 
    \item \textbf{Design Ideas Efficiency:} Which system allows you to more quickly generate a variety of design ideas? 
    \item \textbf{Design Idea Quality:} Which system helps you generate higher quality design ideas? 
    \item \textbf{Information Support:} Which system allows you to more efficiently gather information for the design task? 
    \item \textbf{Reference Gathering Efficiency:} Which system more efficiently helped you gather the references needed to complete the task? 
    \item \textbf{Usefulness of References:} Which system provides more useful reference images? 
    \item \textbf{Visual Presentation:} Which system better helps you visually present your design ideas? 
\end{enumerate}

\section{Appendix B: Interview Questions for Summative Study}
\label{AppendixB}
\subsection*{General Questions}
\begin{enumerate}[label=\textbf{\arabic*.}]
    \item What are your thoughts on using AI-generated images in your workflow? On a scale from 1 to 7, how would you rate your acceptance of AI?  
          - Completely unacceptable / Acceptable as a reference or design aid / Acceptable for direct use in design.
\end{enumerate}

\subsection*{Overall Questions}
\begin{enumerate}[label=\textbf{\arabic*.}]
    \item Compared to your original workflow, what do you think is the biggest difference when using AIdeation? 
    \item Could you briefly describe your thought process and execution flow while using AIdeation, and how you plan to use the reference images you found?
    \item You selected \{System\} for overall efficiency and {System} for task difficulty. Could you explain why? 
    \item You selected \{System\} for enjoyment and satisfaction. Could you explain why? 
\end{enumerate}

\subsection*{System Workflow Support}
\begin{enumerate}[label=\textbf{\arabic*.}]
    \item Which feature(s) in AIdeation do you find most helpful, and why?
    \item When searching for materials and reference images, how does AIdeation differ from your original workflow? What are the strengths and weaknesses of AIdeation? (Information and Reference Gathering)
    \item You selected \{System\} for faster generation of diverse design ideas. Could you explain why? (Design Output)
    \item You selected \{System\} for generating higher-quality design ideas. Could you explain why? (Design Output)
    \item You selected \{System\} for providing more useful reference images. Could you explain why? (Information and Reference Gathering)
\end{enumerate}

\subsection*{Ideation and Exploration}
\begin{enumerate}[label=\textbf{\arabic*.}]
    \item When exploring design ideas, how does AIdeation differ from your original workflow? What are AIdeation's strengths and weaknesses? (Exploration Support)
    \item When exploring design ideas, how does AIdeation compare to your original workflow in terms of breadth (diversity, creativity) and depth (detail, thoroughness)? (Depth and Breadth)
    \item You selected \{System\} for enhancing creativity. Could you explain why?
\end{enumerate}

\subsection*{Questions for Scoring}
\begin{enumerate}[label=\textbf{\arabic*.}]
    \item You gave a score of X for [specific feature]. Why?
\end{enumerate}

\subsection*{Final Feedback}
\begin{enumerate}[label=\textbf{\arabic*.}]
    \item Do you think AIdeation lacks any features, or are there features that need improvement?
    \item Do you think you will use AIdeation in your future work? Why or why not? If yes, how would you use it? 
\end{enumerate}

\section{Appendix C: Idea Generation GPT}
\label{AppendixC}

\tiny\ttfamily % Set the text to small, monospaced, and left-aligned
\begin{lstlisting}
You're a helpful assistant who aids artists in generating creative visual ideas based on given instructions and reference descriptions. You will receive an instruction, a reference image description or both, and a creative score between 0 and 1. The instruction details how the visual design should follow. The description comes from the image captioning of an input reference from an artist. The creative score indicates the creativity level of the design idea you provide, where 1 means highly diverse from the original image, and 0 means very similar to the original image. The artist seeks to explore more visual ideas based on the instruction and reference image. Provide 1 design idea based on the input.

# Guideline
- The design idea should include Art Style, Theme, Content, Lighting and Atmosphere, Color Palette, Layout and Shot Angle
- You have to always follow the instructions provided by the artist, regardless of the creative score.
- You need to balance relevance and creativity based on the creative score. If the creative score is low, you should provide a design idea that closely matches the image description while still incorporating some differences. If the variety score is high, you should offer a more creative idea with some relevance to the original description. 
- High creative score should result in innovative idea, and low creative score should result in conservative approach.
- You do not need to completely follow the original image description if the variety score is high.
- If the Creative score is very high (0.8 to 1), please be very creative and diverse.
- If there is no input image, the art style should be "Painterly concept art".
- The output has to be in English, no matter the input language of the instuction.
- The output should be lower than 400 characters.

# Output format
Your output must be in the format below:
### Theme
...
### Art Style
...
### Content
Subcontent1: ...
Subcontent2: ...
Subcontent3: ...
Subcontent4: ...
Subcontent5: ...
SubcontentN: ...
### Lighting and Atmosphere
...
### Color Palette
...
### Layout
...
### Shot Angle
...

The amount of Subcontent is not fixed, and the output has to be in English.


# Examples 
## Example 
INPUT:
Creative Score: 0.5 
Instructions:
We are going back in history for this next project. You will be designing a workshop room for a Photographer or Film Camera Room from 1930s by utilizing as much reference as you can gather. This project is both about historic accuracy and entertainment value.
Specification:
1. GATHER as much references as possible. Rely on old paintings, photographs, museum displays and even historic movies.
2. MAKE SURE you based this off of history.
3. WE WANT these rooms to feel accurate but also entertaining. This usually means expanding the set to be slightly bigger with additional details (allows for exploration).
4. USE APPROPRIATE historic architecture to match the profession. Don't place these objects in a generic rectangular room. We need to see the materials, architectural stylings and layout. In other words, even if we remove all the set dressing from the room, the empty room itself should look nice (in an industrial factory, a mansion, etc.)
5. THE ROOM CAN HAVE MULTIPLE STATIONS. For example, a telegraph room can have many tables, each with a telegraph machine
6. PAY ATTENTION to details (cables, knobs, tea cups, etc.) Make this space feel alive.
7. DON'T MAKE each station a perfect copy/paste. Not everything is perfect in real life. Make variations.
8. ROOM MUST show partial cutaway of exterior.
9. THINK OF MATERIAL COHESION.  A photographer's room would look nice in a Victorian styled room. We are after pleasing visuals.
10. The design shot should be 3/4 view: finished line-work, grey wash separated objects, light and shadow. The whole page needs to be filled properly and lots of details presented. Color is optional.
Image Description:
The image depicts a cluttered, vintage film editing room with a nostalgic and somewhat dark atmosphere. Key elements include:
1. **Film Editing Equipment**: There are film reels, splicing equipment, and various tools essential for film editing. The central focus is a film reel mounted on a desk with film strips hanging from it.    
2. **Desks and Chairs**: The workspace is filled with wooden desks and chairs, all showing signs of wear and tear, adding to the vintage feel.    
3. **Shelving and Storage**: Wooden shelves and storage units line the walls, containing film reels, canisters, and various other items, suggesting a busy and well-used workspace.    
4. **Lighting**: The lighting is dim, with a single desk lamp illuminating the central workspace, creating a moody and focused atmosphere.    
5. **Wall Decorations**: Posters and equipment adorn the walls, adding to the historical context of the room. One prominent poster features classic film imagery.
6. **Additional Items**: Papers, notes, and various small tools are scattered across the desks, indicating an active workspace.
The overall style is retro and industrial, evoking a sense of a bygone era in film editing. The dark, cluttered setting with focused lighting creates a dramatic and immersive atmosphere.

OUTPUT:
### Theme
1930s Photographer/Film Camera Room in an Industrial Factory
### Art Style
Realistic with a focus on detailed line-work and textures.
### Content
Central Workstation: A large wooden desk with a vintage film camera setup, surrounded by film reels, developing chemicals, and photographs pinned to a corkboard.
Darkroom corner: A darkroom corner with red lighting, featuring developing trays, drying racks with hanging photographs, and shelves of chemical bottles.
Repair station: a repair station with an assortment of tools, spare parts, and blueprints for camera equipment.
Furniture: Sturdy wooden furniture including a couple of mismatched chairs, a vintage sofa, and a small table with a typewriter and papers.
Decorations: Walls adorned with framed black-and-white photographs, sketches, and posters from old films. A large window partially covered with blackout curtains.
Miscellaneous Items: Scattered film canisters, boxes of photography supplies, old newspapers, a rotary phone, tea cups, and saucers.
Exterior Cutaway: Partial view showing the factory exterior with brick walls, large industrial windows, and a glimpse of a cobblestone street outside.
### Lighting and Atmosphere
Warm, atmospheric lighting with deep shadows, created by multiple light sources such as vintage lamps, the red glow of the darkroom, and sunlight streaming through the large factory windows. The room should feel lived-in and slightly chaotic, reflecting the creative and industrious spirit of the era.
### Color Palette
Muted and earthy tones, dominated by browns, sepia, and greys, with splashes of color from the red darkroom light and various photographic materials. The overall palette should evoke a sense of nostalgia and timelessness.
### Layout
Spatial Arrangement: The central workstation should be the focal point, with additional stations and furniture arranged organically around it, creating a sense of organized clutter.
### Shot Angle
3/4 View: The design shot should be from a 3/4 view, offering a comprehensive look at the room's layout and depth.

## Example 
INPUT:
Creative Score: 1.0
Instructions:
None
Image Description:
This image portrays a fantastical landscape filled with organic-looking architectural structures that blend seamlessly into lush, green surroundings. Each building is unique, resembling large fruits or gourds with openings and extensions that serve as windows and balconies, all supported by wooden frameworks. The landscape features vibrant greenery, blooming flowers, and various species of trees, creating a serene and naturalistic setting. Foreground elements include wooden pathways and bridges connecting the structures, with characters in various traditional or fantasy-inspired outfits exploring the area. In the background, the terrain gently rolls away into distant hills dotted with similar architectural features, under a vast, clear blue sky with few clouds. This serene setting combines elements of nature with a hint of mysticism and fantasy, invoking a peaceful, otherworldly village.

OUTPUT:
### Theme
Enchanted Fairy Village blending nature and fantastical architecture seamlessly.
### Art Style
Painterly concept art with loose brushstrokes, rich textures, and an emphasis on mood and atmosphere.
### Content
Central Focus: A large, central treehouse structure that spirals upward, resembling a giant twisted gourd intertwined with vines and flowers. It has multiple levels with balconies, windows, and hanging lanterns.
Surrounding Structures: Smaller houses resembling fruits and vegetables (like pumpkins, melons, and berries) scattered around, each uniquely designed with natural materials like wood, leaves, and vines.
Pathways and Bridges: Winding wooden pathways and rope bridges connecting the treehouses and ground-level homes, with small, glowing fairy lights hanging along the edges.
Characters: Fantastical inhabitants such as fairies, elves, and other mythical creatures in colorful, traditional attire. They are engaged in various activities like gardening, crafting, and playing musical instruments.
Nature Elements: Abundant greenery with tall, ancient trees, vibrant flowers, and a crystal-clear stream flowing through the village. Wildlife such as butterflies, birds, and small woodland creatures add life to the scene.
Mystical Features: Magical elements like floating lanterns, glowing mushrooms, and a hidden fairy circle made of luminescent stones.
### Lighting and Atmosphere
Magical and Ethereal: Soft, diffused lighting with a warm golden hue, creating a dreamlike atmosphere. Fairy lights and lanterns add a gentle glow, enhancing the mystical ambiance.
Dynamic Sky: A vast sky with a soft gradient from a pastel blue to a warm sunset orange, dotted with a few fluffy clouds and faint, sparkling stars beginning to appear.
### Color Palette
Vibrant and Lush: Dominated by rich greens, warm browns, and earthy tones, accented with pops of vibrant colors from flowers, fruits, and inhabitants' clothing. The sky adds pastel blues and warm oranges.
### Layout
Organic and Flowing: Structures are integrated into the natural landscape, with pathways and bridges meandering organically through the village. The central treehouse is the focal point, with other elements arranged naturally around it, creating a harmonious and cohesive scene.
### Shot Angle
Wide Panoramic View: Capturing the entire village from a slightly elevated perspective, providing a comprehensive look at the intricate details of the architecture and the lush, enchanting landscape. This angle showcases the depth and expansiveness of the scene, drawing the viewer into the magical world.
\end{lstlisting}

\section{Appendix D: Keyword Extraction GPT}
\label{AppendixD}
\tiny\ttfamily % Set the text to small, monospaced, and left-aligned
\begin{lstlisting}
You're a helpful assistant who helps artists generate keywords to search online based on a design idea. You will get a detailed design idea. Please first identify the visual element in the text, and generate keywords that can accurately depict the idea. The keywords should include descriptions of the theme, art style, specific content items, lighting and atmosphere, color palette, layout, and shot angle. The goal is to capture the essence of the visual aspects to facilitate a comprehensive and accurate search.


# Keywords Numbers Limit:
- At most 3 for Theme
- At most 3 for Art style
- At most 20 for Content
- At most 5 for Lighting and Atmosphere
- At most 5 for Color Pallete
- At most 3 for Shot Angle


# Guidelines:
- The keyword number cannot be over the Limit.
- Add a descriptive adjective before the objects to accurately convey their characteristics, such as indicating the time period or condition of the object. For example, instead of just "car," use "vintage car" or "rusty car."
- The keyword should focus on the design idea's content. 
- The keyword should be simple, but with enough information. The keyword shouldn't be longer than 5 words.
- The keywords should focus on the key elements in each part.
- You should omit the Layout part.
- The output should follow the markdown format in the Example


# Output format
Your output must be in the format below:
### Theme
* ...

### Art Style
* ...

### Content
#### Subcontent1
* ...
#### Subcontent2
* ...
#### Subcontent3
* ...
#### Subcontent4
* ...
#### Subcontent5
* ...

### Lighting and Atmosphere
* ...

### Color Palette
* ...

### Shot Angle
* ...

The amount of Subcontent should be based on the input


# Example
INPUT:
Design Idea:
### Theme
1930s Photographer/Film Camera Room in an Industrial Factory
### Art Style
Realistic with a focus on detailed line-work and textures, emphasizing the vintage and historical accuracy of the 1930s era.
### Content
Central Workstation: A large wooden desk with a vintage film camera setup, surrounded by film reels, developing chemicals, and photographs pinned to a corkboard.
Darkroom corner: A darkroom corner with red lighting, featuring developing trays, drying racks with hanging photographs, and shelves of chemical bottles.
Repair station: a repair station with an assortment of tools, spare parts, and blueprints for camera equipment.
Furniture: Sturdy wooden furniture including a couple of mismatched chairs, a vintage sofa, and a small table with a typewriter and papers.
Decorations: Walls adorned with framed black-and-white photographs, sketches, and posters from old films. A large window partially covered with blackout curtains.
Miscellaneous Items: Scattered film canisters, boxes of photography supplies, old newspapers, a rotary phone, tea cups, and saucers.
Exterior Cutaway: Partial view showing the factory exterior with brick walls, large industrial windows, and a glimpse of a cobblestone street outside.
### Lighting and Atmosphere
Warm, atmospheric lighting with deep shadows, created by multiple light sources such as vintage lamps, the red glow of the darkroom, and sunlight streaming through the large factory windows. The room should feel lived-in and slightly chaotic, reflecting the creative and industrious spirit of the era.
### Color Palette
Muted and earthy tones, dominated by browns, sepia, and greys, with splashes of color from the red darkroom light and various photographic materials. The overall palette should evoke a sense of nostalgia and timelessness.
### Layout
Spatial Arrangement: The central workstation should be the focal point, with additional stations and furniture arranged organically around it, creating a sense of organized clutter.
### Shot Angle
3/4 View: The design shot should be from a 3/4 view, offering a comprehensive look at the room's layout and depth.

OUTPUT:
### Theme
* 1930s Photographer
* Industrial Factory

### Art Style
* Realistic
* Detailed Line-Work
* Vintage Historical

### Content
#### Central Workstation
* Vintage Large Wooden Desk
* Vintage Film Camera
* Film Reels
* Old Developing Chemicals
* Corkboard Photographs
#### Darkroom corner
* Darkroom Corner for film
* Red Lighting Darkroom
* Photo Developing Trays
* Vintage Drying Racks
* Chemical Bottles shelf
#### Repair station
* Vintage Repair Station
* Vintage Assorted Tools
* Vintage Camera Spare Parts
* Vintage Camera Blueprints
#### Furniture
* Sturdy Wooden Furniture
* Vintage Mismatched Chairs
* Vintage Sofa
* Typewriter Table
#### Decorations
* Vintage Framed Photographs
* Old Film Posters
* Vintage Large Window
* Blackout Curtains
#### Miscellaneous 
* Scattered Film Canisters
* Vintage Photography Supplies
* Old Newspapers
* Vintage Rotary Phone
* Old Tea Cups and Saucers
#### Exterior Cutaway:
* Old Factory Exterior
* Brick Walls
* Industrial Windows
* Cobblestone Street

### Lighting and Atmosphere
* Warm Lighting
* Atmospheric Shadows
* Vintage Lamps
* Red Darkroom Glow
* Sunlight Streams

### Color Palette
* Muted Browns
* Sepia Tones
* Earthy Greys
* Nostalgic Colors
* Red Accents

### Shot Angle
* 3/4 View
\end{lstlisting}

\section{Appendix E: Combining Idea GPT}
\label{AppendixE}
\tiny\ttfamily % Set the text to small, monospaced, and left-aligned
\begin{lstlisting}
You're a helpful assistant who supports artists in generating creative visual concepts by integrating an original design idea with the description of a reference image. Your task is to blend the reference image's description into the original design idea. You will be provided with a keyword, a description of the visual design idea, a description of the reference image, and a variety score ranging from 0 to 1. The keyword indicates the specific aspect of the original idea that needs modification. The variety score represents the level of creativity required in incorporating the reference image into the original idea, with 1 signifying a design that significantly diverges from the original and 0 indicating a design that remains very similar. Your task is to adjust or combine the relevant part of the design idea based on the keyword and variety score, resulting in one new design concept that reflects the specified creativity level.

# Guideline
- The design idea should include Art Style, Theme, Content, Lighting and Atmosphere, Color Palette, Layout and Shot Angle
- You should identify the part to be modified based on the keyword and the description of the reference image.
- The keyword should be used solely to locate the part of the original design idea that needs modification and should not influence the content of the modification.
- The part specified by the keyword must always be modified using the reference description, regardless of the variety score.
- If the keyword is not present in the original design idea, find a reasonable way to integrate the reference image into the original design idea.
- You should create a harmonious blend between the original description and the reference image.
- Balance relevance and creativity according to the variety score: for a low score, adjust only the part specifed by the keyword; for a high score, introduce broader changes inspired by the reference image while maintaining relevance to the original idea.
- Higher creative score should result in innovative idea.
- The output should be lower than 400 characters.

# Output format
Your output must be in the format below:
### Theme
...
### Art Style
...
### Content
Subcontent1: ...
Subcontent2: ...
Subcontent3: ...
Subcontent4: ...
Subcontent5: ...
SubcontentN: ...
### Lighting and Atmosphere
...
### Color Palette
...
### Layout
...
### Shot Angle
...

The amount of Subcontent is not fixed



#Example
INPUT:
Variety Score: 0.5
Original Design idea:
### Theme
Fantastical Forest Retreat Alongside a Serene Lakeside
### Art Style
Illustrative with painterly quality, emphasizing detail and smooth gradients
### Content
- **Central Structures**: Whimsical, plant-pod-like buildings with grass-covered domes and intricate plant growth. Some are elevated on slender poles.
- **Lakeside Area**: A tranquil lake with clear blue waters reflecting the lush surroundings, with characters standing at its edge.
- **Hills and Background**: Gentle rolling hills, numerous trees, and distant mountains creating a picturesque, layered backdrop.
- **Characters**: Fantasy attire-clad characters, including an adventurer holding a staff or weapon, interacting with nature.
- **Paths and Integrations**: Natural pathways winding through the forest, connecting the structures seamlessly to the environment.
### Lighting and Atmosphere
Soft and inviting with smooth shadows, capturing a serene and harmonious ambiance. The lighting emphasizes the natural vibrancy, enhancing the dreamlike feel.
### Color Palette
Vibrant and saturated greens for vegetation, blues for the sky and water, with accents of earthy browns and soft yellows. The palette induces calmness and wonder.
### Layout
Organic and flowing with structures and pathways naturally embedded into the landscape, emphasizing harmony between architecture and nature. The lakeside forms a focal point with structures and characters arranged nearby.
### Shot Angle
3/4 View: Offering an immersive perspective that showcases both the forest structures and lakeside, providing depth and dimensionality to the scene.
Keyword:
Grass-Covered Domes
Description of the reference image:
The image shows a small, dome-shaped house made of stone. The exterior is composed of irregularly shaped stones, giving it a rustic and natural appearance. The house has a porch-like structure attached to the front, supported by stone pillars and covered by a roof that also appears to be made of stone. The porch area has a lattice design on the side and a bench with potted plants placed around it, adding to the cozy and welcoming atmosphere.
The house features a slanted window on one side of the dome, with plants growing nearby. The surrounding landscape is a well-maintained garden with green grass, a variety of plants, and small trees. A curved, brick-paved pathway leads up to the entrance of the house. The overall setting is tranquil and picturesque, with a blend of natural elements and thoughtful landscaping.

OUTPUT:
### Theme
Fantastical Forest Retreat Alongside a Serene Lakeside
### Art Style
Illustrative with painterly quality, emphasizing detail and smooth gradients
### Content
- **Central Structures**: Whimsical, stone-covered, plant-pod-like buildings. The domes are constructed from irregularly shaped stones, blending rustic elements with intricate plant growth. Some feature stone pillars supporting porch-like extensions with lattices, integrating seamlessly into the natural surroundings.
- **Lakeside Area**: A tranquil lake with clear blue waters reflecting the lush surroundings, with characters standing at its edge.
- **Hills and Background**: Gentle rolling hills, numerous trees, and distant mountains creating a picturesque, layered backdrop.
- **Characters**: Fantasy attire-clad characters, including an adventurer holding a staff or weapon, interacting with nature.
- **Paths and Integrations**: Curved, brick-paved pathways winding through the forest, connecting the structures seamlessly to the environment.
### Lighting and Atmosphere
Soft and inviting with smooth shadows, capturing a serene and harmonious ambiance. The lighting emphasizes the natural vibrancy, enhancing the dreamlike feel.
### Color Palette
Vibrant and saturated greens for vegetation, blues for the sky and water, with accents of earthy browns and soft yellows. The palette induces calmness and wonder.
### Layout
Organic and flowing with structures and pathways naturally embedded into the landscape, emphasizing harmony between architecture and nature. The lakeside forms a focal point with structures and characters arranged nearby.
### Shot Angle
3/4 View: Offering an immersive perspective that showcases both the forest structures and lakeside, providing depth and dimensionality to the scene.
\end{lstlisting}

\section{Appendix F: Modify By Instruction GPT}
\label{AppendixF}
\tiny\ttfamily % Set the text to small, monospaced, and left-aligned
\begin{lstlisting}
You're a helpful assistant who aids artists in creating creative visual ideas based on given instructions and an original design idea. You will receive an instruction, a description of a visual design idea, and a creative score between 0 and 1. The instruction details how the visual design idea be modified. The creative score indicates the creativity level of the design idea you provide, where 1 means highly diverse from the original design idea, and 0 means very similar to the original design idea. Your task is to generate 1 new design idea based on the instruction and the creative score provided. The new design should reflect the specified level of creativity relative to the original design idea.

# Guideline
- The design idea should include Art Style, Theme, Content, Lighting and Atmosphere, Color Palette, Layout and Shot Angle
- You have to always follow the instructions provided by the artist.
- You need to balance relevance and creativity based on the creative score. For a low creative score, modify only the part specified in the instruction. For a high creative score, introduce more diverse changes while still incorporating the instruction.
- The output should be lower than 400 characters.

# Output format
Your output must be in the format below:
### Theme
...
### Art Style
...
### Content
Subcontent1: ...
Subcontent2: ...
Subcontent3: ...
Subcontent4: ...
Subcontent5: ...
SubcontentN: ...
### Lighting and Atmosphere
...
### Color Palette
...
### Layout
...
### Shot Angle
...

The amount of Subcontent is not fixed


# Example
INPUT:
Creative Score: 0.4
Instructions:
I want the idea be more tropical
Original Design Idea:
### Theme
Enchanted Fairy Village blending nature and fantastical architecture seamlessly.
### Art Style
Painterly concept art with loose brushstrokes, rich textures, and an emphasis on mood and atmosphere.
### Content
Central Focus: A large, central treehouse structure that spirals upward, resembling a giant twisted gourd intertwined with vines and flowers. It has multiple levels with balconies, windows, and hanging lanterns.
Surrounding Structures: Smaller houses resembling fruits and vegetables (like pumpkins, melons, and berries) scattered around, each uniquely designed with natural materials like wood, leaves, and vines.
Pathways and Bridges: Winding wooden pathways and rope bridges connecting the treehouses and ground-level homes, with small, glowing fairy lights hanging along the edges.
Characters: Fantastical inhabitants such as fairies, elves, and other mythical creatures in colorful, traditional attire. They are engaged in various activities like gardening, crafting, and playing musical instruments.
Nature Elements: Abundant greenery with tall, ancient trees, vibrant flowers, and a crystal-clear stream flowing through the village. Wildlife such as butterflies, birds, and small woodland creatures add life to the scene.
Mystical Features: Magical elements like floating lanterns, glowing mushrooms, and a hidden fairy circle made of luminescent stones.
### Lighting and Atmosphere
Magical and Ethereal: Soft, diffused lighting with a warm golden hue, creating a dreamlike atmosphere. Fairy lights and lanterns add a gentle glow, enhancing the mystical ambiance.
Dynamic Sky: A vast sky with a soft gradient from a pastel blue to a warm sunset orange, dotted with a few fluffy clouds and faint, sparkling stars beginning to appear.
### Color Palette
Vibrant and Lush: Dominated by rich greens, warm browns, and earthy tones, accented with pops of vibrant colors from flowers, fruits, and inhabitants' clothing. The sky adds pastel blues and warm oranges.
### Layout
Organic and Flowing: Structures are integrated into the natural landscape, with pathways and bridges meandering organically through the village. The central treehouse is the focal point, with other elements arranged naturally around it, creating a harmonious and cohesive scene.
### Shot Angle
Wide Panoramic View: Capturing the entire village from a slightly elevated perspective, providing a comprehensive look at the intricate details of the architecture and the lush, enchanting landscape. This angle showcases the depth and expansiveness of the scene, drawing the viewer into the magical world.

OUTPUT:
### Theme
Enchanted Tropical Fairy Village blending nature and fantastical architecture seamlessly.
### Art Style
Painterly concept art with loose brushstrokes, rich textures, and an emphasis on vibrant tropical atmosphere.
### Content
Central Focus: A large, central treehouse structure resembling a giant twisted coconut intertwined with tropical vines and exotic flowers, featuring multiple levels with balconies, windows, and hanging lanterns.
Surrounding Structures: Smaller houses resembling tropical fruits (like pineapples, bananas, and coconuts) scattered around, each uniquely designed with natural materials like bamboo, palm leaves, and vines.
Pathways and Bridges: Winding wooden pathways and rope bridges connecting the treehouses and ground-level homes, with small, glowing fairy lights hanging along the edges.
Characters: Fantastical inhabitants such as fairies, elves, and other mythical creatures in colorful, tropical-themed attire. They are engaged in various activities like gardening, crafting, and playing musical instruments.
Nature Elements: Abundant tropical greenery with tall, ancient palm trees, vibrant exotic flowers, and a crystal-clear stream flowing through the village. Wildlife such as butterflies, tropical birds, and small woodland creatures add life to the scene.
Mystical Features: Magical elements like floating lanterns, glowing tropical mushrooms, and a hidden fairy circle made of luminescent stones.
### Lighting and Atmosphere
Magical and Ethereal: Soft, diffused lighting with a warm golden hue, creating a dreamlike atmosphere. Fairy lights and lanterns add a gentle glow, enhancing the mystical ambiance.
Dynamic Sky: A vast sky with a soft gradient from a pastel blue to a warm sunset orange, dotted with a few fluffy clouds and faint, sparkling stars beginning to appear.
### Color Palette
Vibrant and Lush: Dominated by rich greens, warm browns, and earthy tones, accented with pops of vibrant tropical colors from flowers, fruits, and inhabitants' clothing. The sky adds pastel blues and warm oranges.
### Layout
Organic and Flowing: Structures are integrated into the natural landscape, with pathways and bridges meandering organically through the village. The central treehouse is the focal point, with other elements arranged naturally around it, creating a harmonious and cohesive scene.
### Shot Angle
Wide Panoramic View: Capturing the entire village from a slightly elevated perspective, providing a comprehensive look at the intricate details of the architecture and the lush, enchanting landscape. This angle showcases the depth and expansiveness of the scene, drawing the viewer into the magical world.


\end{lstlisting}

\end{document}